\documentclass{aastex63}

\usepackage{natbib}
\usepackage{CJK}
\usepackage{longtable}
\usepackage{amsmath}
\usepackage{multirow}
\usepackage{subfigure}
\usepackage{graphicx}
\usepackage{txfonts}
\usepackage{epsfig}
\usepackage{anyfontsize}
\usepackage{color,xcolor}

\makeatother

\submitjournal{ApJS}

\shorttitle{Molecules in CRL\,2688}
\shortauthors{Qiu et al.}

\begin{document}
\begin{CJK*}{UTF8}{gkai}

\title{Molecules in the carbon-rich protoplanetary nebula CRL\,2688}

\correspondingauthor{Yong Zhang, Jiangshui Zhang}
\email{Email: zhangyong5@mail.sysu.edu.cn, jszhang@gzhu.edu.cn}

\author[0000-0002-9829-8655]{Jian-Jie Qiu(邱建杰)}
\affiliation{School of Physics and Astronomy, 
                Sun Yat-sen University, 2 Da Xue Road, 
                Tangjia, Zhuhai 519000, 
                Guangdong Province, China}

\author[0000-0002-1086-7922]{Yong Zhang(张泳)}
\affiliation{School of Physics and Astronomy, 
                Sun Yat-sen University, 
                2 Da Xue Road, 
                Tangjia, Zhuhai 519000, 
                Guangdong Province, China}
\affiliation{Laboratory for Space Research, The University of Hong Kong, Hong Kong, China}

\author{Jiang-Shui Zhang(张江水)}
\affiliation{Center For Astrophysics, 
                 Guangzhou University, 
                 230 Wai Huan Xi Road, 
                 Guangzhou Higher Education Mega Center, 
                 Guangzhou 510006, 
                 Guangdong Province, China}
\author{Jun-ichi Nakashima(中岛淳一)}
\affiliation{School of Physics and Astronomy, 
                Sun Yat-sen University, 
                2 Da Xue Road, 
                Tangjia, Zhuhai 519000, 
                Guangdong Province, China}

\begin{abstract}

We present { observations} of the carbon-rich protoplanetary nebula (PPN) CRL\,2688 made with the Institut de Radioastronomie Millim{\'e}trique (IRAM) 30 m telescope in the 3\,mm and 2\,mm bands. In total, 196 transition lines belonging to 38 molecular species and isotopologues are detected, among which,
to our best knowledge,  { 153} transition lines and { 13} species are the first report for this object.
Additionally, in order to contribute to future research, we have collected observational data on the molecular lines of CRL\,2688 from the literature and compiled them into a single unified catalog.
We find that the molecular abundance of CRL\,2688 cannot be explained by the standard model of a circumstellar envelope.
The implications of metal-bearing molecules on circumstellar chemistry are discussed.

\end{abstract}

 
\section{Introduction} 
\label{sec:introduction}

Mass loss of asymptotic giant branch (AGB) stars fuels the recycling of materials in the galaxy
\citep[see, e.g.,][for reviews]{Herwig05,Ziurys06,Hofner18}. 
Suffering from complex dynamic and chemical processes, 
the resultant circumstellar envelopes (CSEs) are active { sites} for the formation of substantial amount of dust and molecular gas, making them bright  infrared (IR) and radio sources.
Along the evolution beyond AGB, the CSEs gradually expand, resulting in decreasing density, and 
the central stars would become visible at a certain moment \citep{Bloecker95}. 
With further evolution, the gas in CSEs are partly or fully dissociated and ionized by intense ultraviolet (UV) radiation from the central stars, and emits strong ionic emission lines at optical wavelengths, marking the beginning of a planetary nebula (PN) stage. Passing through the PN evolution, 
the circumstellar matters finally disperse into and contaminate the interstellar medium (ISM). 
Recently, there are increasing observational evidences showing that some molecules are able to survive in the PN evolution and then seed the chemistry in diffuse and molecular clouds \citep{Tenenbaum09,Schmidt16,Schmidt18,zhanga17}. 

The variations of molecular composition during the CSE evolution is a hot topic in astrochemistry \citep{Cernicharo11}. 
So far, more than 70 kinds of molecular species have been detected in circumstellar environments \citep[and references therein]{McGuire18,McGuire21}. 
Many of these detections greatly benefited from molecular line surveys performed with advanced single-dish telescopes.
However, spectral line surveys have been overwhelmingly concentrated on the brightest nearby carbon star IRC+10216 \citep[etc.]{Cernicharo00,Cernicharo10,He08,Patel11,Gong15,Zhang17}. 
In comparison, line surveys of other CSEs, although existing, are far sparser, which would have the risk of 
leading to biased views of circumstellar chemistry. 
It is evidently necessary to extend such observations to CSEs at different evolutionary stages.

Protoplanetary nebulae (PPNs) represent a transient phase before the PN stage when the AGB winds cease \citep{Kwok93}, during which the physical conditions of the CSEs change drastically on spatial and temporal scales. 
Given their relative short evolutionary timescale ($\sim$10$^{3}$\,yr), the PPNs that are suitable for line-survey observations are rare, and thus the chemical processes in this transition phase are far from fully understood. 
CRL\,2688 and CRL\,618 are two ideal PPNs for line-survey observations \citep{pardo07, Park08, Zhang13}. 
Both are carbon rich, but CRL\,618 appears to be closer to the PN stage \citep{Bujarrabal88}. 
The spectra of the two targets exhibit dramatically different molecular species, indicating that quick chemical reactions have taken place during the PPN stage. 
Detailed chemical models of PPNs are practically limited to CRL\,618. 
\citet{Howe92} theoretically investigated the chemistry of PPNs using a interacting stellar winds model, and found that a few small molecules and cations can be formed in the shocked regions, while the molecules originally forming in the AGB stage are quickly destroyed unless clumpy structures exist. 
Based on a model of a dense, warm gaseous slab, irradiated by an intense UV radiation,
\citet{Woods03a} concluded that complex organic chemistry can be driven in the torus
of late PPNs, which can qualitatively account for the radio observations of CRL\,618.
\citet{Cernicharo04} modelled the chemistry in the neutral layers of CRL\,618 and pointed out that large carbon-bearing molecules can be quickly produced by photochemistry.
A general chemical model of late-AGB/early-PPN stage objects was constructed by
\citet{Woods05}, who suggested that the molecule-rich PPNs should be attributed to the existence of a dense torus. 
More observations of PPNs at different evolutionary stages are necessary to constrain these chemical models.

This paper is devoted to investigate the molecular species in CRL\,2688, a prototypical bipolar reflection nebula surrounding a post-AGB star. 
This nebula exhibits a wealth of structures, including a pair of bipolar lobes, an equatorially dense dusty cocoon, a pair of nonuniform annular holes coaxial with the polar axis, and a large number of roughly round arcs \citep{Sahai98b}.
The interferometric observations in the CO $J=2 \rightarrow 1$ line has revealed two distinct high-velocity outflows along the optical axis and close to the equatorial plane, respectively \citep{Cox00}.
A high-velocity outflow is revealed by bright H$_{2}$ emission arising from shocked gas \citep{Sahai98a}. 
Table~\ref{Table1} summarizes the basic parameters of CRL\,2688 and its central star. 
The distances estimated by several authors have a large uncertainty. 
CRL\,2688 is a bright IR source, and most of IR emission originates from the central region with a diameter of 1$\rlap{.}\arcsec$5 and a temperature of about 150\,K \citep{Ney75a}. 
The central dust-obscured post-AGB star shows an F5\,Ia spectrum \citep{Crampton75}. 
Assuming a distance of 1\,kpc, \citet{Ney75a} obtained a total bolometric luminosity of $1.8 \times 10^{4}\,L_{\odot}$. 
Based on a distance of 420\,pc and an optical extinction of A$_{V}$\,mag of 0.8 mag \citep{Ueta06}, \citet{Balick12} derived the luminosity and effective temperature of the central star
to be $5500\pm1100$\,$L_{\odot}$ and 7000\,K, respectively. 
Measuring the overall motions of the lobes at 2 $\mu$m based on the $HST$ NICMOS images from 1997 to 2002, \citet{Ueta06} determined the nebular dynamical age of roughly 350\,yr. 
Using the same approach, but based on the $HST$ images at 0.6\,$\micron$ observed from 2002 to 2009, \citet{Balick12} obtained a dynamical age of about 250\,yr.

One of the key motivations of this work is to investigate the gas-phase molecules in sources exhibiting the so-called 21\,$\mu$m and 30\,$\mu$m unidentified IR features \citep[see][for a recent review]{Volk20}. 
The rare 21\,$\mu$m feature is observed exclusively in carbon-rich PPNs,
indicating its transient nature.
It is always accompanied with the 30\,$\mu$m feature.
\citet{Buss93} found that the IRAS/LRS spectra of CRL\,2688 revealed a weak feature near 20\,$\mu$m, which was later confirmed by
\citet{Omont95}. 
However, the {\it ISO}/SWS spectrum does not show any feature around this wavelength \citep{Volk20}, casting a doubt whether CRL\,2688 is a  21\,$\mu$m source. 
The 21\,$\mu$m sources also exhibit unusually strong 3.4--3.5\,$\mu$m emission, and this is the case for CRL\,2688.
If the 21\,$\mu$m feature is indeed lack in CRL\,2688, \citet{Geballe92} proposed that this PPN may be in a transition between the `21\,$\mu$m' and normal phases. 
Strong 30\,$\mu$m feature has been discovered in the {\it ISO}/SWS spectrum of CRL\,2688 \citep{Hony02}.  \citet{Zhang20aa} found that the relative abundance between carbon-chain and Si-bearing molecules are highly similar between CRL\,2688 and a well-established strong 21\,$\mu$m source, and then inferred that gas-phase molecules may hold key clues about the carrier of the 21\,$\mu$m feature.

The paper is organized as follows. 
In Section \ref{sec2}, we introduce the observations and data reductions. 
In Section \ref{sec3}, we present the results of a abundance analysis. 
In Section \ref{sec4},  we compare the molecules in CRL\,2688 and other CSEs, as well as observations and
modelling results. 
A brief summary is given in Section \ref{sec5}.

\section{Observations and data reductions}
\label{sec2}

The observations toward the central region of CRL\,2688 (RA: 21:02:18.27, DEC: 36:41:37.0, J2000) were performed with the Institutde Radioastronomie Millim{\'e}trique (IRAM) 30m-diameter telescope at the Pico Veleta Observatory (Spain)
 in 2017 January 14. 
We used the Eight Mixer Receiver (EMIR) with dual-polarization and the Fourier Transform Spectrometers (FTS) backend to cover frequency ranges from 75.7--83.5 GHz and 91.4--99.2 GHz in the 3 mm window, and 160.8--168.6 GHz and 176.5--184.3 GHz in the 2 mm window. 
The instantaneous frequency coverage per sideband and polarization is 4 GHz with a frequency channel spacing of 195 kHz. 
The standard wobbler switching mode with a $\pm$ 110$''$ offset at a rate of 0.5 Hz beam throwing was applied. 
The on-source integration times of each band are about two hours.
Pointing was checked frequently with strong millimeter emitting quasi-stellar objects.
The typical system temperatures are 84--112 K in the 3 mm band and 110--540 K in the 2 mm band. 

The antenna temperature $(T_{\rm A})$ is converted to the main beam temperature $(T_{\rm mb})$ with the expression of $T_{\rm mb}=T_{\rm A}\times\frac{F_{\rm eff}}{B_{\rm eff}}$, 
where the forward efficiency ($F_{\rm eff}$) and main-beam efficiency ($B_{\rm eff}$) of the telescope at each band are listed in Table~\ref{Table2}. 
The data are reduced using the CLASS software of the GILDAS package\footnote{http://www.iram.fr/IRAMFR/GILDAS}. 
The spectra at the same frequency coverage were co-added to improve the signal-to-noise ratio 
{according to the weights proportional to $1/\sigma^2$, where $\sigma$ is the root-mean-square uncertainty in
the continuum.}
The identification of each transition line was made by using the information from the NIST database recommended rest frequencies for observed interstellar molecular microwave transitions\footnote{http://pml.nist.gov/cgi-bin/micro/table5/start.pl} \citep{Lovas04}, 
the Cologne Database for Molecular Spectroscopy catalog\footnote{https://cdms.astro.uni-koeln.de} \citep{Muller01,Muller05}, 
and splatalogue database for astronomical spectroscopy\footnote{https://splatalogue.online//advanced.php}.
To validate the detection and identification of a given line, we checked the other transition lines from the same species that are available in the current observations and those in literature.

In order to get a more complete view of the chemistry in CRL\,2688, this work was complemented with observational data reported in the literature. 
We have collected as many results as possible from previous molecular radio observations of CRL\,2688 using single-dish telescopes, except for a few old observations where the line intensity was not given correctly.
For the past reports that did not explicitly give the half-power bandwidths (HPBW) of used telescopes, we take the values derived by the expression of 
$\rm {HPBW}=75531$\rlap{.}\arcsec$2
\left[\frac{\nu}{{\rm GHz}}\right]^{-1}\left[\frac{D}{{\rm m}}\right]^{-1}$, 
where $\nu$ is the rest frequency of the transition lines in units of GHz and $D$ is the diameter of the telescope in units of m. The results obtained from the present observation and those collected from literature are summarized in Table~\ref{Table3}.

\section{Results}
\label{sec3}

\begin{figure}
\gridline{\fig{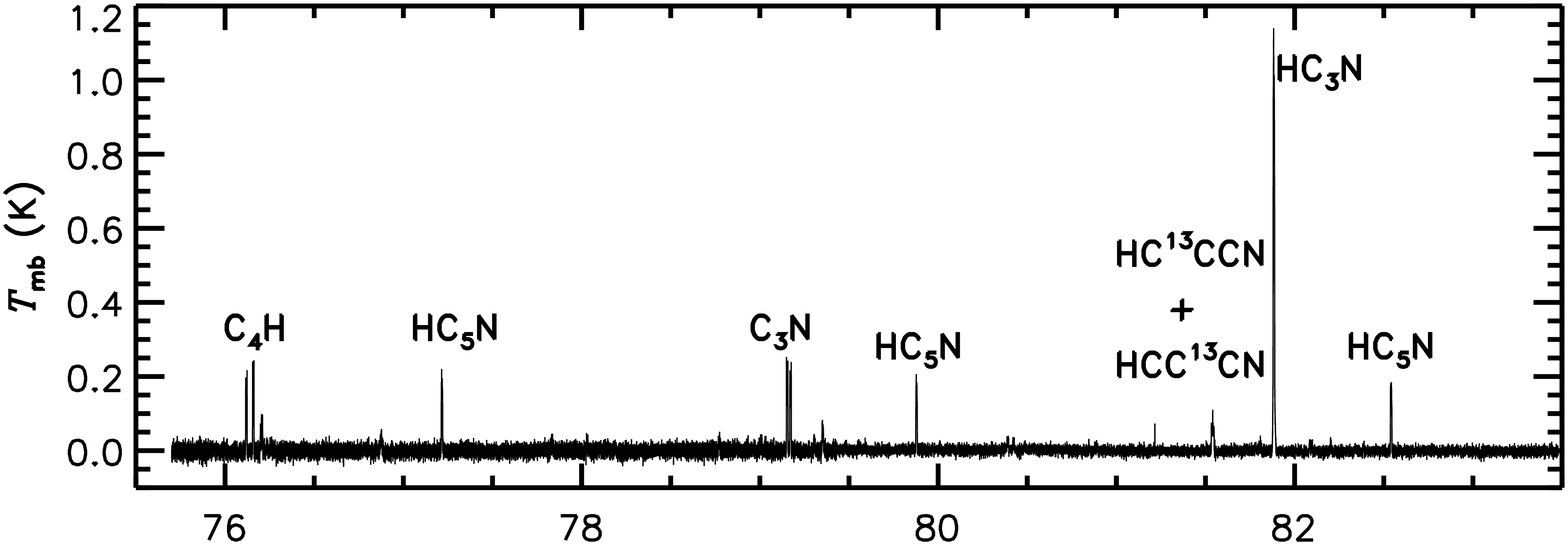}{0.9\textwidth}{}
         }
\vspace{-1.4cm}
\gridline{\fig{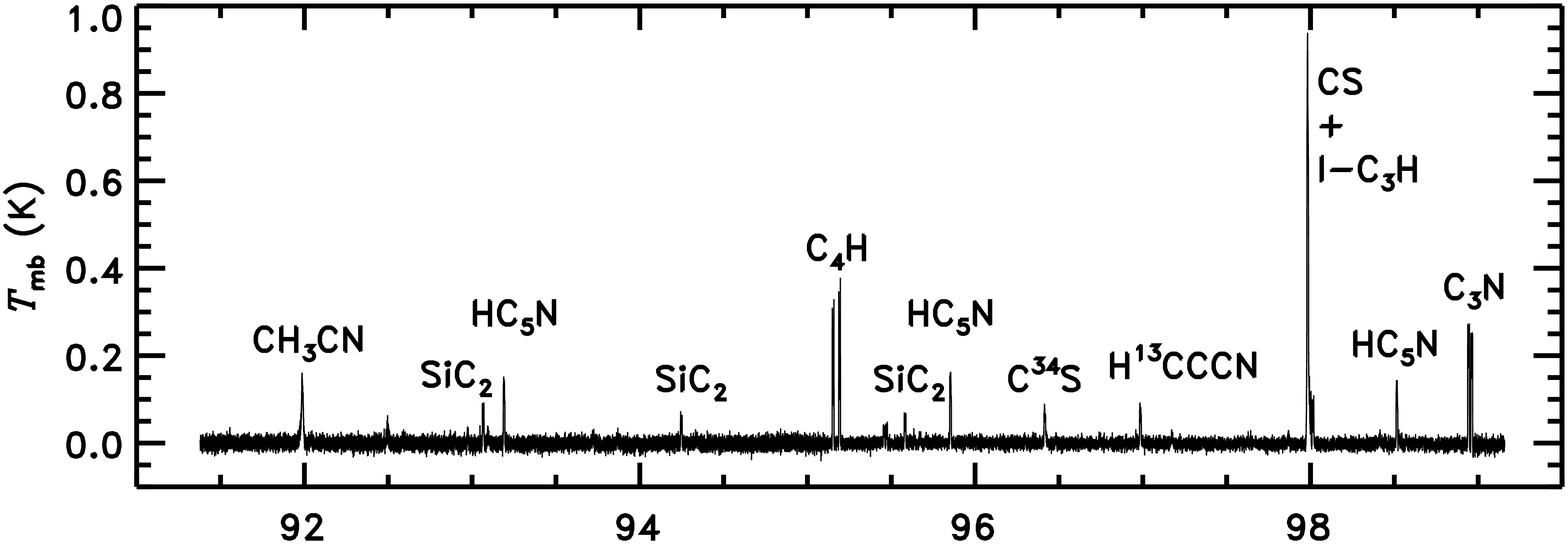}{0.9\textwidth}{}
         }
\vspace{-1.4cm}
\gridline{\fig{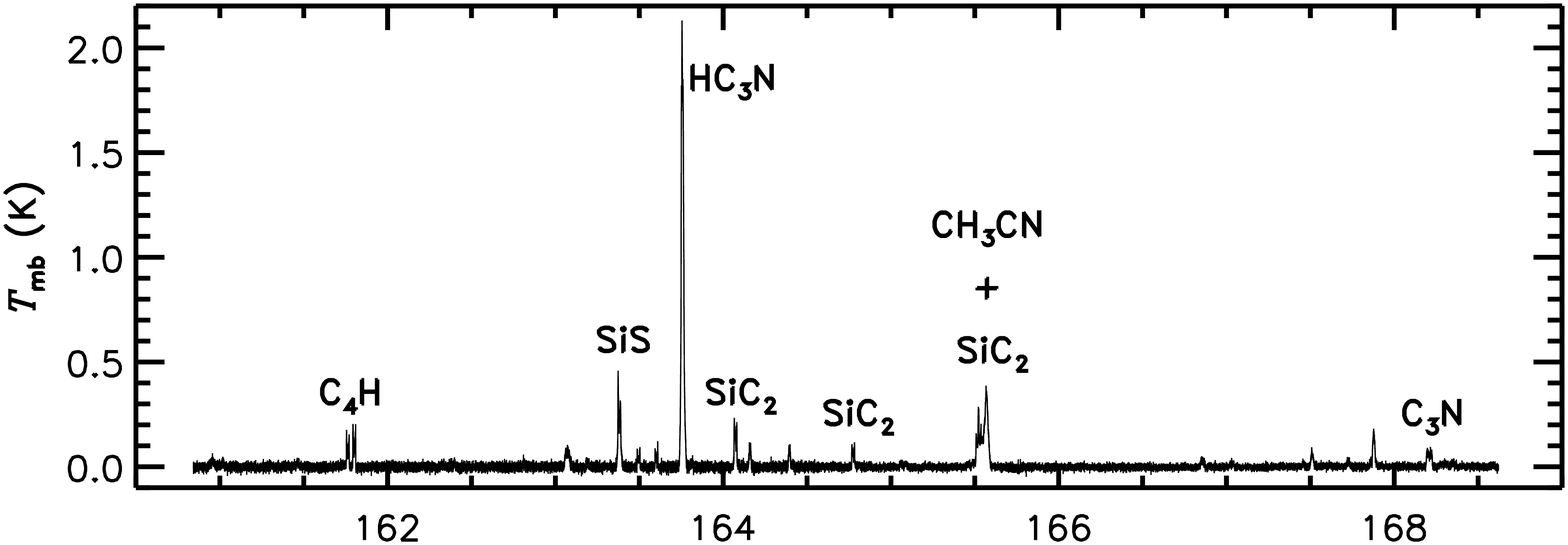}{0.9\textwidth}{}
         }
\vspace{-1.4cm}
\gridline{\fig{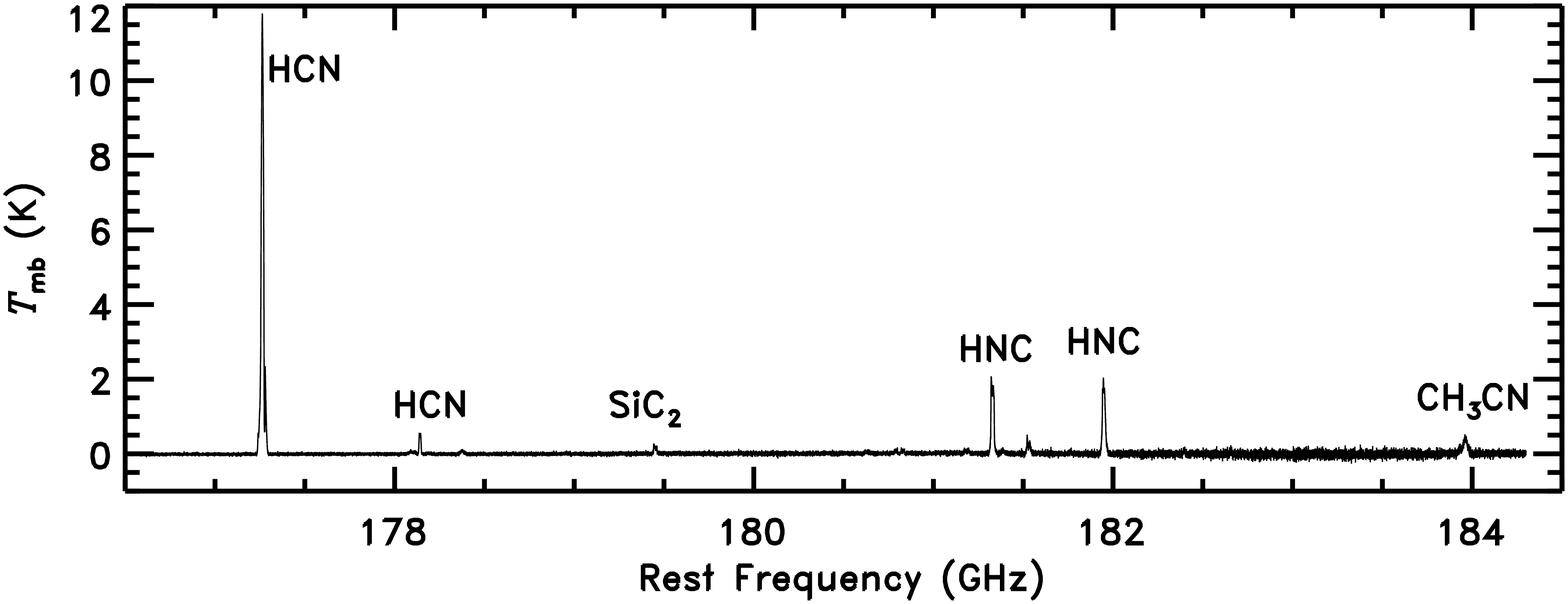}{0.9\textwidth}{}
         }
\vspace{-0.8cm}
\caption{   
 Overview of the full spectra of CRL\,2688 obtained in current observations with strong lines marked.
\label{Figure1}
}
\end{figure}

The spectra of the entire frequency range observed are shown in Fig.~\ref{Figure1}, in which the identified strong lines are marked.
We totally detect 196 emission lines, among which four remain unidentified, and the others are attributed to 24 molecular species and 14 different isotopologues, as shown in Fig.~\ref{Figure2} and listed in Table~\ref{Table3}. 
Most of the molecules, except SiS, NS, PN, and AlF, are carbon-bearing. 
This is consistent with the rich carbon nature of CRL\,2688. 
We note that the detections of C$^{33}$S, Si$^{34}$S, $^{29}$SiS, NS, $^{26}$MgNC, Si$^{13}$CC, $^{29}$SiC$_{2}$, $^{13}$CCH, H$^{13}$C$^{15}$N, C$_3$S, $c$-C$_{3}$H, HCC$^{13}$CN, $c$-C$_{3}$H$_{2}$, and C$^{13}$CCCH have not been previously reported for this object in literature although some of them have been detected in other CSEs of evolved stars. 
The features with a central intensity of larger than 3$\sigma$ are regarded as real detections. 
The spectra also reveal a few weak features with an intensity of less than 3$\sigma$, which have wavelengths closed to those of lines from usual circumstellar molecules, and here we classify them as tentative detection. 
The spectra show a line density of about 6 lines per GHz.
The velocity-integrated intensities are obtained by fitting the lines using the stellar-shell model in the CLASS package. 
For the lines that are too faint to be fitted, the intensities are obtained by directly integrating $T_{\rm mb}$ over a certain velocity interval (see Table~\ref{Table3}).
The fitting results are shown in Fig.~\ref{Figure3} (the complete figure set is available in the online version), 
which are ordered according to the molecular sizes and the rest frequencies of the transition lines. 
In the following, the results are presented species by species, which are followed by a comparison with previous observations.
Also present are the molecular abundances derived from 
 an analysis of rotation diagrams and radiative transfer modelling.

\subsection{Individual species}

\subsubsection{Carbon monosulfide -- CS}
The $J=2 \rightarrow 1$ lines of CS and its $^{13}$C, $^{33}$S, and $^{34}$S isotopologues are detected, as shown in Fig.~\ref{Figure3}.
The CS\,$J=2\rightarrow1$ line shows a broad line wing. 
The blue wing is blended with the $l$-C$_{3}$H $^{2}\Pi_{1/2}~ J=9/2 \rightarrow 7/2$ line. 
The $^{13}$CS\,$J = 2 \rightarrow 1$ line shows a double-topped line profile, and is partly blended with the C$_{3}$S\,$J = 16 \rightarrow 15$ line.
All these lines are fitted by using the stellar-shell model. 
High spatial-resolution observations with the Nobeyama Millimeter Array (NMA) show
that the CS\,$J = 2 \rightarrow 1$ line exhibits symmetrical peaks on both sides of the optical dark lane, indicating a shock-induced sulfur-bearing molecule chemistry \citep{Kasuga97}.
Compared to the NMA observations \citep[see Fig.~4 in][]{Kasuga97}, our observations show
great improvement of revealing the line profile. 
The CS\,$J = 2 \rightarrow 1$ line shows an asymmetric profile with a stronger red peak. 
Our observations suggest that the local standard of rest velocities ($v_{\rm LSR}$) of the red peak is $-33$\,km s$^{-1}$, slightly differing from that given by NMA ($-29$\,km s$^{-1}$), although the ranges of this line are the same in $v_{\rm LSR}$ space.
This might be due to the higher spectral resolution and higher sensitivity in our work.
The $J=2 \rightarrow 1$ lines of CS isotopologues show a similar profile with the CS\,$J =2 \rightarrow 1$ line.
Both CS and C$^{34}$S lines exhibit two emission components at velocity ranges from $-60$ to $-50$\,km s$^{-1}$ and from $-68$ to $-65$\,km s$^{-1}$, respectively.
The NMA observations show that they are associated with the regions located at about 2$\arcsec$ from the center to the northwest. 

\subsubsection{Silicon monosulfide -- SiS}
\label{SiSprofile}
SiS in CRL\,2688 was firstly discovered with the Nobeyama Radio Observatory (NRO) 45 m telescope by \cite{Fukasaku94}. 
So far, the SiS $J=5 \rightarrow 4$, $6 \rightarrow 5$, $13 \rightarrow 12$, and $14 \rightarrow 13$ lines and the Si$^{33}$S $J=8 \rightarrow 7$ line have been detected in this object \citep{Fukasaku94, Highberger03a, Zhang13}. 
The four SiS and isotopic lines, SiS $J=9 \rightarrow 8$ and $10 \rightarrow 9$, $^{29}$SiS $J=10 \rightarrow 9$, and Si$^{34}$S $J=10 \rightarrow 9$, lie within our frequency coverage.
All are clearly detected, as shown in  Fig.~\ref{Figure3}. 
The double-peaked profiles suggest extended ($>$14$\arcsec$) and optically thin natures.
There are intriguing similarities between the profiles of SiS and CS lines; both exhibits 
a broad component distributed from $v_{\rm LSR}=-70$ to 0\,km s$^{-1}$ and
the strengthened red peak, probably suggesting that sulfur-bearing molecules are formed through
a common chemical pathway.

\subsubsection{Nitrogen sulfide -- NS}

We detected two NS lines with $>$3$\sigma$.  Figure~\ref{Figure3} shows the fitting result to
the NS $^{2}\Pi_{1/2}~ J=7/2 \rightarrow 5/2~ l=e$ line. The other tentatively detected lines of NS is blended with other lines and cannot
be fitted well. The NS line is too faint to accurately retrieve its profile, but it seems to be asymmetric
probably due to the blend of hyperfine components.
The NS $^{2}\Pi_{3/2}~ J=7/2 \rightarrow 5/2$ line at 162.67 GHz is also within our frequency coverage, but is not clearly detected due presumably to its intrinsic weakness.

\subsubsection{Phosphorous mononitride -- PN}

The  PN\,$J = 2 \rightarrow 1$ line is  detected at a $4.5\sigma$ level, as shown
in Fig.~\ref{Figure3}. This is the only PN line lying within
our frequency coverage. 
{ This molecule has been previously discovered by \citet{Milam08} 
in this object, IRC+10216, and an oxygen-rich supergiant star VY Canis Majoris, 
 using the Arizona Radio Observatory (ARO) 12 m telescope.
}

\subsubsection{Aluminum fluoride -- AlF}
\label{sect3.5}
AlF was firstly discovered in CRL\,2688 with the IRMA 30 m telescope by \citet{Highberger01} through
the $J=4 \rightarrow 3$, $5 \rightarrow 4$, and $8 \rightarrow 7$ lines.
Two AlF lines lie within our frequency coverage, one of which
($J=5 \rightarrow 4$) is detected, as shown in Fig.~\ref{Figure3}.
This line exhibits an approximately flat-topped  profile, probably suggesting that AlF is located
within a central compact region. 
The $J=3 \rightarrow 2$ line at 98.9\,GHz is 
unfortunately located at the edge of the spectrometer where the signal to noise is relatively low.

\subsubsection{Sodium cyanide -- NaCN}
\label{sect3.6}
NaCN in CRL\,2688 was firstly detected with the ARO 12 m telescope by \cite{Highberger01} at the 2 mm band.
Nine transition lines of NaCN are clearly detected in our observations, as shown in Fig.~\ref{Figure3}.
Eight of them are newly detected.
The NaCN $J_{K_a,K_c} = 10_{1,11} \rightarrow 10_{1,10}$ line has been detected
by \citet{Highberger03a}, which shows a consistent intensity and profile with ours.
The NaCN\,$J_{K_a,K_c} = 5_{0,5} \rightarrow 4_{0,4}$ line is partly blended with the HC$_{7}$N\,$J =69 \rightarrow 68$ line
(see Fig.~\ref{Figure3}), and thus two stellar-shell components are used to decompose its profile.
Most of the transition lines show a double-peaked profile, indicating that 
NaCN emission is optically thin and the emission region is resolved by the telescope beam.

\subsubsection{Magnesium isocyanide -- MgNC}
\label{sect3.7}
\begin{figure}
\gridline{\fig{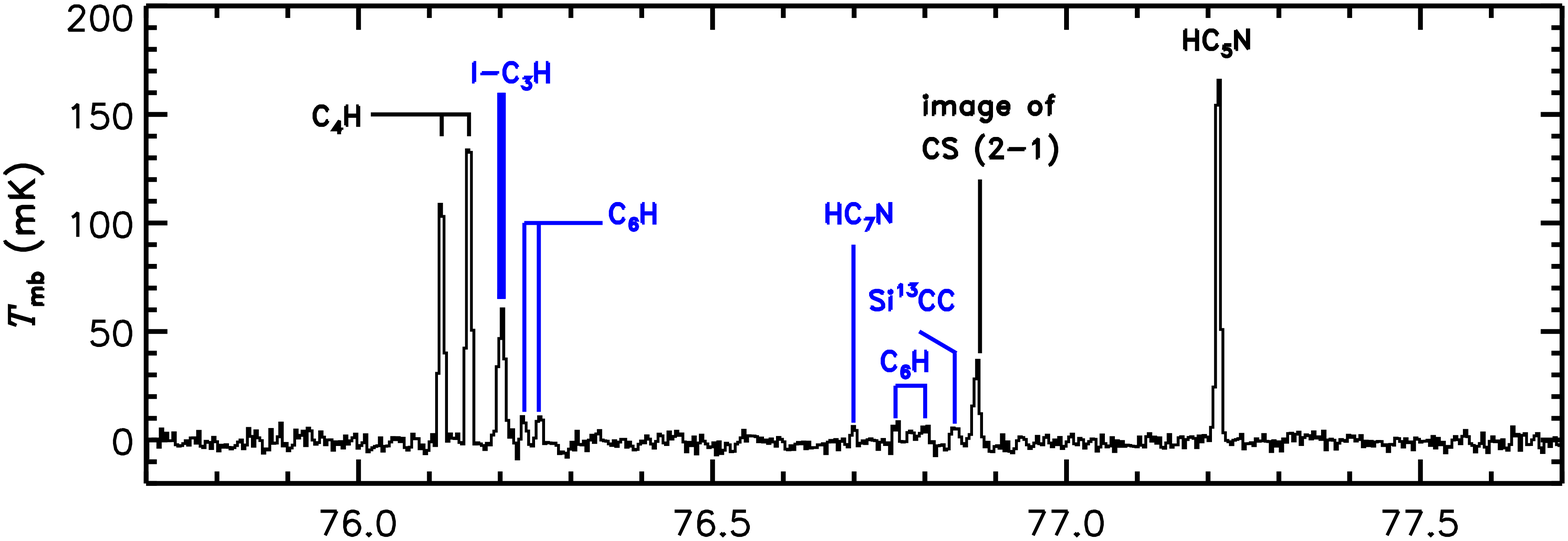}{0.9\textwidth}{}
         }
\vspace{-1.4cm}
\gridline{\fig{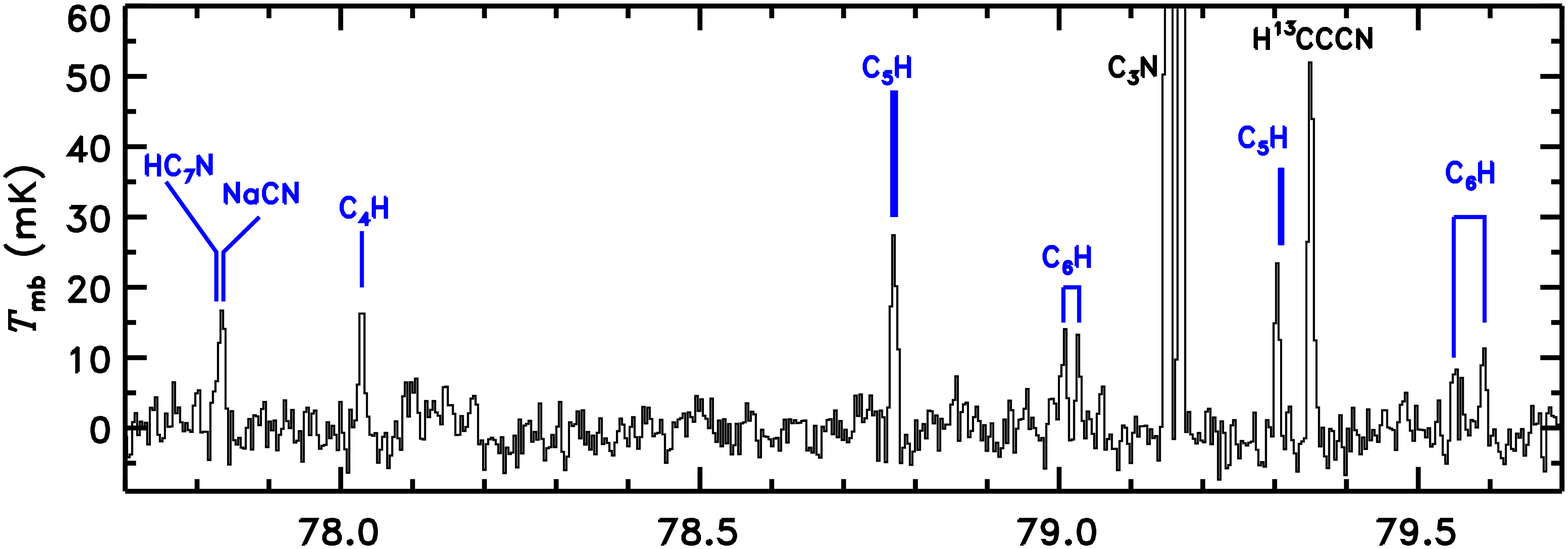}{0.9\textwidth}{}
         }
\vspace{-1.4cm}
\gridline{\fig{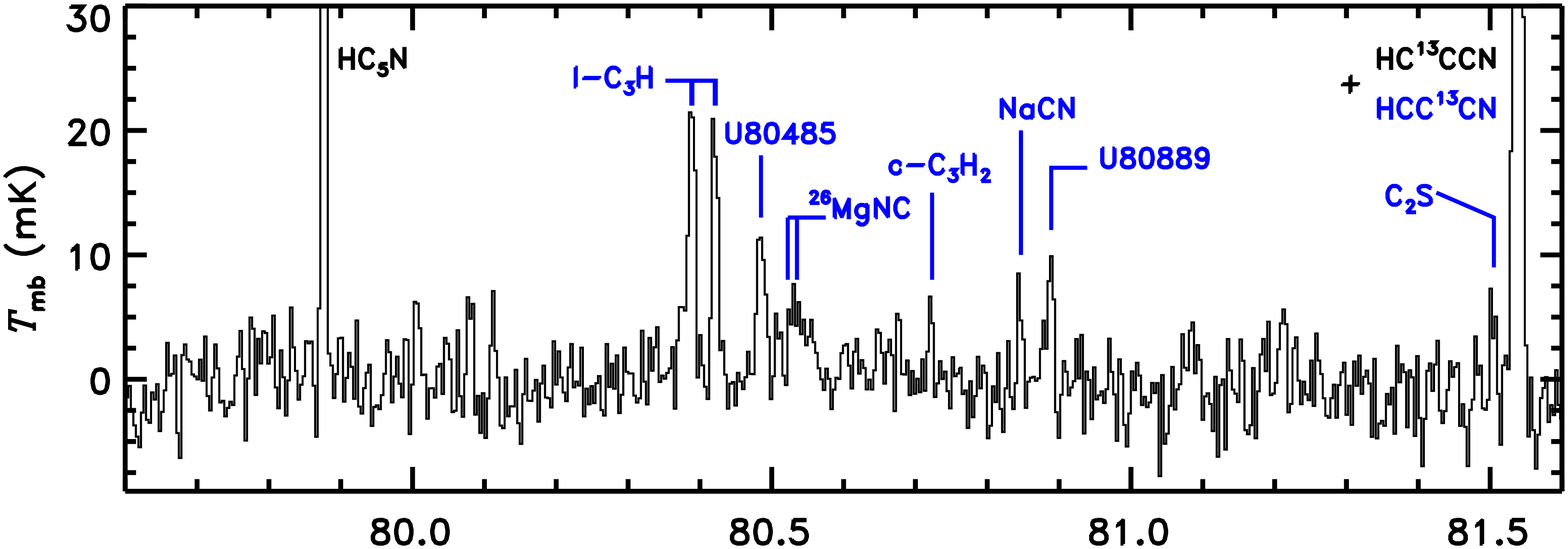}{0.9\textwidth}{}
         }
\vspace{-1.4cm}
\gridline{\fig{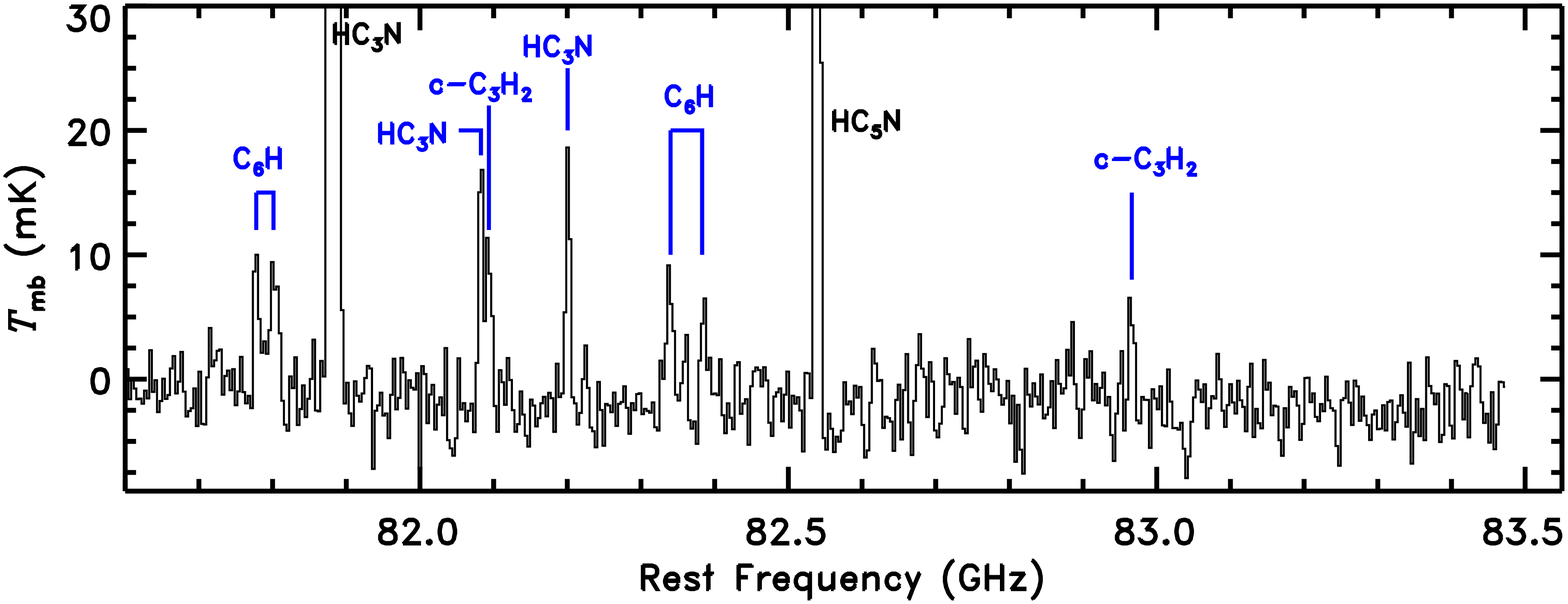}{0.9\textwidth}{}
         }
\caption{    
    Spectra of our observations with all lines marked.
    The RMS are 6.6, 11.0, 10.4, and 24.7 mK after smoothing to a frequency resolution of 13.5, 10.9, 6.5, and 5.8 MHz at the rest frequency of 78.0, 96.9, 163.1, and 182.0 GHz, respectively. Newly detected  transition lines are indicated in blue.
\label{Figure2}
}
\end{figure}

MgNC in CRL\,2688 was firstly detected with the ARO 12 m telescope by \cite{Highberger01} through
the $N=11 \rightarrow 10$, $12 \rightarrow 11$, and $13 \rightarrow 12$ lines. 
We detect three new transition lines,  $N=8 \rightarrow 7$, $14 \rightarrow 13$, and $15 \rightarrow 14$
 as shown in Fig.~\ref{Figure3}.  The double-peaked line profiles of the MgNC\,$J =8 \rightarrow 7$ line indicate an optically thin nature.
 We also marginally detect the weaker $N=7 \rightarrow 6$ and $8 \rightarrow 7$ transition lines of $^{26}$MgNC (see Fig.~\ref{Figure2} and Fig.~\ref{Figure3}).
 The $^{26}$MgNC\,$J=7 \rightarrow 6$ line is blended with an unidentified line (U-line) at 80.5\,GHz, as shown in Fig.~\ref{Figure2}. We 
 notice that this U-line was also detected in IRC+10216 
 (see the NIST database).

\subsubsection{Silacyclopropynylidene -- SiC$_{2}$}
The detection of SiC$_2$ in CRL\,2688 was firstly reported with the IRMA 30 m telescope by \cite{Bachiller97} for the $J_{K_{a},K_{C}}=6_{0,6} \rightarrow 5_{0,5}$ line.
We detected nine transition lines from SiC$_{2}$
and one isotopic transition from Si$^{13}$CC, as shown in Fig.~\ref{Figure3}. 
Except for two transition lines at 165.5 GHz that are heavily blended with the
 CH$_{3}$CN\,$J = 9 \rightarrow 8$ line (see Fig.~\ref{Figure3}), all
have a relatively high signal to noise ratio and exhibit a symmetric double-peaked profile.
Moreover, four  $^{29}$SiC$_{2}$
isotopic transition lines are marginally detected, as shown in Fig.~\ref{Figure2}, but cannot be well fitted using the stellar-shell model assuming spherical symmetry.

\subsubsection{Dicarbon sulfide -- C$_{2}$S}
Two weak C$_{2}$S transition lines are detected, as shown in Fig.~\ref{Figure3}.
The $(N, J)=(7,8) \rightarrow (6,7)$ line is partly blended with one vibrationally excited transition line of C$_{4}$H.
\citet{Park08} reported a detection of the C$_2$S\ $(N, J)=(8,8) \rightarrow (7,7)$ line 
with a flux density of 3.5\,Jy. However, this line was not revealed in the
more sensitive spectrum of  \citet{Zhang13}. 
Based on our measurements and the intrinsic strength ratio
between the $(N,J)=(7,8) \rightarrow (6,7)$ and $(8,8) \rightarrow (7,7)$ lines
($1:1.16$), we estimated the flux density of the latter line to be less than 0.012\,Jy, 
confirming the speculation of \citet{Zhang13} that this line was unlikely detected
by \citet{Park08}.

\subsubsection{Ethynyl radical -- $^{13}$CCH}
\cite{Huggins84} first reported the detection of C$_{2}$H through the
 $(N,J)=(1,3/2) \rightarrow (0,1/2)$ line in CRL\,2688
using the  NRAO 11 m telescope.
Although our spectral range does not cover any C$_{2}$H lines, we
detected two $^{13}$CCH isotopic transition lines with hyperfine components, as shown in Fig.~\ref{Figure3}. 
They are obviously optically thin.

\subsubsection{Hydrogen cyanide -- HCN, hydrogen isocyanide -- HNC, and formyllum cation -- HCO$^{+}$}
The $J=2 \rightarrow 1$ rotational transition lines of HCN, H$^{13}$C$^{15}$N, HNC, and HCO$^{+}$ and two vibrationally excited transition lines of HCN at the $\nu_2=1$ state are detected, as shown in Fig.~\ref{Figure3}.
The HCN $J =2 \rightarrow 1$
line is blended with the much weaker 
 HCN $\nu_{2}=1~ J=2 \rightarrow 1~ l=1e$ line.
 \citet{Zhang13} reported a tentative detection of the HCO$^{+}$ $J=1 \rightarrow 0$ line.
 The existence of a small amount of  HCO$^{+}$ in CRL\,2688 is confirmed by our detection of the weak 
 HCO$^{+}$ $J=2 \rightarrow 1$ line, although it 
 is partly blended with the HC$_{5}$N $J=67 \rightarrow 66$ line.

 Similar to those of sulfur-bearing molecules, the HCN and HNC lines exhibit
 a broad and asymmetric component with a stronger blue wing.
 A prominent absorption feature, ranging in velocity from $-50$ to $-60$ km\,s$^{-1}$,
 is revealed in the broad wing of the HCN $J=2 \rightarrow 1$ line.
 The absorption ranging from $-50$ to $-55$ km\,s$^{-1}$ has been detected
in the CO $J=1 \rightarrow 0$, $3 \rightarrow 2$, and $4 \rightarrow 3$
, HCN $J =4 \rightarrow 3$, H$^{13}$CN $J=4 \rightarrow 3$, and CS\,$J=7 \rightarrow 6$
 lines \citep{Kawabe87,Jaminet92, Young92,Cox00}. Our observations reveal
that this feature is composed of two components, ranging from
 $-50$ to $-55$\,km\,s$^{-1}$ and $-55$ to $-60$\,km\,s$^{-1}$,
 respectively.
 Based on a Voigt fitting \citep{Ginsburg11}, we 
 obtained the column densities of the two HCN absorption components to be
   5.39 $\times$ 10$^{13}$ and 1.51 $\times$ 10$^{13}$ cm$^{-2}$, respectively.

\subsubsection{Cyanoethynyl radical -- C$_{3}$N}
The first detection of 
 C$_{3}$N in CRL\,2688 was 
 presented by \citet{Lucas86}.
 We detected the $N=8 \rightarrow 7$, $10 \rightarrow 9$, $17 \rightarrow 16$,  and $18 \rightarrow 17$ lines, 
 as shown in Fig.~\ref{Figure3}.
 The C$_{3}$N $N=8 \rightarrow 7$ and $10 \rightarrow 9$ lines show an optical thin double-peak profile,
 while the a parabolic profile is revealed for the $N=17 \rightarrow 16$ lines,
  suggesting that the higher-$N$ lines might arise from the inner compact
 region that cannot be resolved by the telescope beam.
The two  $N=18 \rightarrow 17$ lines are heavily blended with each other, and cannot
be well fitted using the stellar-shell model.

\subsubsection{Propynylidyne radical -- $l$-C$_{3}$H and Cyclopropenylidene radical -- $c$-C$_{3}$H}

 $l$-C$_{3}$H has been discovered in CRL\,2688 \citep{Zhang13} .
 In this work,
 we detected eight vibrationally excited transition lines of $l$-C$_{3}$H,
 as shown in Fig.~\ref{Figure3}. Each line splits into two hyperfine structure.
 All show an optical thin double-peak profile.

The $c$-C$_{3}$H $N_{K_{-1},K_{+1}} = 2_{1,2} \rightarrow 1_{1,1}$ line, including two
hyperfine structure components at 91.5 GHz
is detected with a  low signal to noise ratio (see Fig.~\ref{Figure2}).
$c$-C$_{3}$H  has no other lines within our frequency coverage,
except for the line at 91.7 GHz. However, the spectrum around  91.7 GHz
is too noisy to allow us to confirm the secure detection.
 We thus infer that the abundance of $c$-C$_{3}$H is much lower than
 its linear counterpart.

\subsubsection{Cyanoacetylene -- HC$_3$N}
 HC$_3$N is detected through the $J=9 \rightarrow 8$, $18 \rightarrow 17$, and $20 \rightarrow 19$ lines, 
 as shown in Fig.~\ref{Figure3}. These lines consistently exhibit three peaks and a broad wing. 
 We also detect six  $\nu_{7}=1$ vibrationally excited lines of HC$_3$N and
 several isotopic lines from 
  H$^{13}$CCCN, HC$^{13}$CCN, and HCC$^{13}$CN (see Fig.~\ref{Figure3}).
The transition lines of HC$^{13}$CCN and HCC$^{13}$CN 
 from the same energy levels are partly blended with each other.
 We are not able to adequately fit the profile of the  $J=18 \rightarrow 17$ and $20 \rightarrow 19$ lines of HC$^{13}$CCN and HCC$^{13}$CN using the stellar-shell model.

\subsubsection{Cyclopropenylidene -- $c$-C$_{3}$H$_{2}$}
The first detection of 
$c$-C$_{3}$H$_{2}$ in a PPN (CRL\,618) was made by
\citet{Bujarrabal88}, who also observed CRL\,2688.
But they did not report any $c$-C$_{3}$H$_{2}$ line
in CRL\,2688 although some lines are obviously lying in their spectral coverage. 
We detected three $c$-C$_{3}$H$_{2}$ lines,  as shown in Fig.~\ref{Figure3}.
The double-peaked profile suggests that they are likely to be optically thin.

\subsubsection{{ Butadiynyl radical} -- C$_4$H}
 C$_{4}$H in CRL\,2688 was detected for the first time by \citet{Lucas86}
through two $N=10 \rightarrow 9$ transition lines. But it was only tentatively seen in CRL\,618 by \citet{Bujarrabal88}.
All the eight rotational lines lying within our spectral coverage are detected with a prominent double-peaked profile,
as shown in Fig.~\ref{Figure3}.
In addition, we detected  six and two vibrationally excited transition lines within the 
$^{2}\Pi_{3/2}$ and $^{2}\Pi_{1/2}$ states, respectively, as shown in Fig.~\ref{Figure3}.
The C$_{4}$H $\nu_7=1~ ^{2}\Pi_{3/2}~ J=33/2 \rightarrow 31/2$ line 
is significantly blended with
 the C$^{13}$CCCH $N=17 \rightarrow 16~ J=35/2 \rightarrow 33/2$ line (see Fig.~\ref{Figure3}), and cannot be well fitted with the stellar-shell model. Due to the low signal to noise ratio, it is hard to tell whether the profile of these vibrationally excited lines is  double-peaked or flat-topped.

\subsubsection{Pentynylidyne Radical -- $l$-C$_{5}$H}
The first detection of $l$-C$_{5}$H in CRL\,2688
was made by  \citet{Highberger01} through the  $^{2}\Pi_{1/2}~ J=65/2 \rightarrow 63/2~ l=e~ \&~ f$ line.
Several vibrationally excited transition lines of $l$-C$_{5}$H at $^{2}\Pi_{3/2}$ and $^{2}\Pi_{1/2}$ states were subsequently detected \citep{Highberger03a,Zhang13}.
We detected six vibrationally excited lines of $l$-C$_{5}$H,
as shown in Fig.~\ref{Figure3}.
Each line is split into $l=e$ and $f$ hyperfine structure components.
Other six vibrationally excited lines
 of $l$-C$_{5}$H are lying within our spectral coverage but not
being detected due presumably to their intrinsic weakness; their intensity upper limits are estimated to be 10\,mK.

\subsubsection{Methyl cyanide -- CH$_{3}$CN}
 \citet{Matthews83} failed to detect 
CH$_{3}$CN in CRL\,2688, which was then discovered by
 \citet{Bujarrabal88} through the $J=8 \rightarrow 7$ line.
 More transition lines were subsequently detected by
 \citet{Zhang13}. We detected the $J=5 \rightarrow 4$, $9 \rightarrow 8$, and $10 \rightarrow 9$ lines of CH$_{3}$CN,
as shown in Fig.~\ref{Figure3}. 
Each line includes several hyperfine structure components.
The CH$_{3}$CN $J=9 \rightarrow 8$ line is blended with the SiC$_{2}$\,$J_{K_a,K_c}=7_{4,3} \rightarrow 6_{4,2}$ and $7_{4,4} \rightarrow 6_{4,3}$ lines (see Fig.~\ref{Figure3}).

\subsubsection{Cyanodiacetylene -- HC$_5$N}

The observations of \citet{Bujarrabal88} using the IRAM 30 m telescope resulted in an upper limit of the
HC$_5$N   abundance in CRL\,2688. This molecule then  detected
by \citet{Truong88} with the Effelsberg 100 m telescope through the $J=8 \rightarrow 7$ and $9 \rightarrow 8$ 
 lines at 21 and 24\,GHz, respectively. 
We detected nine transition lines of  HC$_5$N\,$J=29 \rightarrow 18$, $30 \rightarrow 29$, $31 \rightarrow 30$, $35 \rightarrow 34$, $36 \rightarrow 35$, $27 \rightarrow 36$, $61 \rightarrow 60$, $62 \rightarrow 61$, and $63 \rightarrow 62$, as  shown in  Fig.~\ref{Figure3}.
Generally, multiple peaks are found in the line profiles, specially for the low-$J$  transition lines.

\subsubsection{Hexatriynyl radical -- C$_{6}$H}
C$_{6}$H was marginally
detected in CRL\,2688 with the ARO 12 m telescope by \citet{Highberger01}, and then the detection of the line at 102 GHz was confirmed \citet{Highberger03a} using the IRMA 30 m telescope.
Six and five vibrationally excited transition lines of C$_{6}$H, respectively in the $^{2}\Pi_{3/2}$ and
$^{2}\Pi_{1/2}$ states, are detected in our observations as shown in Fig.~\ref{Figure3}. The profiles of 
the eight strongest C$_{6}$H lines are well fitted by using the stellar-shell model,  suggesting an optically thin nature.

\subsubsection{Cyanotriacetylene -- HC$_{7}$N}

The first detection of HC$_7$N in CRL\,2688 was made by \citet{Nguyen84} using the Effelsberg 100 m telescope through the $J=21 \rightarrow 20$  line.
 The present observation detected three new lines, $J=68 \rightarrow 67$, $71 \rightarrow 70$, and $83 \rightarrow 82$,
 as shown in Fig.~\ref{Figure3}.  The $J=69 \rightarrow 68$
line is only marginally detected because of a blending with 
 the NaCN $J_{K_a,K_c} = 5_{0,5} \rightarrow 4_{0,4}$ line  (see Fig.~\ref{Figure3}).

\subsubsection{Unidentified lines}
Four U-lines that cannot be assigned to known  transition lines of any molecular species are newly detected, increasing the number of U-lines in CRL\,2688 to 13. It is clear that the frequency spans of these U-lines cannot be accounted for by a single rotation constant, suggesting that they are unlikely to arise from the same species. 
Probably they are associated with unknown circumstellar molecules or internal rotations of
polymer molecules, which call for further experimental investigation. 
For comparison, we examine the U-lines in the spectrum of IRC+10216 ranging from 129.0 to 172.5 GHz presented by \citet{Cernicharo00}, but do not find any match. We cannot rule out the possibility that some of them are artificial 
features.

\subsection{A comparison with previous line survey observations of CRL\,2688}
Our spectral range has overlaps with those of \citet{Park08} and \citet{Zhang13} 
in frequency coverages, allowing us to make a comparison between
these measurements.
Within the overlapped frequency range from 91.38--99.16 GHz, the CS $J=2 \rightarrow 1$ line is the only line discovered by
\citet{Park08}, while we unambiguously detect 35 lines. 
All the 23 lines observed by \citet{Zhang13} in the overlapped frequency range from 75.70--83.48 GHz
are visible  in our spectra, in which we totally detect 102 lines. The HPBWs of the IRAM 30 m and ARO 12 m telescopes are different (69$\rlap{.}\arcsec$4--64$\rlap{.}\arcsec$0 and 26$\rlap{.}\arcsec$9--24$\rlap{.}\arcsec$8, respectively, for the overlapped frequency range), resulting in
different main beam temperatures for the the same line between ours and that measured by \citet{Zhang13}.
Through a comparison between these measurements, we can roughly estimate the
size of the region where the molecular lines originate. The same technique has been used by \citet{Takano19} and \citet{Qiu20} to investigate the molecular emission regions in extragalactic objects.

With the assumption of a Gaussian intensity distribution,
the velocity-integrated intensities measured by different telescopes should satisfy the relationship
 $I_{\rm mb}(\theta _{\rm b}^2 + \theta _{\rm s}^2) = I^{'}_{\rm mb}(\theta _{\rm b}^{'2} + \theta _{\rm s}^{2})$, 
where ($I_{\rm mb},\theta_{\rm b}$) and ($I_{\rm mb}^{'},\theta _{\rm b}^{'}$) represent the intensities and beam sizes
corresponding to the IRAM 30 m and ARO 12 m telescopes, respectively, and $\theta _{\rm s}$ represents { the Gaussian full width at half-maximum (FWHM) of the source size}.
To estimate $\theta _{\rm s}$, 
we define a parameter for the $i$th line
\begin{equation}
f_i=\left|\frac{I_{{\rm mb},i}(\theta _{\rm b}^2 + \theta _{\rm s}^2)}{I^{'}_{{\rm mb},i}(\theta _{\rm b}^{'2} + \theta _{\rm s}^{2})}-1\right| \times\frac{1}{\sigma_i}, 
\label{equation1}
\end{equation}
where $\sigma_i$ is the uncertainty of $I_{{\rm mb},i}/I_{{\rm mb},i}^{'}$. 
Then an $F$ factor is introduced by $F=\sum_{i}f_{i}$. Here, $f_i$ is the residual of the ratio between the  velocity-integrated intensities measured by IRAM 30 m and ARO 12 m telescopes considering the beam dilution weighted by 1/$\sigma_i$. 
The source size $\theta_{\rm s}$ can be obtained by minimizing the $F$ factor.
We find that the  $F$ factor reaches the minimum value if the source size is 
 15$\rlap{.}\arcsec$5, as shown in the upper panel of Fig.~\ref{Figure4}.
The CO mapping of CRL\,2688 in the $J=1 \rightarrow 0$ line shows that 
the emission is dominantly from the central core with a radius of 10$\arcsec$--13$\arcsec$ 
 \citep{Kawabe87}, while the $^{13}$CO image of the $J=1 \rightarrow 0$ line
reveals a high-brightness central core, a  high-velocity component near the center, and a less bright extended envelope with a diameter of about 15$\arcsec$ \citep{Yamamura95}. 
These observations are in reasonable agreement with our size estimation based on single-dish data.
The $F$ factor cannot reach zero (see Fig.~\ref{Figure4}), probably suggesting that the molecular
gas has a stratified structure
 and/or the brightness distributions deviate from the Gaussian assumption. 
To further examine this, we plot in the lower panel of  Fig.~\ref{Figure4} the
$[I_{\rm mb}(\theta _{\rm b}^2 + \theta _{\rm s}^2)]/[ I^{'}_{\rm mb}(\theta_{\rm b}^{'2} + \theta _{\rm s}^{2})]$ ratios
for all the lines lying within the overlapped frequency range, where $\theta_{\rm s}=15\rlap{.}\arcsec$5.
The deviations from one in this plot is an indication that different molecules may 
have different spatial distributions, and thus the obtained source size only represent the average value.

{ Alternatively, we could assume that the source has a disk geometry, where
the brightness distribution is constant within a diameter of $\theta_{\rm d}$.
Then we have $I_{\rm mb}/(1-2^{-\theta^2_{\rm d}/\theta^2_{\rm b}})=I^{'}_{\rm mb}/(1-2^{-\theta^2_{\rm d}/\theta^{'2}_{\rm b}})$. Following a similar way as described
above, but defining $f_i$ by
\begin{equation}
f_i=\left|\frac{I_{{\rm mb},i}/(1-2^{-\theta^2_{\rm d}/\theta^2_{\rm b}})}{I^{'}_{{\rm mb},i}/(1-2^{-\theta^2_{\rm d}/\theta^{'2}_{\rm b}})}-1\right| \times\frac{1}{\sigma_i}, 
\label{equation1b}
\end{equation}
we find $\theta_{\rm d}=24\rlap{.}\arcsec$7 when $F$ approaches the minimum
(see the upper panel of Fig.~\ref{Figure4}). 
This means that we would obtain the same beam-filling factors if assuming
$\theta_{\rm d}=24\rlap{.}\arcsec$7 for a disk distribution or assuming
$\theta_{\rm s}=15\rlap{.}\arcsec$5 for a Gaussian distribution. 
Therefore, the
$[I_{\rm mb}(1-2^{-\theta^2_{\rm d}/\theta^{'2}_{\rm b}})]/[I^{'}_{\rm mb}(1-2^{-\theta^2_{\rm d}/\theta^2_{\rm b}})]$
ratios for $\theta_{\rm d}=24\rlap{.}\arcsec$7 are equal to
the $[I_{\rm mb}(\theta _{\rm b}^2 + \theta _{\rm s}^2)]/[ I^{'}_{\rm mb}(\theta_{\rm b}^{'2} + \theta _{\rm s}^{2})]$ ratios
for  $\theta_{\rm s}=15\rlap{.}\arcsec$5, as shown
in the lower panel of  Fig.~\ref{Figure4}.
It is natural that $\theta_{\rm s}$ is smaller than $\theta _{\rm d}$ since the former is
the FWHM of the surface brightness profile while the latter represents the extent 
of the source.
}

\subsection{Rotation diagram, column Densities, and fractional Abundances}
\label{sect3.4}
The column densities ($N$) and excitation temperatures  ($T_{\rm ex}$) of 22 molecules and isotopologues, including 
CS, $^{13}$CS, C$^{33}$S, C$^{34}$S, HCN, HNC, HCO$^{+}$, SiS, C$_2$S, SiC$_2$, $l-$C$_{3}$H, C$_3$N, HC$_{3}$N,  H$^{13}$CCCN, HC$^{13}$CCN,  HCC$^{13}$CN, C$_{4}$H, $c$-C$_3$H$_2$, C$_{6}$H, CH$_{3}$CN, HC$_{5}$N, 
and HC$_{7}$N, are estimated by means of the rotation diagram.
Under the assumption of local thermal equilibrium (LTE) and negligible optical depth, the level populations can be
expressed with
\begin{equation}
{\rm ln} \frac{N_{\rm u}}{g_{\rm u}} = {\rm ln} \frac{8\pi k \, \nu^{2} \, \int T_{\rm S} \, dv}{hc^{3} \, A_{\rm ul} \, g_{\rm u}}\,  =  {\rm ln} \frac{N}{Q(T_{\rm ex})} - \frac{E_{\rm u}}{kT_{\rm ex}},
\label{equation2}
\end{equation}  
where $k$ and $h$ are the Boltzmann constant and Planck constant,
$\nu$ is the rest frequency in Hz, 
$Q(T_{\rm ex})$, $A_{\rm ul}$, $g_{\rm u}$ are the partition function, spontaneous emission coefficient, and upper state degeneracy, respectively.
Under the assumption of a Gaussian distribution of the surface brightness,
the source brightness temperature $T_{\rm S}$ is derived 
after correcting for beam dilution,
$T_{\rm S}=T_{\rm mb}(\theta_{\rm b}^{2} + \theta_{\rm s}^{2})/\theta_{\rm s}^{2}$,
where $\theta_{\rm b}$ is the antenna HPBW and $\theta_{\rm s}$ is the source size.
$\int T_{\rm S}\,d \nu$ is the velocity-integrated intensity. 
$Q(T_{\rm ex})$, $A_{\rm ul}$, $g_{\rm u}$, and $E_{\rm u}/k$ 
are taken from  the CDMS catalog.
Following the previous works  \citep{Park08, Zhang13}, we simply assume a
source size of 20$''$ for the calculations of all the molecules.
The resultant  
rotation diagrams are presented in Fig.~\ref{Figure5}, and the results are listed in Table~\ref{Table4}.

If the optical depth ($\tau$) poses an effect, the  dots in the rotation diagram might deviate from
a linear distribution. If this is the case, an optical depth correction factor of $C_{\tau}=\tau/(1-e^{-\tau})$
needs to be introduced to construct the rotation diagrams \citep{Goldsmith99}.
However, such a deviation cannot be clearly seen in Fig.~\ref{Figure5}.
Using the population diagram method developed by  \citet{Goldsmith99}, we estimate 
$\ln C_{\tau}<0.03$, suggesting that the effect of  optical depth is negligible.

We use the expression given by \cite{Olofsson97} to derive
the molecular abundances with respect to molecular hydrogen ($f_{X}$), 
\begin{equation}
f_{\rm X} = 1.7 \times 10^{-28} \frac{v_{\rm e} \theta_{\rm b}D}{\dot{M}_{\rm H_2}} \frac{Q(T_{\rm ex})\nu^{2}}{g_{\rm u}A_{\rm ul}} \frac{e^{E_{\rm l}/kT_{\rm ex}}\int T_{\rm mb}dv}{\int^{x_e}_{x_i}e^{-4x^2{\rm ln2}}dx},
\label{equation3}
\end{equation}  
where $v_{\rm e}$ is the expansion velocity in km\,s$^{-1}$, 
$D$ is the distance in pc (assumed to be 1\,kpc hereafter), 
$\dot{M}_{\rm H_{2}}$ is the mass-loss rate in $M_{\odot}$\,yr$^{-1}$, 
$\nu$ is the rest frequency in GHz, 
$E_{\rm l}$ is the energy of the lower level, 
$\int T_{\rm mb} dv$ is the velocity-integrated intensity in K\,km\,s$^{-1}$,
$x_{e,i} = R_{e,i}/(\theta_{\rm b}D)$ with $R_{e}$ and $R_{i}$ for the outer and inner radii of the shell, 
and $\theta_{\rm b}$, $Q(T_{\rm ex})$, $g_{\rm u}$, $A_{\rm ul}$, and $k$ are the same as in Equation \ref{equation1}.
To simplify the calculations, we assume the values of $x_{e}$ and $x_{i}$ to be 1 and 0 for all the species, respectively. 
For the molecules that have more than one transition line detected, we adopt the average abundances weighted by their velocity-integrated intensities.
As mentioned in \citet{Zhang09}, the abundances is  essentially insensitive to 
 $T_{\rm ex}$ in the case of $E_{\rm l} \ll kT_{\rm ex}$.
 The results are listed in Table~\ref{Table4}.
 It should be noted that if the transition lines are optically thick,  
  the derived $f_{\rm X}$ values should be considered as lower limits.

\subsection{The fractional abundances derived from radiative transfer method}
\label{LIME}

In order to determine the fractional abundances ($f_X$ in Table~\ref{Table4}), we have assumed that all the lines originating 
from a given molecule are characterized by a single excitation temperature, which is generally valid if the molecule is located within a narrow layer. 
However, CRL 2688 exhibits a moderate radial temperature gradient \citep{Truong90}.
For the abundance computation of  molecules lying within a region with wide temperature variation, 
a more sophisticated approach is to use a radiative transfer model. As such, we redetermine the
fractional abundances using the three-dimensional  non-local-thermodynamical (NLTE) radiative transfer code LIME  \citep{Brinch10}.
We limited the calculations to a few strong lines, aiming to examine the systematic uncertainties in the determination of $f_X$.

For the modelling, we assume a spherically symmetric structure with a velocity law of
$v(r) = v_{\rm exp}(1-{r_{0}}/{r})^{0.5}$, where $r_0$ is the inner radius of the envelope 
and is taken to be $10^{15}$\,cm \citep{Truong90}. 
Such a velocity law is the solution of the equations of the stellar wind driven by 
a force decreasing with $r^{-2}$.
The terminal velocity $v_{\rm exp}$ is taken from  Table~\ref{Table3}. In most parts of the envelope
($r>10^{16}$\,cm), $v(r)$ is very close to $v_{\rm exp}$. We
thus simply assume a H$_2$ number density law of
$n(r)=\dot{M}/(4\pi m_{\rm H_2} r^2 v_{\rm exp})$, where
$m_{\rm H_2}$ is the mass of a H$_2$ molecule, and the 
mass-loss rate $\dot{M}$ is taken to be $1.7\times10^{-4}$\,$M_{\odot}$\,yr$^{-1}$
\citep{Truong90} . Even though the assumed mass-loss rate may be slightly larger than the current value of CRL~2688, we consider that it would be acceptable as a first approximation. Moreover, we apply a radial distribution 
 of the kinetic temperature obtained by
 \citet{Truong90} through high-resolution  mapping of the
CO $J = 1 \rightarrow 0$ and $2 \rightarrow 1$ lines.
The molecular data,  including the energy states, Einstein coefficients, and collisional rates,  are taken from the LAMDA database\footnote{http://www.strw.leidenuniv.nl/~moldata} \citep{Schoier05}.
The output spectral resolution is set to be 1\,km\,s$^{-1}$.

Fixing the distance between CRL\,2688 and earth at 1\,kpc and allowing the inner radius ($R_i$) and outer radius ($R_e$) of individual molecules to vary, we determine the fractional
abundance by fitting the LIME model to the observations.
The fittings for the CS $J =2 \rightarrow 1$, SiS $J=10 \rightarrow 9$, PN $J=2 \rightarrow 1$, SiC$_{2}$ $J_{K_a,K_c}=4_{0,4} \rightarrow 3_{0,3}$, HCN $J=2 \rightarrow 1$, HNC $J=2 \rightarrow 1$, HCO$^{+}$ $J=2 \rightarrow 1$, HC$_{3}$N $J=9 \rightarrow 8$, $c$-C$_{3}$H$_{2}$ $J_{K_a,K_c}=2_{0,2} \rightarrow 1_{1,1}$, and CH$_{3}$CN $J_{K}=5_{3} \rightarrow 4_{3}$ lines are shown in
Fig~\ref{Figure3}. 
The LIME models appear to match the main components of the observed line profiles reasonably well,
but fail to reproduce
 the broad wings of CS, SiS, HCN, and HNC lines.
An interferometric observation has revealed that HCN and HC$_3$N emission primarily traces dense gas in
polar and equatorial bipolar outflows \citep{dl20}. Therefore, the mismatches between the
models and the observations may suggest a deviation from a spherical geometry.
In order to reproduce the present observations more closely, it would be necessary to assume a more realistic geometry, which is possibly determined by high-angular resolution interferometry. However, such the observation is beyond the scope of the present research. 
Table~\ref{Table5} summarizes our fitting results, showing that the agreement
between the molecular abundances derived from the LIME models and $f_X$ is better
than one order of magnitude.

\subsection{Isotopic abundance ratios}
\label{iso}

Isotopic abundance ratios reflect the  nucleosynthesis in the stellar interior 
and mixing processes \citep{Herwig05}.
We determine the isotopic abundance ratios of C, S, Si, and Mg elements in CRL\,2688 based 
on the intensities of the corresponding molecular transition lines.
The results, along with those of the Sun and IRC+10216 \citep{Lodders03,Kahane00}, are
given in Table~\ref{Table6}.
The  $^{12}$C/$^{13}$C  ratio  
 is believed to be one of the most useful tracers
of the relative degree of primary to secondary processing in stars, which
drops during the first dredge-up and is enhanced by the third dredge-up.
Our measurements suggest  a  $^{12}$C/$^{13}$C ratio of 15--40 in CRL\,2688, slightly
lower than that in IRC+10216 and significantly lower than the solar value. This can be attributed to
 the more evolved nature of CRL\,2688.
 Strikingly, the  $^{12}$C/$^{13}$C ratios derived from HC$_3$N lines
 tend to increase with increasing $J$ (see Table~\ref{Table6}).
 { Chemical fractionation may alter the isotopic ratio.}
 The reaction, $^{13}$C$^{+}$ + CN $\rightarrow$ $^{13}$CN + $^{12}$C$^{+}$, preferentially
 takes place under low temperature \citep{Watson76, Langer92, Milam05}. As CN is the key parent molecule of
 cyanopolyynes, the enhanced $^{13}$CN would leads to a higher abundance of 
 the $^{13}$C isotopologues of HC$_3$N in the cool region. 
 If the gas temperature of the CSE has a negative gradient form the inner side to the outer side,
 the low-$J$ lines preferentially arise from the outer low-temperature region, respective to
 the  high-$J$ lines. 
 { This offers a seemingly plausible explanation for the observations that 
  the HC$_3$N  low-$J$ lines result in a lower   $^{12}$C/$^{13}$C ratio. 
 However, chemical fractionation is efficient only at an extremely low kinetic temperature.
 Through theoretical calculations,
\citet{Milam05} found that for a molecular cloud with a kinetic temperature ranging from 10--100\,K,
the effect of fractionation causes negligible alteration of the  $^{12}$C/$^{13}$C ratio
during the typical life timescale ($10^6$\,yr).
 Although PPNs are denser than molecular clouds, tending to decrease the reaction timescale, their lifetime
 is much shorter ($\sim$10$^3$\,yr). Moreover, PPNs are energetic sources with a kinetic temperature
 of much higher than 10\,K. Therefore, 
 chemical fractionation is unlikely to affect the  $^{12}$C/$^{13}$C ratio in CRL\,2688.
Isotope-selected photodissociation could also affect the    $^{12}$C/$^{13}$C 
isotopologue ratio.
For instance, $^{12}$CO is more abundant than $^{13}$CO, and thus is more self-shielding from the interstellar UV field,
resulting in an increasing $^{12}$C/$^{13}$C ratio.  
HC$_3$N is the product of photochemistry of HCN and CN, which thus would result
in a radial $^{12}$C/$^{13}$C gradient for HC$_3$N if their isotopologues have
different self-shielding factors.
However, unlike CO, HCN and CN are dissociated mainly in the continuum rather than bands, suggesting that their isotopologues are equally affected by UV radiation \citep{sa17}.
As a result, we can rule out 
the isotope-selected photodissociation as the cause of 
the   $^{12}$C/$^{13}$C  gradient  observed for HC$_3$N.
 }
 
 \citet{Giesen20} found that the C$^{13}$CC/$^{13}$CCC isotopic ratio
 in the massive star-forming region Sgr\,B2(M) is higher than the
 statistically expected value, which is probably due to
 the lower zero energy of C$^{13}$CC relative to   $^{13}$CCC 
 making the position-exchange reaction of converting  $^{13}$CCC to C$^{13}$CC
 favorable. We find that the column densities of
 HC$^{13}$CCN, H$^{13}$CCCN, and HCC$^{13}$CN are quite consistent with
 each other, suggesting that such an effect is not significant 
 for HC$_{3}$N isotopolgues.

 Presumably, the isotopic abundance ratios of very heavy elements are not altered during the AGB evolution. 
 We do not find a significant difference of the S, Si, and Mg isotopic abundance ratios between CRL\,2688, 
 IRC+10216, and the Sun, being consistent with the prediction of stellar theory.
 The only exception is the CS $J=2 \rightarrow 1$ line which gives a relatively low 
   $^{32}$S/$^{34}$S ratio.  But considering that this line is likely to be optically thick,
   the resultant S isotopic ratio only represents a lower limit.

\section{Discussion}
\label{sec4}

\subsection{Temporal variations}

Evidences of temporal variations of molecular lines in IRC+10216 were presented by
\citet{Cernicharo14}, who found that most of the abundant molecules, except SiC$_2$, exhibit
strong variations in line intensities
and the variations are more pronounced for high-$J$ lines. This can be attributed to
the variations in the IR pumping rate. In this scenario, optically thick lines,
such as low-$J$ lines of SiC$_2$, are insensitive to the IR pumping. Such studies are
relevant to the reliability of the intensities of molecular lines in AGB envelopes
as the flux calibrator and the tracer of mass loss rate.

CRL\,2688 is a commonly used standard source for flux calibration.
The brightness of its lobes has been found to increase by $-0.22$\, V magnitude 
during 1994--2008 \citep{HL10}. \citet{KS00} found that the radial velocity of the envelope varies with time.
Presumably, the rapid variable stellar brightness and circumstellar dynamics may lead to
temporal variations of the intensities of some molecular lines in the CSE.
The C$_{4}$H $J=21 \rightarrow 19$, C$_{4}$H $J=19 \rightarrow 17$, SiC$_{2}$ $J_{K_a,K_c}=4_{2,2} \rightarrow 3_{2,1}$, and H$^{13}$CCCN $J=11 \rightarrow 10$ lines in CRL\,2688 were also observed in May 1985 by \citet{Lucas86}  using
the same observational configurations as ours.
We investigated the intensity variations 
using the data sets spanning over 32 years.
The velocity-integrated intensity ratios of the four lines between
the measurements of \citet{Lucas86} and ours are
$1.6\pm0.1$, $1.7\pm0.1$, $0.6\pm0.2$, and $1.2\pm0.1$, respectively.
It is notable that the two C$_{4}$H lines show a consistently increasing velocity-integrated intensity ratio,
whereas no significant variation is found for 
the SiC$_{2}$ and H$^{13}$CCCN lines.  Although we cannot  conclusively 
attribute this to the luminosity variations,
it would be interesting to study the different variational behaviors of these molecular lines in the future.

\subsection{A comparison with previous line survey observations of IRC+10216}

A $\lambda$ = 2 mm line survey of IRC+10216, the prototype of carbon-rich CSEs, has been presented by
\citet{Cernicharo00} with the IRAM 30 m telescope. An overlap of their and our spectra occurs in the 
spectral range from  160.8 to 168.6 GHz. To investigate 
the alteration of physical and chemical environments during the AGB--PPN evolution,
we make a comparison of the  velocity-integrated  intensities of the lines in the 
overlapping range between IRC+10216 and CRL\,2688. 
The line intensities, normalized such that $I$(HC$_{3}$N\,$J$=18$\rightarrow$17) = 1, 
are shown in Fig.~\ref{Figure6}, which reveals systematic differences.
The $^{13}$C isotopologues  of  HC$_{3}$N
shows slightly strengthened intensities in CRL\,2688 { with an unknown reason} (see Section~\ref{iso}). CRL\,2688 also exhibits enriched cyanides, such as HC$_{5}$N and CH$_{3}$CN, which can be ascribed to the enhancement of CN fragments 
resulted from the photolysis of HCN during the AGB--PPN transition.
The vibrational excitation line of HC$_3$N is largely strengthened in CRL\,2688, suggesting a
higher temperature in the inner regions of the PPN where this line arises.
In CRL\,2688, the lines of C-rich molecules, such as C$_{3}$N, $l$-C$_{3}$H,  
and C$_{4}$H, and of Si-bearing molecules, such as SiS and SiC$_{2}$,
appear to be consistently weaker than those in IRC+10216,
This can be interpreted if IRC+10216 is intrinsically richer in carbon abundance and
the gas-phase refractory-element molecules in CRL\,2688 have been heavily depleted onto grains
during the PPN expansion. The depletion of SiC$_2$ of CRL\,2688 relative
to IRC+10216 is consistent with the 3 mm observations of
\citet{Lucas86}. However, \citet{Lucas86} suggested an enhanced C$_4$H in CRL\,2688, contrary to
our result and that of \citet{Zhang13}.
This might be due to the effect of excitation conditions and spatial distribution of
this molecules, as pointed out by \citet{Zhang13}.
We find that the column density of $l$-C$_{3}$H  is higher than that of  $c$-C$_{3}$H
by a factor of 3 in CRL\,2688, where the  $l$-C$_{3}$H/$c$-C$_{3}$H 
 ratio is close to that of IRC+10216 \citep[$\sim$2;][]{{Cernicharo11}}.
 The higher $l$-C$_{3}$H/$c$-C$_{3}$H 
ratio in IRC+10216 relative to that in the dark cloud TMC-1 has been regarded as an evidence
of atom-neutral reactions 
\citep{ko96}.

\subsection{Modelling and circumstellar chemistry}
\label{CSEmodel}

Chemical modelling is a powerful tool for deepening our understanding of  circumstellar chemistry \citep[e.g.][]{mh94,Agundez20}. The abundance patterns of many AGB molecules can
be satisfactorily reproduced by state-of-the-art astrochemistry models, which, however, seemingly perform less well at explaining the formation of complex molecules in PPNs. New observations continue to
show unexpected results, prompting the development of models. To investigate the shortcomings of
the models in studying PPN chemistry, we compare the observations with model predictions.  
For the modelling, we use the package presented by \citet{McElroy13}, which is based on
an updated UMIST Database for Astrochemistry (RATE12). 
The reaction network consists of 6173 gas-phase reactions and 195 reactions on grain surfaces involving 467 atomic and molecular species. We assume an uniform loss-rate of  3.0 $\times$ 10$^{-5}$ $M_{\odot}$\,yr$^{-1}$ \citep{Lo76} and an spherically symmetrical expansion velocity  of 16.5 km s$^{-1}$ \citep{Crampton75}.
The gas density radially decreases as 1/$r^{2}$ with an inner radius of $1.5\times10^{16}$\,cm ($\sim$1$\arcsec$ at $D=1$\,kpc) where the parent species are injected into the CSE and an outer radius of 1.5 $\times$ 10$^{18}$\,cm ($\sim$100$\arcsec$ at $D=1$\,kpc) where the majority of molecules are dissociated
by a standard interstellar ultraviolet radiation field. The radiation field of the central star is
ignored.

The modelling results are listed in Table~\ref{Table4} and are compared with the observations
in Fig.~\ref{Figure7}. We find that for 19 out of the 26 detected molecular species, the predicted and observed column densities show agreement within one order of magnitude.
An excellent agreement  is shown for the abundance patterns of cyanopolyynes (HC$_{2n-1}$N: $n=1$--4); the only exception is HCN, which exhibits a much lower observed abundance and could be attributed to
the effect of optical depth. For other C-rich species, C$_n$H ($n=2$--6) and C$_n$S ($n=1$--3),
the models can reproduce the observations to a reasonable accuracy; 
there are only slight disagreements for C$_5$H and  and  C$_6$H.
Based on the 20 GHz band data obtained with the 100 m Green Bank Telescope (GBT),
\citet{Gupta09} derived the C$_6$H column density of 
$4.5\times10^{12}$\,cm$^{-2}$, which is about six times lower than our result,
and even worse when comparing with the modelling result.
Thus the disagreements are likely to result from the
uncertainties of the model. The other molecules that cannot be modelled well are
 PN, H$_{2}$CS, H$_{2}$CO, and SiO, which  show  the  observed-to-calculated
 abundance ratios of 826, 42, 151, and 0.001, respectively.
 Figure~\ref{Figure8} shows the modelled abundances of the molecules in CRL\,2688 as a function of radius as well as the observed values. It is obvious that
the observed abundances of PN, H$_{2}$CS, and H$_2$CO are higher than
the modelled results at any radius. 
 
Apart from the uncertainties introduced by observations and data analysis, the possible causes
of the disagreements include poor knowledge of the physical conditions, missing routes
in the modelling, and inaccurate reaction rates. To decouple the affects of uncertain
physical conditions, we compare the  observed-to-calculated abundance ratios of the molecules in IRC+10216, the progenitor object of CRL\,2688. The results are shown in Fig.~\ref{Figure7}. 
The modelled and observed data of IRC+10216 are taken from \citet{McElroy13} and the references
therein. Generally, the agreement of the molecules in IRC+10216  is slightly better than
those in CRL\,2688. However, it is interesting to note that the models overestimate the
abundances PN, H$_{2}$CS, and H$_2$CO in CRL\,2688 and IRC+10216 by about the same degree,
suggesting that significant reaction routes for forming these molecules
might been neglected or undervalued in the modelling.

The enhancement of SiO has been commonly observed in C-rich AGB stars, and can be attributed
to shock-induced chemistry \citep{so06} that has not been taken into account in this model.
CRL\,2688 shows rich shock features caused by fast outflows. However, the largely overestimated 
SiO abundances by the model are against with the predictions of shock chemistry. We infer that 
the model might substantially underestimate the freeze-out of SiO in the PPN stage.
Alternatively, because the fast outflows in CRL\,2688 are highly collimated, the  shock chemistry processes
only in rather small regions and does not play an important role on the enhancement of gas-phase SiO. Severe depletion of SiO and SiS has been also found in another PPN 
 IRAS\,$22272+5435$  \citep{Zhang20aa}, which exhibits strong 21\,$\micron$ feature.
 The CO observations suggest that the material ejection of IRAS\,$22272+5435$
 is dominated by the spherical wind, but a jet is developing and interacting with
 ambient materials \citep{nk12}. CRL\,2688 has well developed jets and is thought
 to just metamorphose from a `21\,$\micron$ phase' \citep{Geballe92}. Therefore,
we reasonably conjecture that the depletion of Si-bearing molecules and the development
of jets may be associated to the phenomenon of the 21\,$\micron$ feature in PPNs.
Sensitive observations of a large PPN sample are required to draw firm conclusion.

\subsection{Metal-bearing molecules}
\label{sec4.1}

CSEs surrounding evolved stars are characterized by unambiguous identification
of metal-bearing (Na-, Mg-, Al-, K-, Ca-, and Fe-bearing) molecules.
These molecules have never been discovered in other types of sources, even in
the rich molecular cloud TMC-1, which might be attributed to severe depletion
onto dust grains \citep{Turner05}.
So far, the detected metal-bearing molecules in IRC+10216 include
NaCl, NaCN, MgCN, MgNC, HMgNC, MgC$_2$H, MgC$_3$N, MgC$_4$H, MgC$_5$H, MgC$_6$H, 
AlNC, AlF, AlCl, KCl, KCN, CaNC, and FeCN
\citep[][and references therein]{Cernicharo19,Pardo21},
indicating a rich gas-phase metal chemistry in C-rich circumstellar environments.
The first identification of metal halides was made by \citet{Cernicharo87}, who also 
reported a tentative detection of AlF, later confirmed by \citet{Ziurys94}.
According to the chemical equilibrium calculations \citep{Tsuji73},
these stable closed-shell diatomic molecules are mostly formed
in the warn inner shell, where the LTE condition holds. 
\citet{kaw93} identified MgNC for the first time in the spectrum of IRC+10216 presented by \citet{gue86}. 
The metal cyanide has an open-shell structure and should
be formed in the outer envelope along with other radicals,
facilitating the successful detection of NaCN in
 IRC+10216 \citep{Guelin93,Turner94}. 
 The different spatial distributions of metal halides and metal cyanides are revealed by line profiles and high spatial resolution maps of IRC+10216 \citep[][and references therein]{Ziurys06}. 
When the CSE expands, materials move from the inner to outer regions. As a result, one could expect that the 
AlF/MgNC abundance
ratio  decreases through the transition from the AGB to the PPN phase. This can be tested by measuring the 
abundance of metal-bearing molecules in CRL\,2688.

Our observations result in the detection of AlF, MgNC, and { NaCN} in CRL\,2688.  All 
the three Metal-bearing molecules have already been discovered by \citet{Highberger01}
in the 1.2 and 2\,mm windows. Apart from the three species, the only metal-bearing
molecule detected in CRL\,2688 is NaCl \citep{Highberger03a}. Unfortunately, there is no
NaCl line lying within our frequency range. Following \citet{Highberger01}, we plot the
rotation diagram of AlF, NaCN, and MgNC by assuming the source sizes to be respectively
5\arcsec, 5\arcsec, and 30\arcsec  for the corrections of the beam-dilution effect, 
as displayed in Fig.~\ref{Figure9}. For this we use the $J=4 \rightarrow 3$, $5 \rightarrow 4$, and $8 \rightarrow 7$ transitions for AlF,
the $N=8 \rightarrow 7$, $14 \rightarrow 13$, and $15 \rightarrow 14$ transitions for MgNC, 
and the 
 $J_{K_a,K_c}=5_{0,5} \rightarrow 4_{0,4}$, $5_{1,4} \rightarrow 4_{1,3}$, $6_{0,6} \rightarrow 5_{0,5}$, $6_{2,4} \rightarrow 5_{2,3}$, $6_{1,5} \rightarrow 5_{1,4}$, $10_{1,9} \rightarrow 9_{1,8}$, $10_{1,11} \rightarrow 10_{1,10}$, $10_{0,11} \rightarrow 9_{0,10}$, and $10_{1,10} \rightarrow 9_{1,9}$  transitions for NaCN. 
 Furthermore, we perform the rotation-diagram analysis by assuming various source sizes, as listed
 in Table~\ref{Table7}.  It is clear that AlF and MgCN have the highest and lowest
  excitation temperatures, respectively, supporting their origins respectively from the inner and outer shells. The excitation temperature of NaCN is closer to that of MgCN, as opposed
  to the theoretical expectation that NaCN is an inner-layer molecule.
 
  For comparison, Figure~\ref{Figure9} and  Table~\ref{Table7} show the results of 
 \citet{Highberger01} who used the  same transitions for AlF,
the $N=11 \rightarrow 10$, $12 \rightarrow 11$, and $13 \rightarrow 12$ transitions for MgNC, and 
the $J_{K_a,K_c}=9_{0,9} \rightarrow 8_{0,8}$, $10_{0,10} \rightarrow 9_{0,9}$, $10_{3,8} \rightarrow 9_{3,7}$, $10_{3,7} \rightarrow 9_{3,6}$, and $10_{2,8} \rightarrow 9_{2,7}$ transitions for NaCN. 
We find that
the results of AlF are consistent with those of \citet{Highberger01},
whereas there a slight and significant disagreement for MgNC and NaCN, respectively.
Under the assumption of a source size of 5$\arcsec$, we obtain the column density of NaCN 
an order of magnitude lower than that by \citet{Highberger01}. However, the difference
can be alleviated if NaCN arises from a more extended shell. Based on observed line profiles,
\citet{Highberger03a} suggested that NaCN in CRL\,2688 should be sited much farther away from the central star than that in IRC+10216. In Fig.~\ref{Figure10}, we show the column densities calculated by using different source sizes for the beam-dilution corrections.
We can see that if the source sizes of AlF and MgNC are 5$\arcsec$ and 20$\arcsec$, respectively,
our results are perfectly consistent with those of \citet{Highberger01}.
However, even for source size of 30$\arcsec$,
the column density of NaCN derived by \citet{Highberger01} is still 6 times larger than our value,
suggesting that the disagreement cannot be completely ascribed to the uncertainty of the source size.

The fractional abundances are determined following the approach described in \citet{Highberger01} and \citet{Highberger03b}. The input parameters and the results are listed in Table~\ref{Table8}. 
The assumed geometries are the same as those used by \citet{Highberger01} and \citet{Highberger03b};
AlF is sited in a spherical envelope while NaCN and MgNC are confined on spherical shells. 
We derive a AlF/MgNC abundance ratio of $\sim$0.7, significantly lower than that in IRC+10216 ($\sim$3), supporting the theoretical expectation of the CSE expansion. 
Further support for the chemical evolution comes from the observations of CRL\,618, the presumable descendant of CRL\,2688, in which MgNC is the only discovered metal-bearing molecule \citep{Highberger03b}.

The fractional abundances of these metal-bearing molecules can provide a lower limit
of the gas-phase elemental abundance. Our calculations show that magnesium locked in MgNC
has a Mg/H ratio of $2\times10^{-8}$ (Table~\ref{Table8}). 
According to the element depletion parameters given by \citet{jen09}, one can estimate that
the gas-phase Mg/H ratio in the ISM is larger than
2 $\times$ 10$^{-6}$. If assuming that solar abundances can well reflect the total elemental abundance and the depletion factor in the PPN is similar to that of the ISM, we could conclude that less than $1\%$ gas-phase magnesium is in the form of MgNC, and thus MgNC may not 
be the dominant Mg-bearing molecule in CRL\,2688.

Fluorine has a cosmic abundance much lower than aluminium, and thus is the key element of AlF. The astrophysical origin of fluorine remains a widely debated problem \citep[see, e.g.,][and  references therein]{ic17,sm18}. Because the only stable fluorine isotope, $^{19}$F, is fragile in hot stellar interiors, the most possible explanation for cosmic
fluorine abundance is that most of the F escapes from the harsh stellar interiors
immediately after its production. The AGB star is an ideal source of F, where
$^{19}$F is synthesized and brought to stellar surface 
during the He-burning thermal pulses and the third dredge-up
\citep{Forestini92,Jorissen92}. The F enhancement in AGB stars has been
confirmed by the observations of F ions in PNs \citep[e.g.][]{Zhang05}.
The most abundant interstellar F-bearing molecule is HF, which, however, is inaccessible
from ground telescopes. CF$^{+}$ has been found to be the second abundant one
in photodissociation regions \citep{ns06}. Our research group has systematically searched for
 CF$^+$ in CSEs, but failed to see any positive detection
 \citep{zk14}. Therefore, AlF provides an unique opportunity to constrain the F
 abundance in CSEs. 
 A discussion on this aspect has been made by 
 \citet{Highberger01}. We reexamine this problem  using our data of CRL\,2688.
 The derived AlF/H$_2$ abundance ratio is 2.7 $\times$ 10$^{-8}$ (Table~\ref{Table8}), suggesting a F/H abundance
 of $>$1.4 $\times$ 10$^{-8}$.
The obtained lower limit of F/H is only 2.4 times lower than the solar value  of \citet{Lodders03}, and
is presumably far below the fluorine abundance in CRL\,2688 since AlF is not the dominant F-bearing molecule
in circumstellar environments.
Although the $\alpha$ element Al in the PPN can be enhanced through the third drudge-up,
  given the fact that available hydrogen is much richer than aluminium  in circumstellar environments 
 \citep[solar Al/H =3 $\times$ 10$^{-6}$;][]{Lodders03}, 
 the abundance of HF
 should be larger than AlF by orders of magnitude. As a result, we infer that CRL\,2688
 has a significant F enhancement. Combined with the C-rich nature of CRL\,2688,
 our results further confirm the third drudge-up in AGB stars
 as the source of F production.

\section{Summary}
\label{sec5}

CRL\,2688 is a prototypical source suitable for studying the chemical alteration of CSEs during
the transition from AGB to PPN phase. It might hold key clues to investigate the
puzzle of the 21\,$\micron$ feature. This paper reports a new observation towards 
this PPN at the 3 mm and 2 mm bands, which results in clear detections
of 196 transition lines arising from 38 molecules. Although no new circumstellar molecule is discovered, 
 { 153} transition lines and  { 13} molecules are new detections in this object. 
We compile a comprehensive list of the gas-phase molecules discovered in CRL\,2688 so far, providing a valuable reference for the community of circumstellar chemistry.
By comparing the line intensities observed by different telescopes, we  estimate 
{ a  Gaussian brightness FWHM of 15$\rlap{.}\arcsec$5 or a source size 
of 27$\rlap{.}\arcsec$7 for a disk geometry, }
although the exact sizes are different from
species to species.
The standard chemical modelling is unable to reproduce
the observed abundances of PN, H$_{2}$CS, and H$_{2}$CO, suggesting that 
some key reaction routes may have been missed in the chemical network.
Through a comparison  between the molecular abundances in IRC+10216, CRL\,2688,
and CRL\,618, we conclude that the differences can be partially attributed to the effect 
of CSE evolution. The three CSEs are the only  sources where metal-bearing molecules are detected. 
The results of the analysis  support the  theoretical expectations
that AlF is confined in the inner shell while MgNC and NaCN are located on extended regions. An enhancement of F
in the PPN is revealed by the fractional abundance of AlF relative to H$_{2}$.

The chemical processes in CSEs are complex. In-depth explorations of this research
field call for high-sensitivity and broad-frequency-coverage observations of a large sample, which, however,
are relatively scarce, and have been limited to only a few bright CSEs. In the future,
such efforts should be vigorously pursued to decouple the effects of various factors, including stellar
and circumstellar properties (stellar mass, metallicity, binary system, mass-loss rate, wind velocity, radiation field, dust formation, shocks, etc.), on the circumstellar
chemistry.

\acknowledgments

We would like to thank an anonymous referee whose comments significantly improved the paper. 
This work was supported by National Science Foundation of China (NSFC, Grant No. 11973099) awarded to YZ. 
JJQ and JSZ thanks the support of NSFC (Grant Nos.  12003080, 12041302, and 12073088), 
the China Postdoctoral Science Foundation funded project (No. 2019M653144), the Guangdong Basic and Applied Basic Research Foundation (No. 2019A1515110588), and the Fundamental Research Funds for the Central Universities (Sun Yat-sen University, No. 2021qntd28 and 19lgpy284). 
We also acknowledge the science research grants from the China Manned Space Project with NO. CMS-CSST-2021-A09 and CMS-CSST-2021-A10.
We wish to express my gratitude to the staff at the IRAM 30 m telescope for their kind help and support during our observations.
IRAM is supported by INSU/CNRS (France), MPG (Germany), and IGN (Spain). 
This work presents results from the European Space Agency (ESA) space mission Gaia. Gaia data are being processed by the Gaia Data Processing and Analysis Consortium (DPAC). Funding for the DPAC is provided by national institutions, in particular the institutions participating in the Gaia MultiLateral Agreement (MLA). The Gaia mission website is \url{https://www.cosmos.esa.int/gaia}. The Gaia archive website is \url{https://archives.esac.esa.int/gaia}.

\clearpage

\clearpage

\clearpage
\renewcommand{\thefigure}{\arabic{figure} (Cont.)}
\addtocounter{figure}{-1}

\begin{figure}
\gridline{\fig{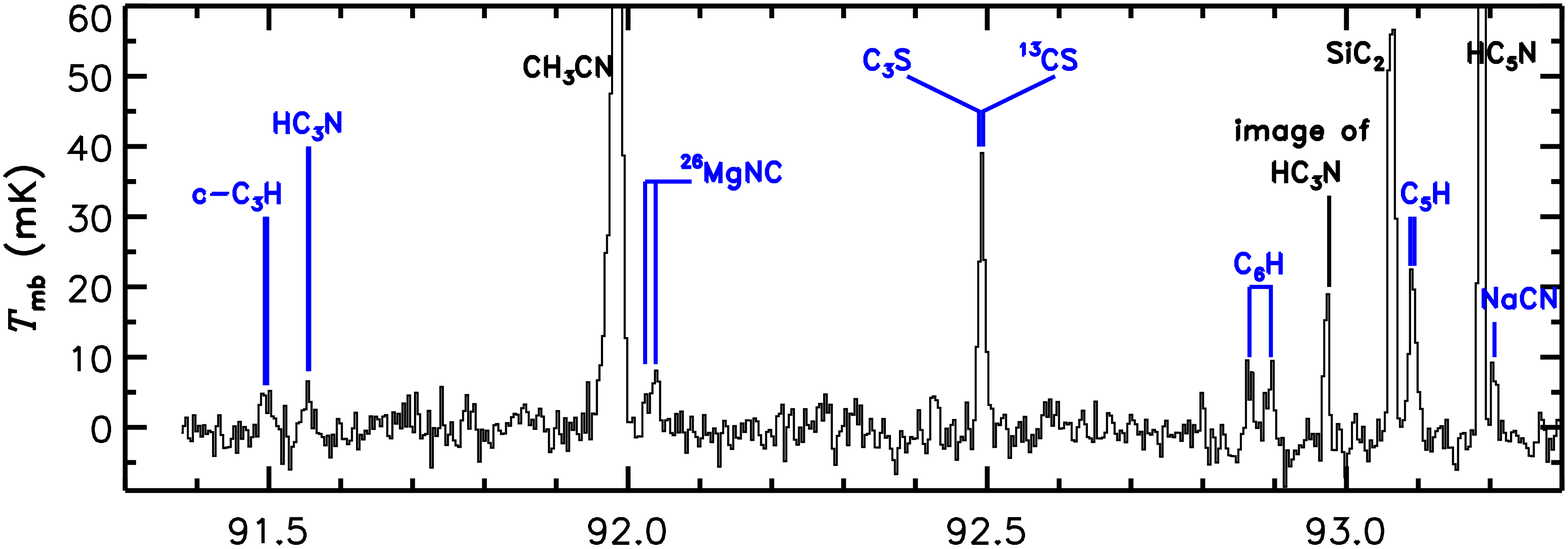}{0.9\textwidth}{}
         }
\vspace{-1.4cm}
\gridline{\fig{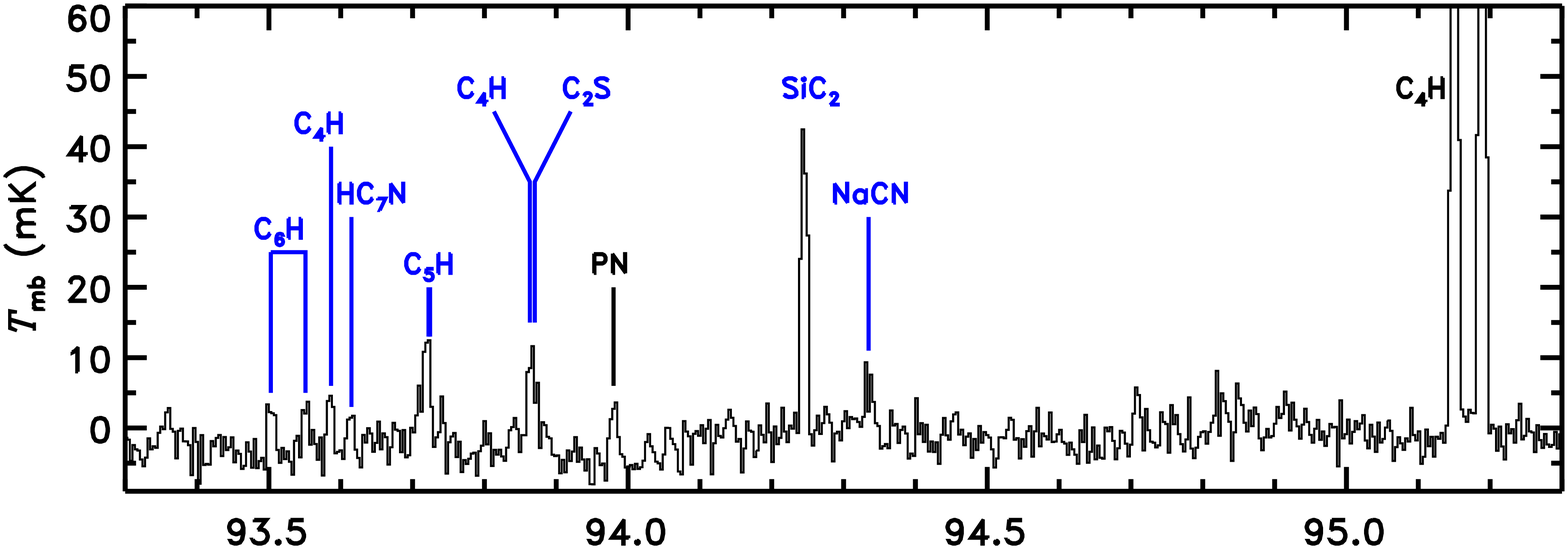}{0.9\textwidth}{}
         }
\vspace{-1.4cm}
\gridline{\fig{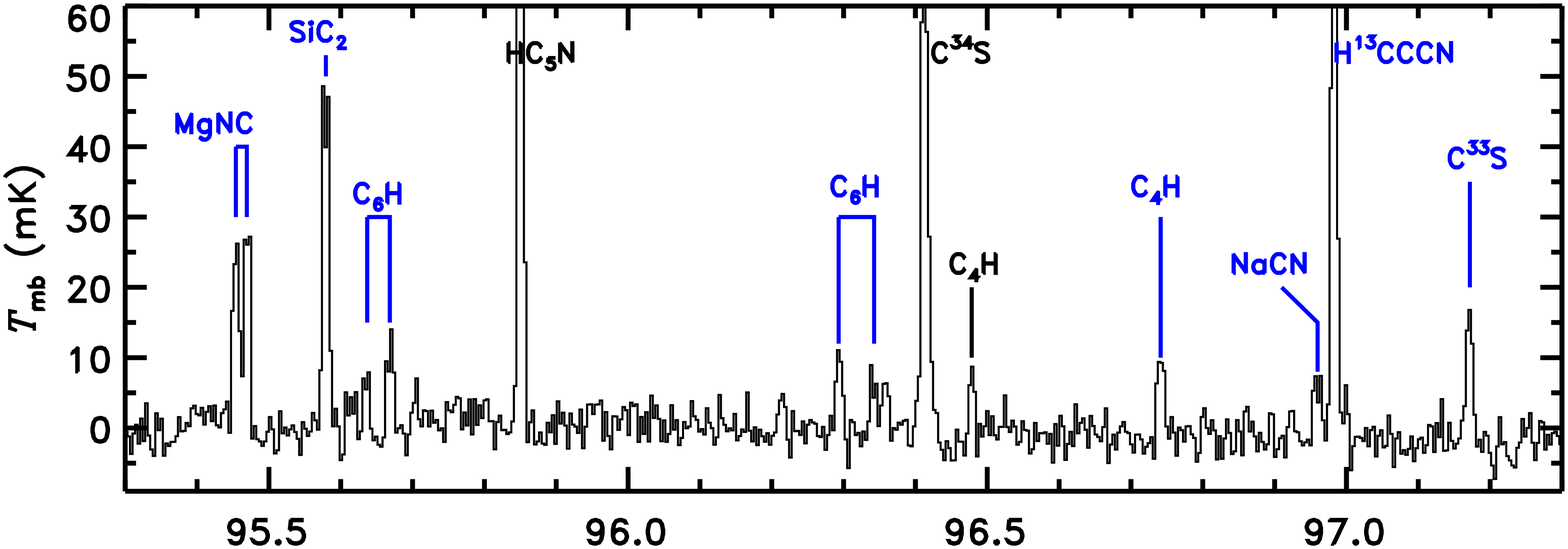}{0.9\textwidth}{}
         }
\vspace{-1.4cm}
\gridline{\fig{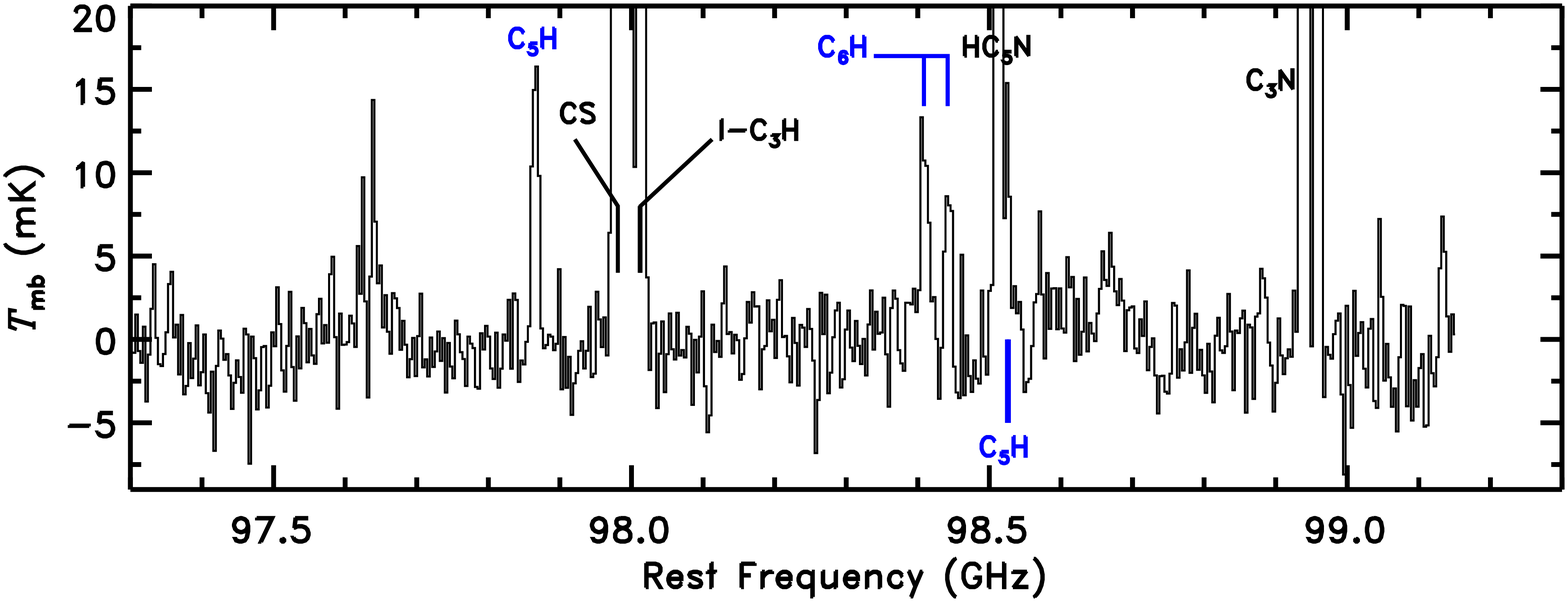}{0.9\textwidth}{}
         }
\caption{    
}
\end{figure}

\clearpage

\addtocounter{figure}{-1}

\begin{figure}
\gridline{\fig{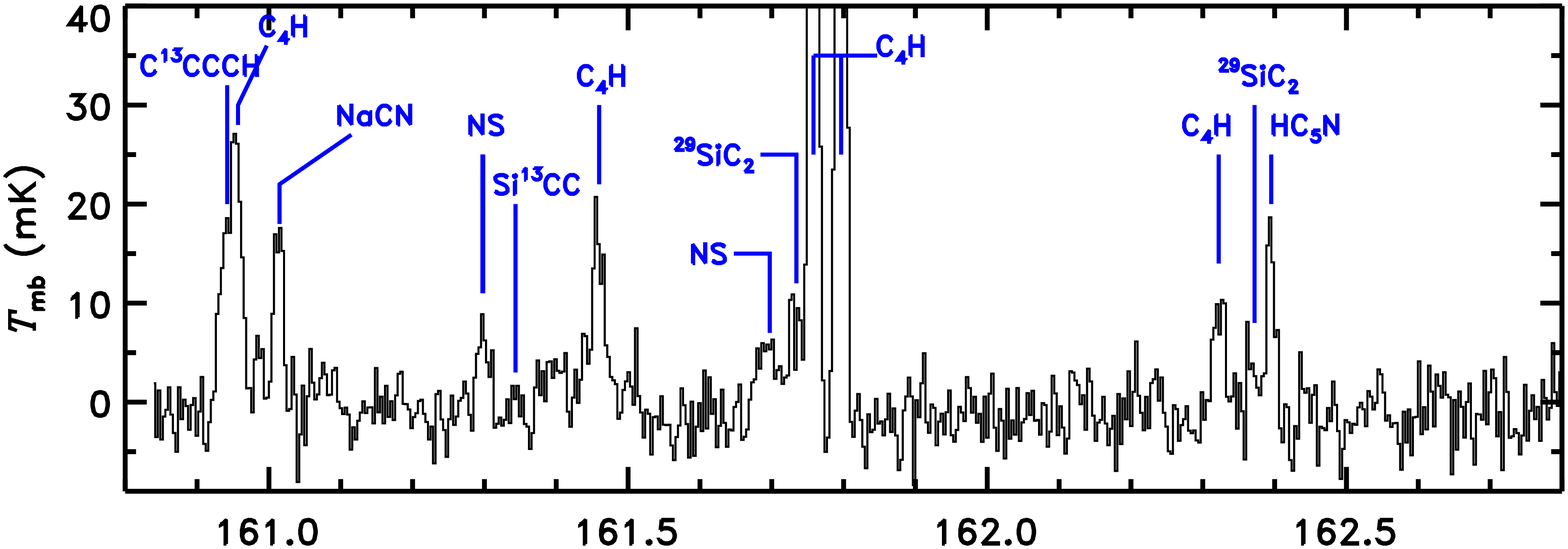}{0.9\textwidth}{}
         }
\vspace{-1.4cm}
\gridline{\fig{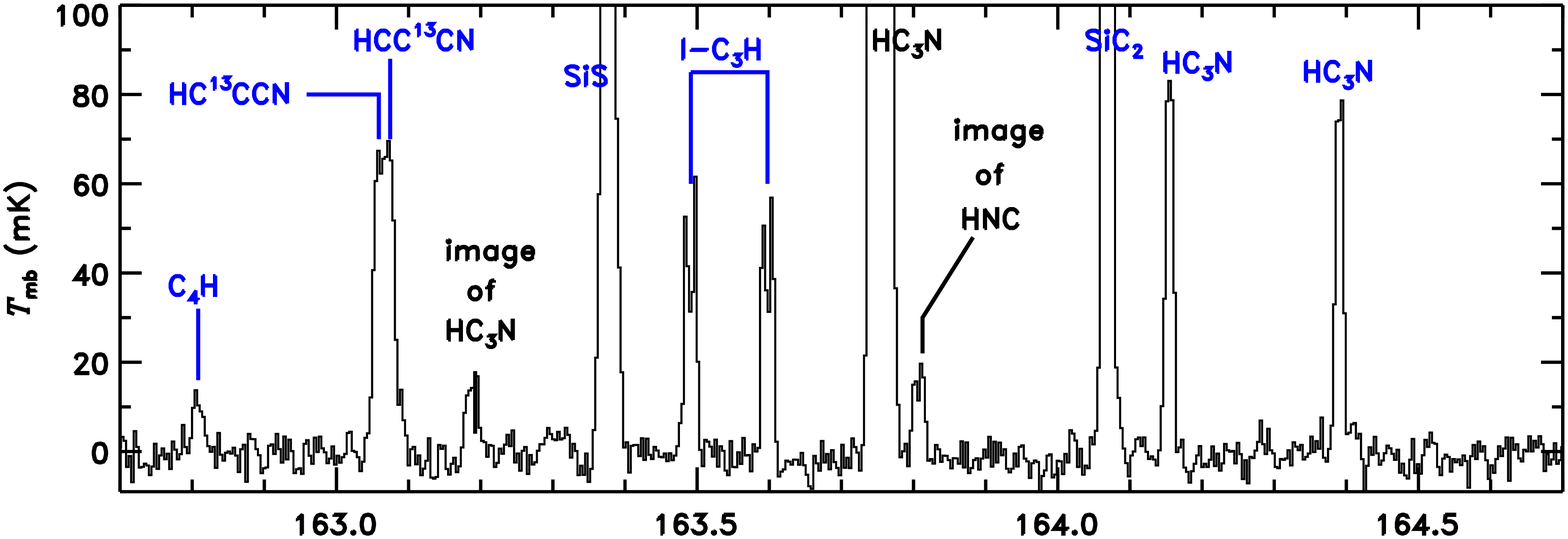}{0.9\textwidth}{}
         }
\vspace{-1.4cm}
\gridline{\fig{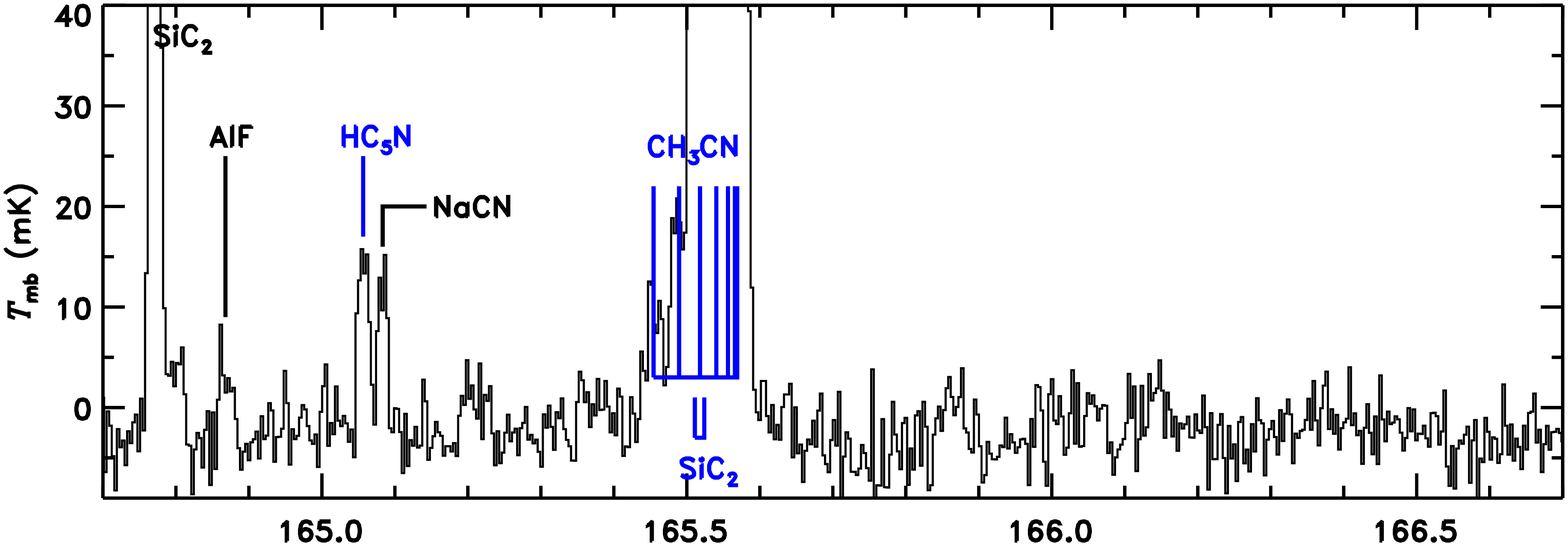}{0.9\textwidth}{}
         }
\vspace{-1.4cm}
\gridline{\fig{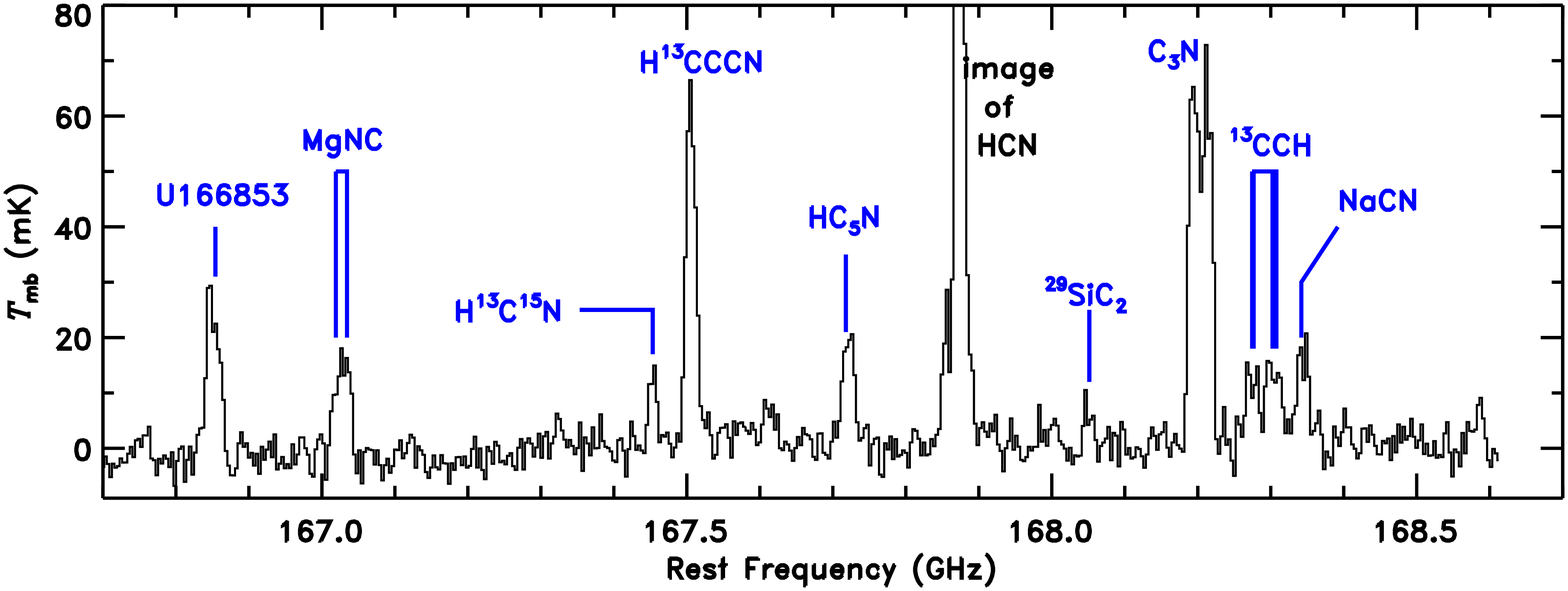}{0.9\textwidth}{}
         }
\caption{    
}
\end{figure}

\clearpage
\addtocounter{figure}{-1}

\begin{figure}
\gridline{\fig{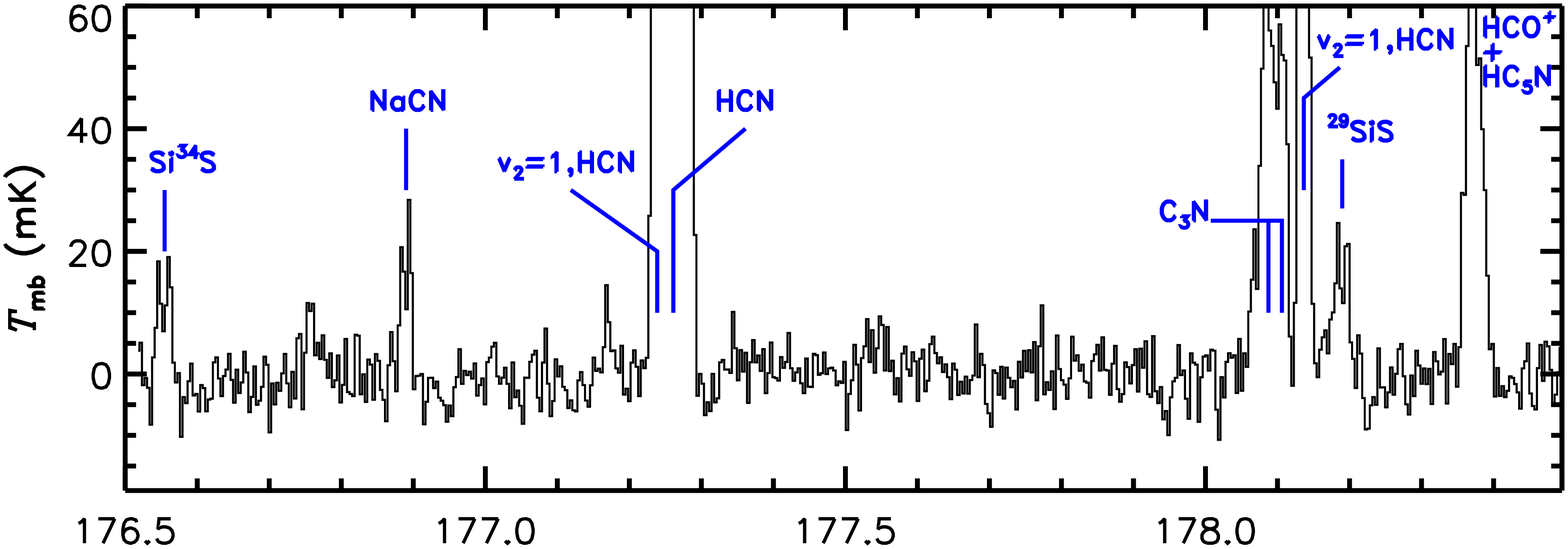}{0.9\textwidth}{}
         }
\vspace{-1.4cm}
\gridline{\fig{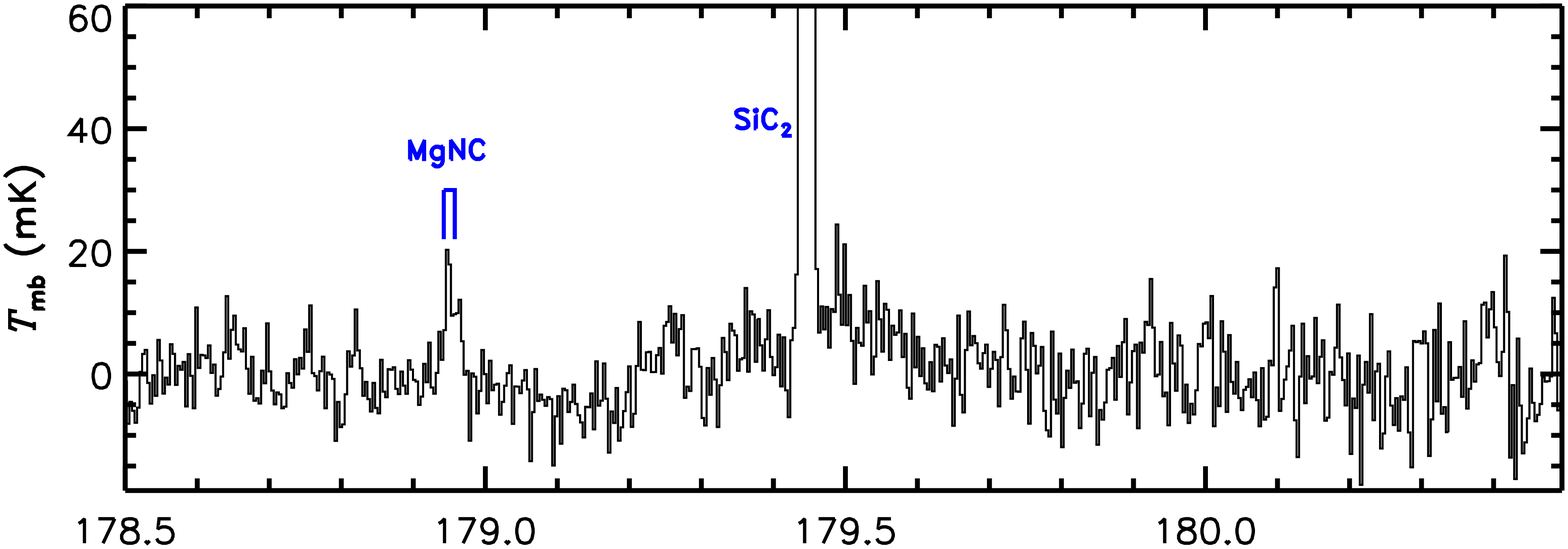}{0.9\textwidth}{}
         }
\vspace{-1.4cm}
\gridline{\fig{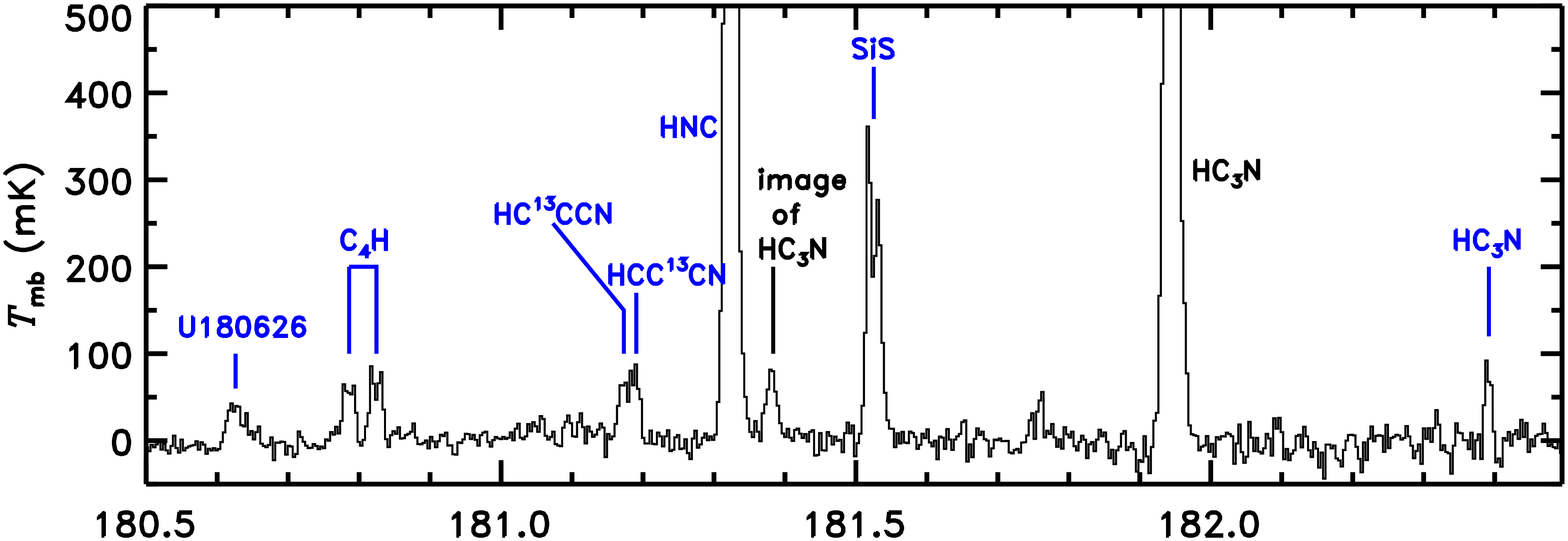}{0.9\textwidth}{}
         }
\vspace{-1.4cm}
\gridline{\fig{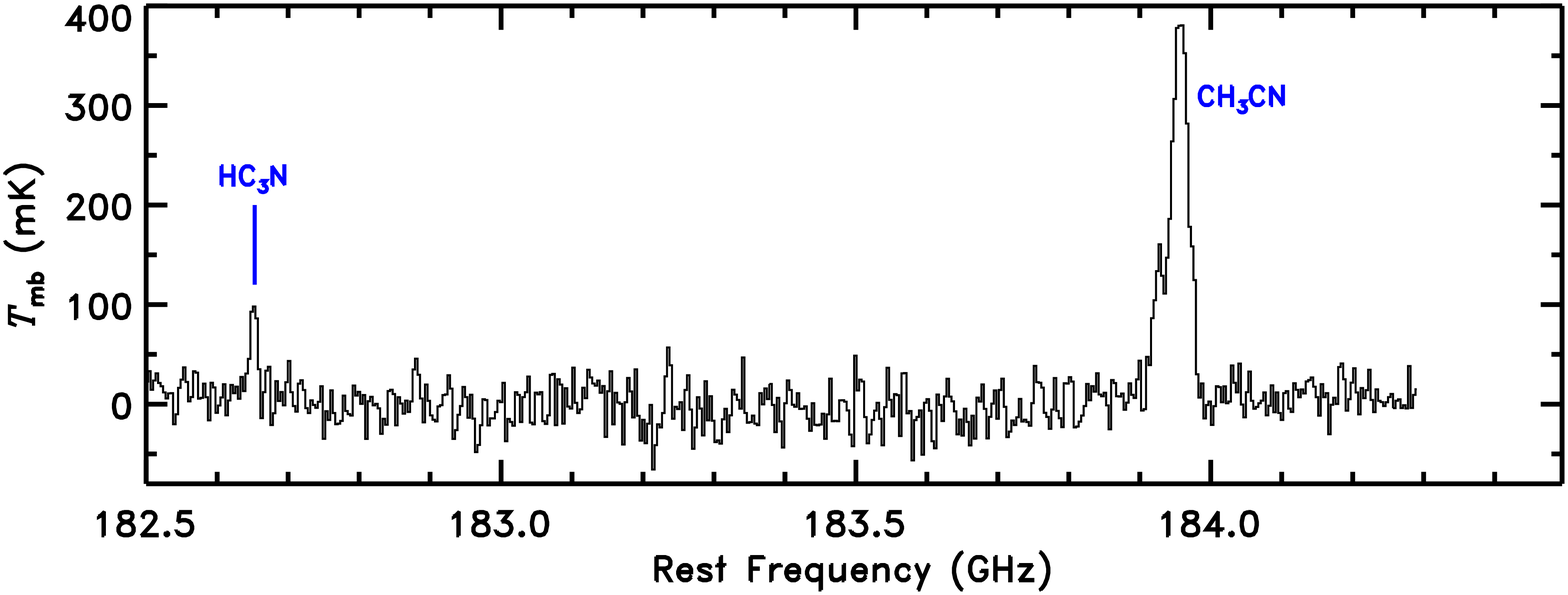}{0.9\textwidth}{}
         }
\caption{    
}
\end{figure}
\clearpage
\renewcommand{\thefigure}{\arabic{figure}}

\begin{figure}
\gridline{\fig{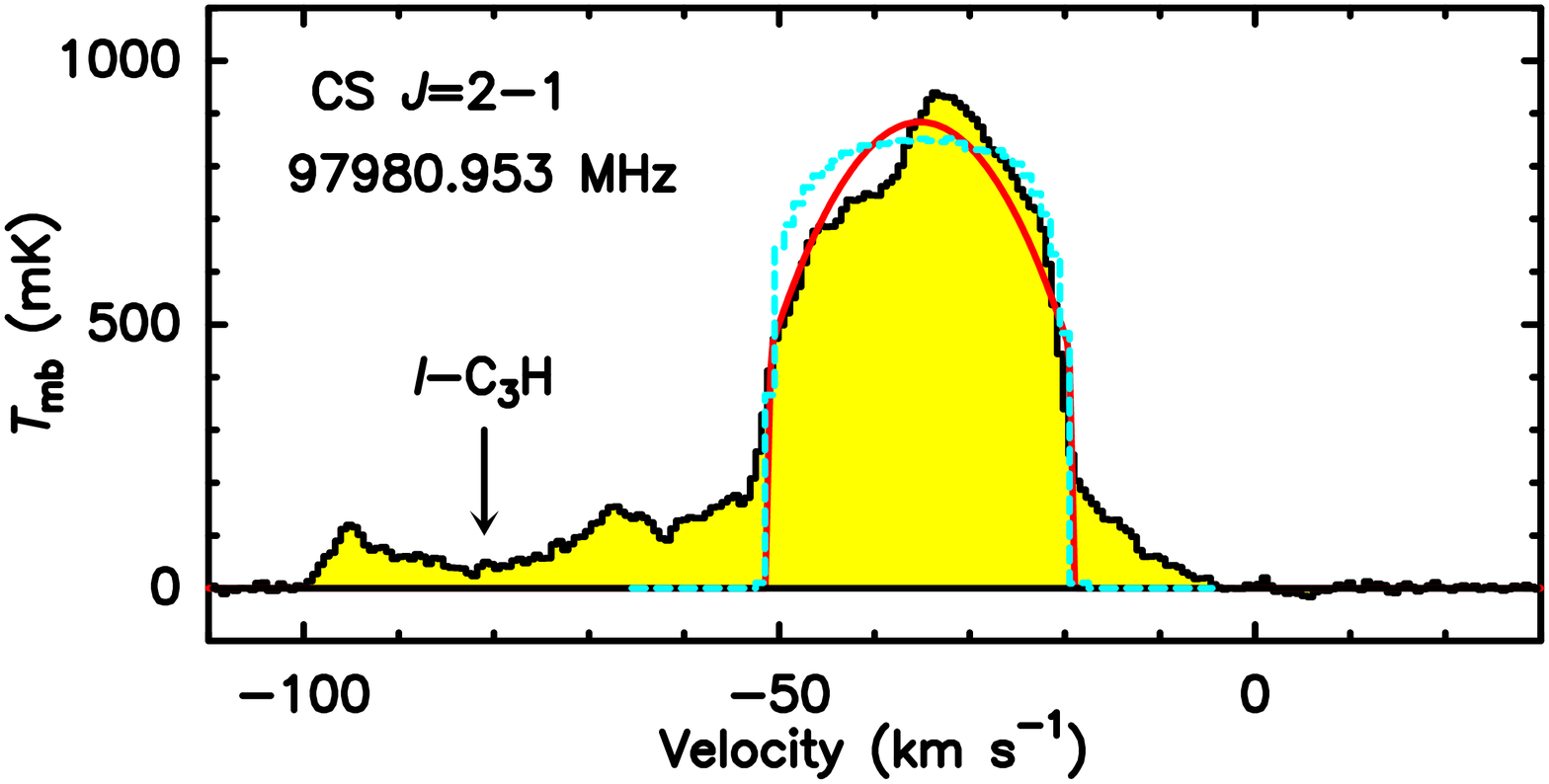}{0.45\textwidth}{}
          \fig{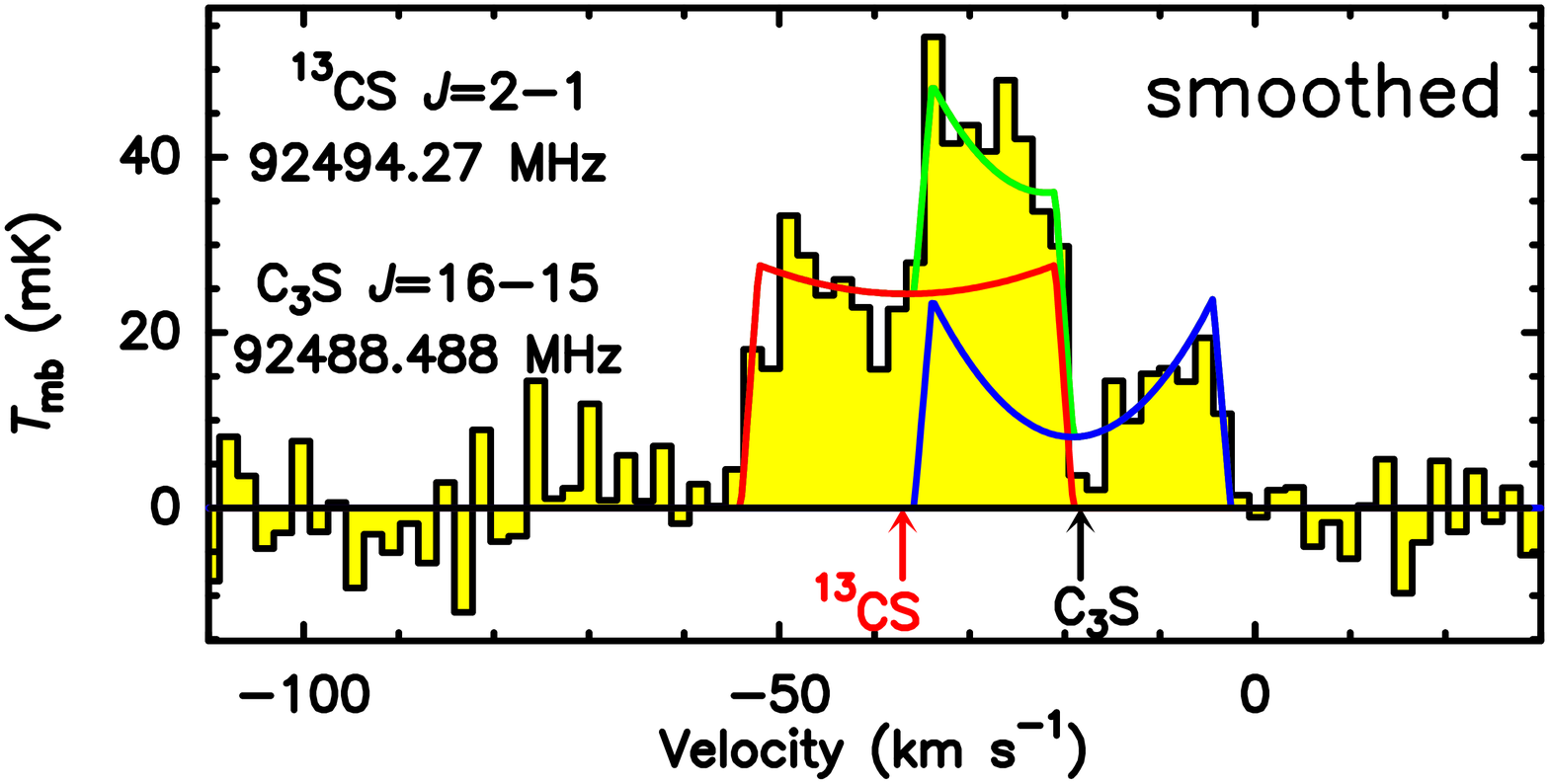}{0.45\textwidth}{}
         }
\vspace{-0.5cm}
\gridline{\fig{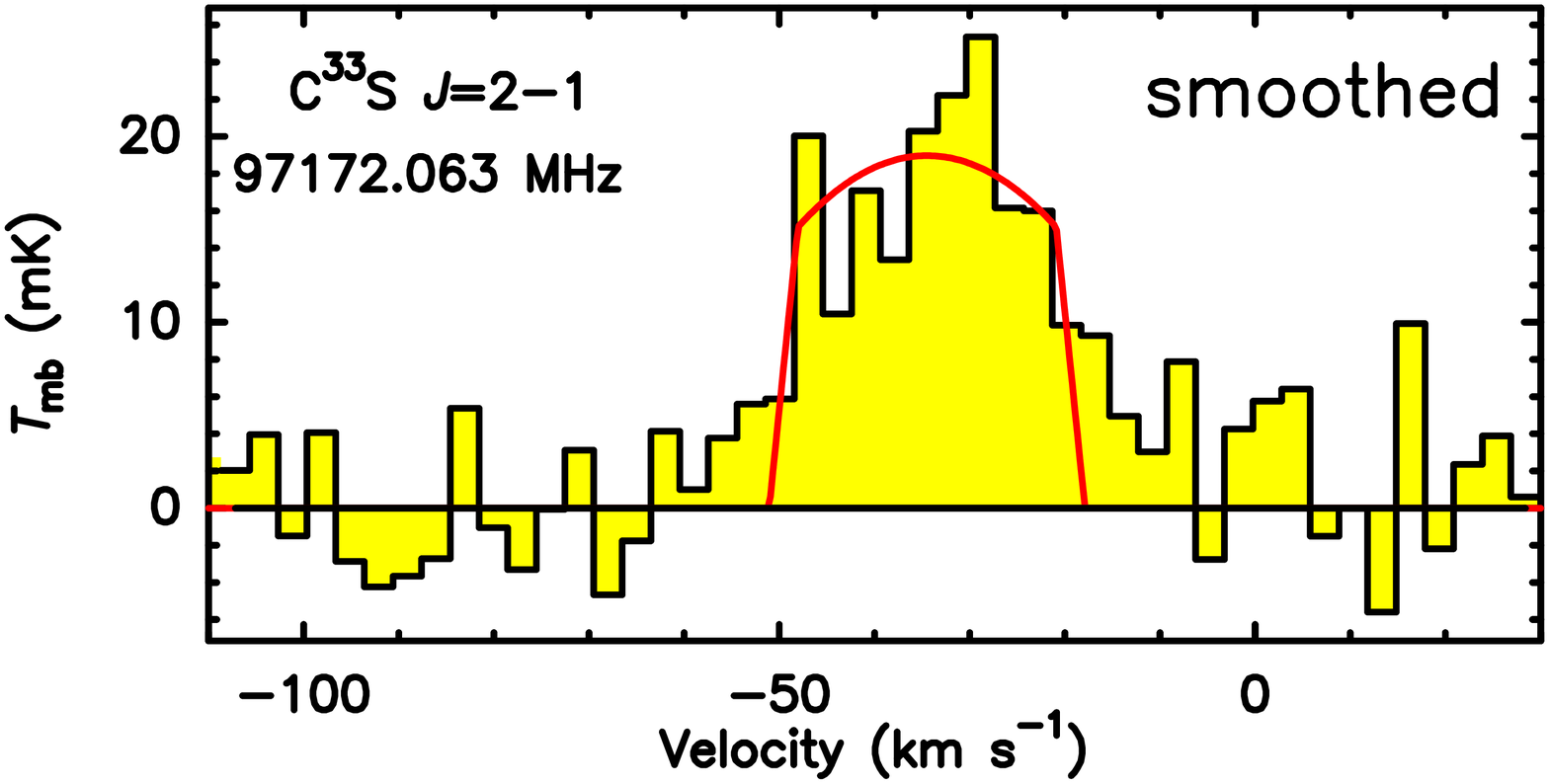}{0.45\textwidth}{}
          \fig{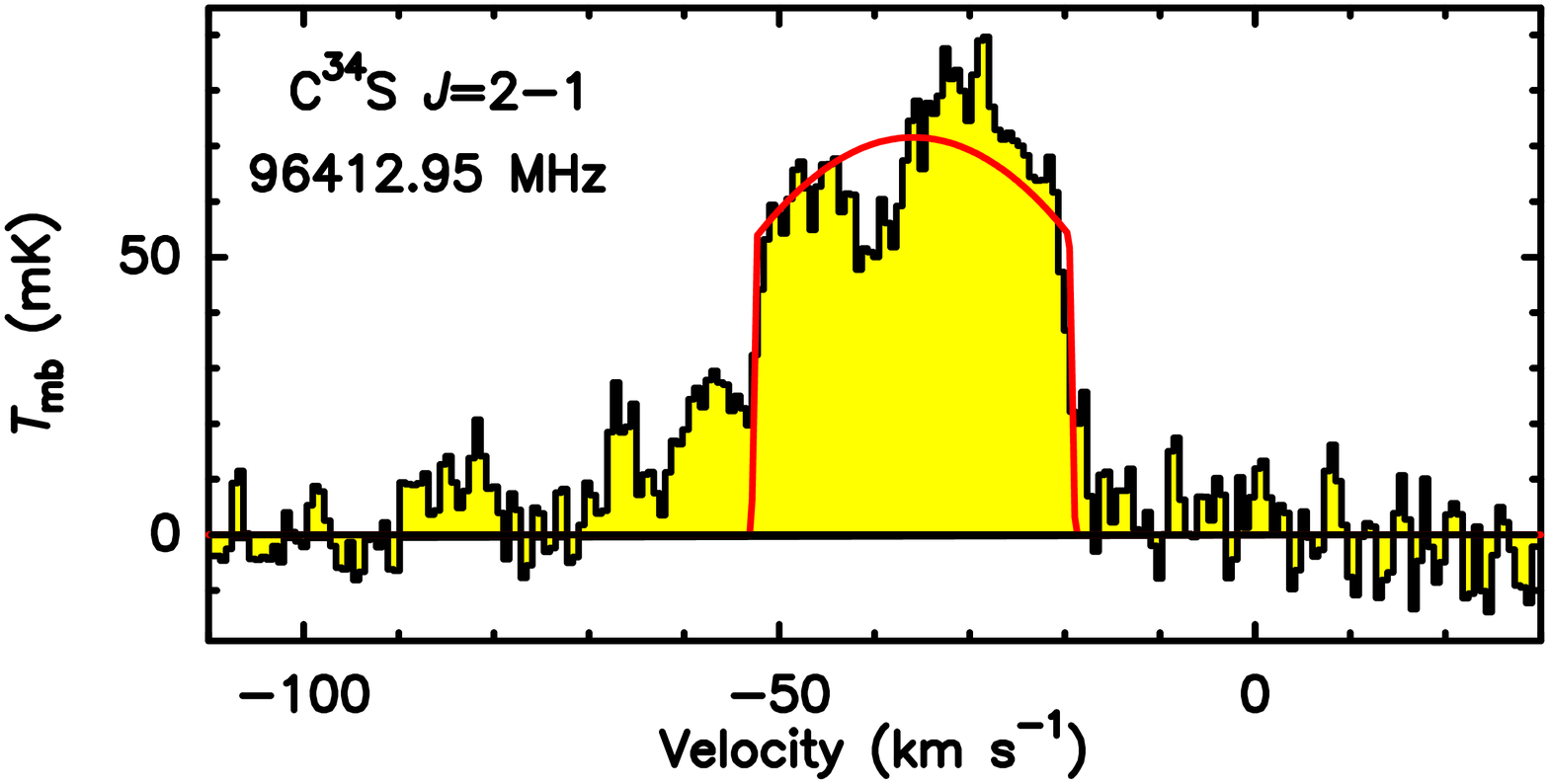}{0.45\textwidth}{}
         }
\caption{  
The emission lines of CS and its $^{13}$C, $^{33}$S, and $^{34}$C isotopologues. The emission line of other molecules can be found in the online figure set.
The red and blue curves represent the stellar-shell fitting of each transition. 
The green curve represents the total stellar-shell fitting of blended transitions. 
The cyan dashed curve represents the line profile resulting from 
the LIME modelling. 
The vertical lines mark the positions and relative intensities of hyperfine components. 
The panels marked with `smoothed' show the smoothed spectra. 
(The complete figure set (116 images) is available in the online journal.)
\label{Figure3}
}
\end{figure}

\figsetstart
\figsetnum{3}
\figsettitle{
The emission lines of molecular species and their isotopologues. The red and blue curves represent the stellar-shell fitting of each transition. 
The green curve represents the total stellar-shell fitting of blended transitions. 
The cyan dashed curve represents the line profile resulting from 
the LIME modelling. 
The vertical lines mark the positions and relative intensities of hyperfine components. 
The panels marked with `smoothed' show the smoothed spectra. (See the published version for the figure set.)
\label{figureset}
}

\figsetend

\clearpage

\begin{figure}
\gridline{\fig{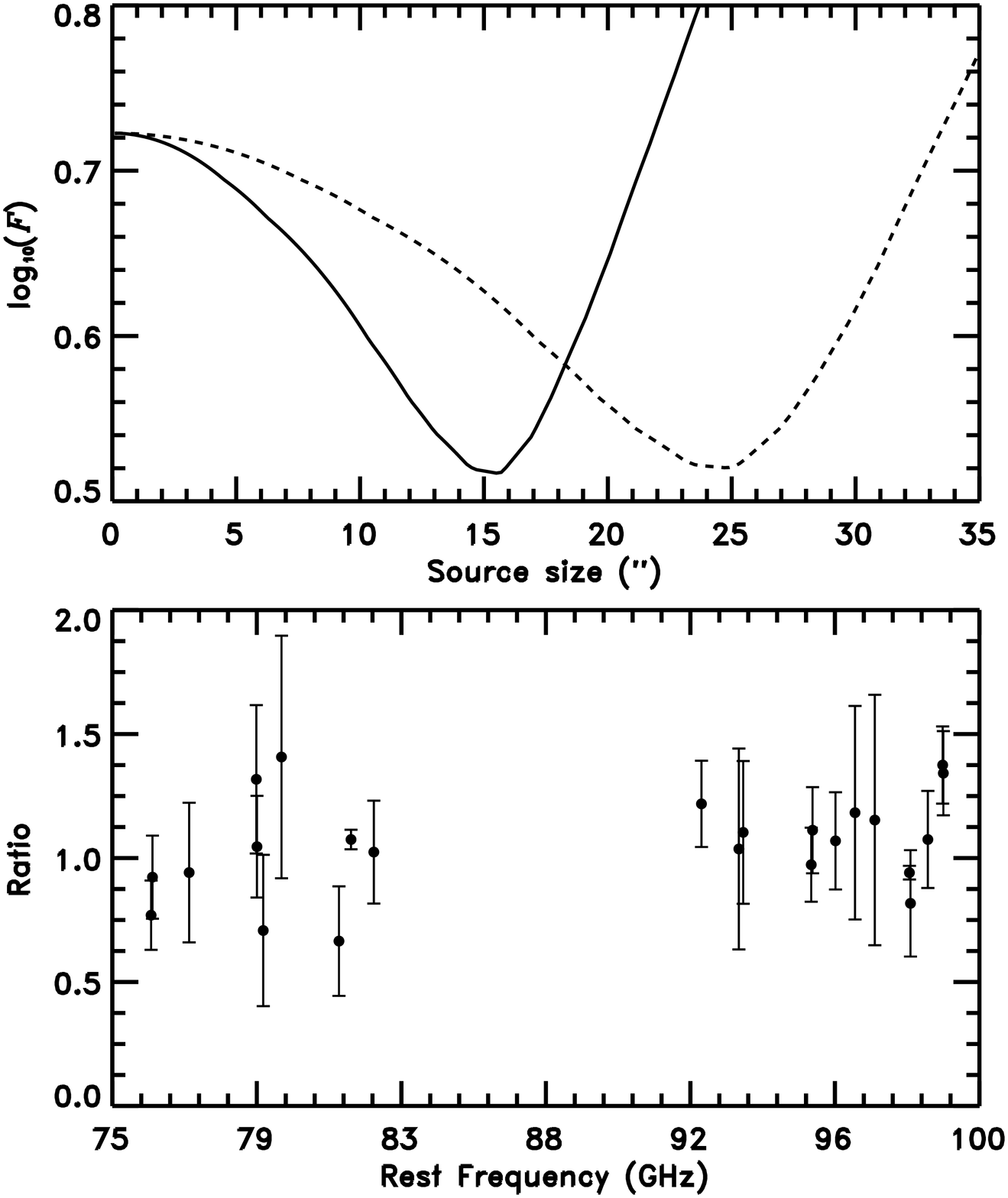}{0.8\textwidth}{}
            }
\caption{{\it Upper}: The $F$ factor versus the source size ($\theta_{\rm s}$ or $\theta_{\rm d}$), where the solid and dashed
curves represent those calculated assuming Gaussian and disk brightness distributions, respectively.  {\it Lower}: The velocity-integrated intensity ratios 
of the lines detected by the IRAM 30 m and ARO 12 m telescopes,
after correcting for the beam-dilution effect. See the text for the details.
\label{Figure4}
}
\end{figure}

\clearpage

\begin{figure}
\gridline{\fig{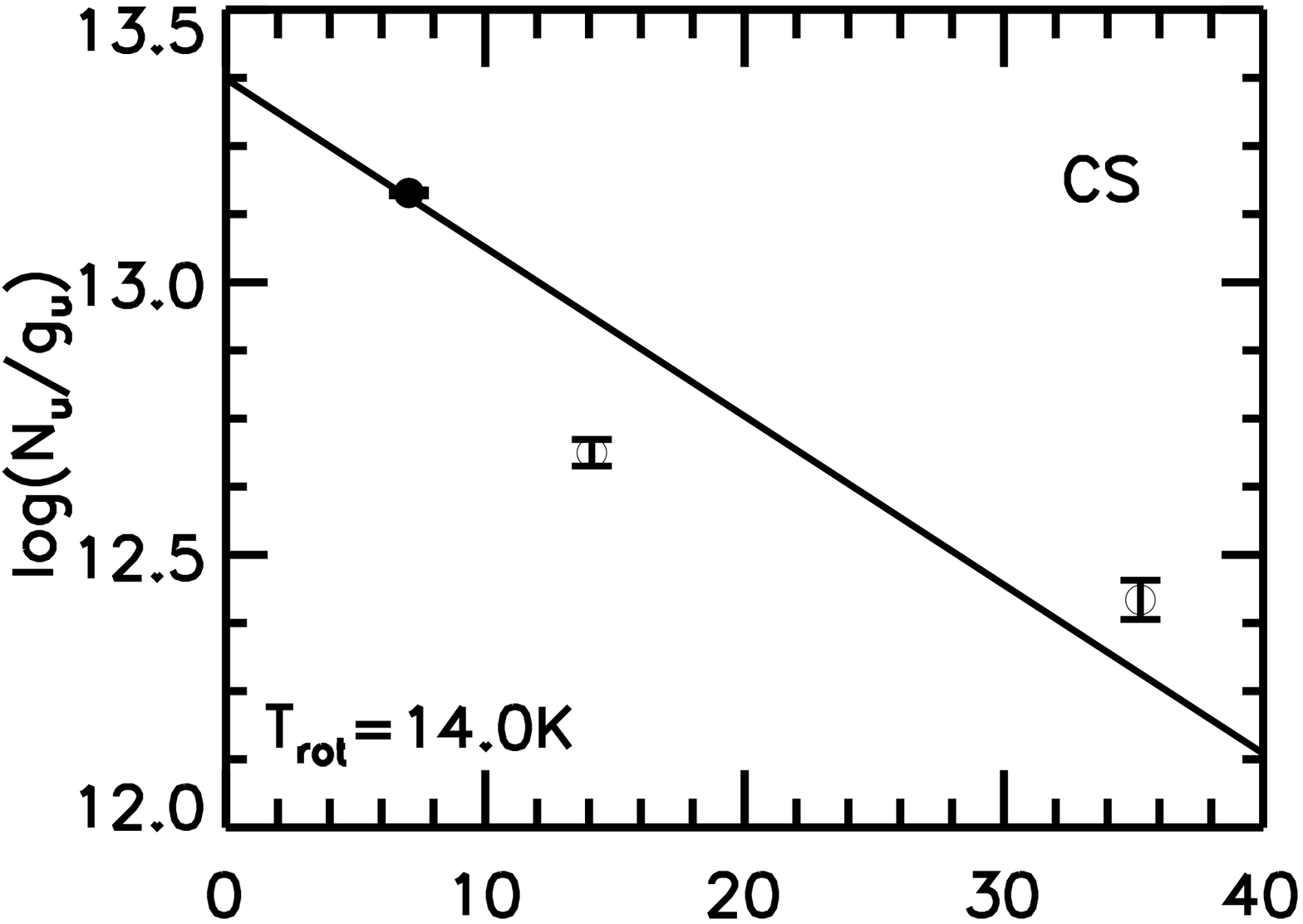}{0.45\textwidth}{}
          \fig{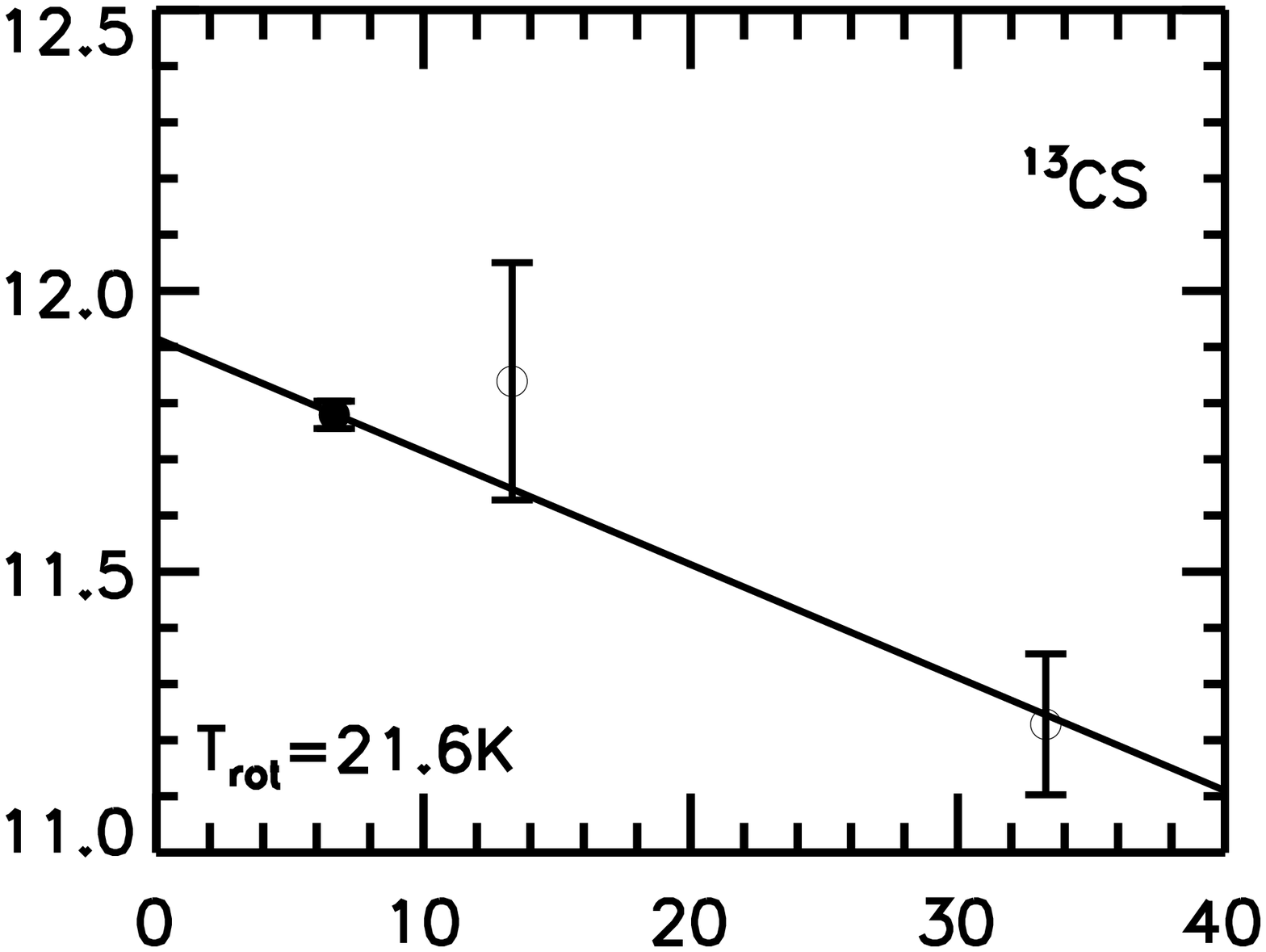}{0.45\textwidth}{}
         }
\vspace{-0.5cm}
\gridline{\fig{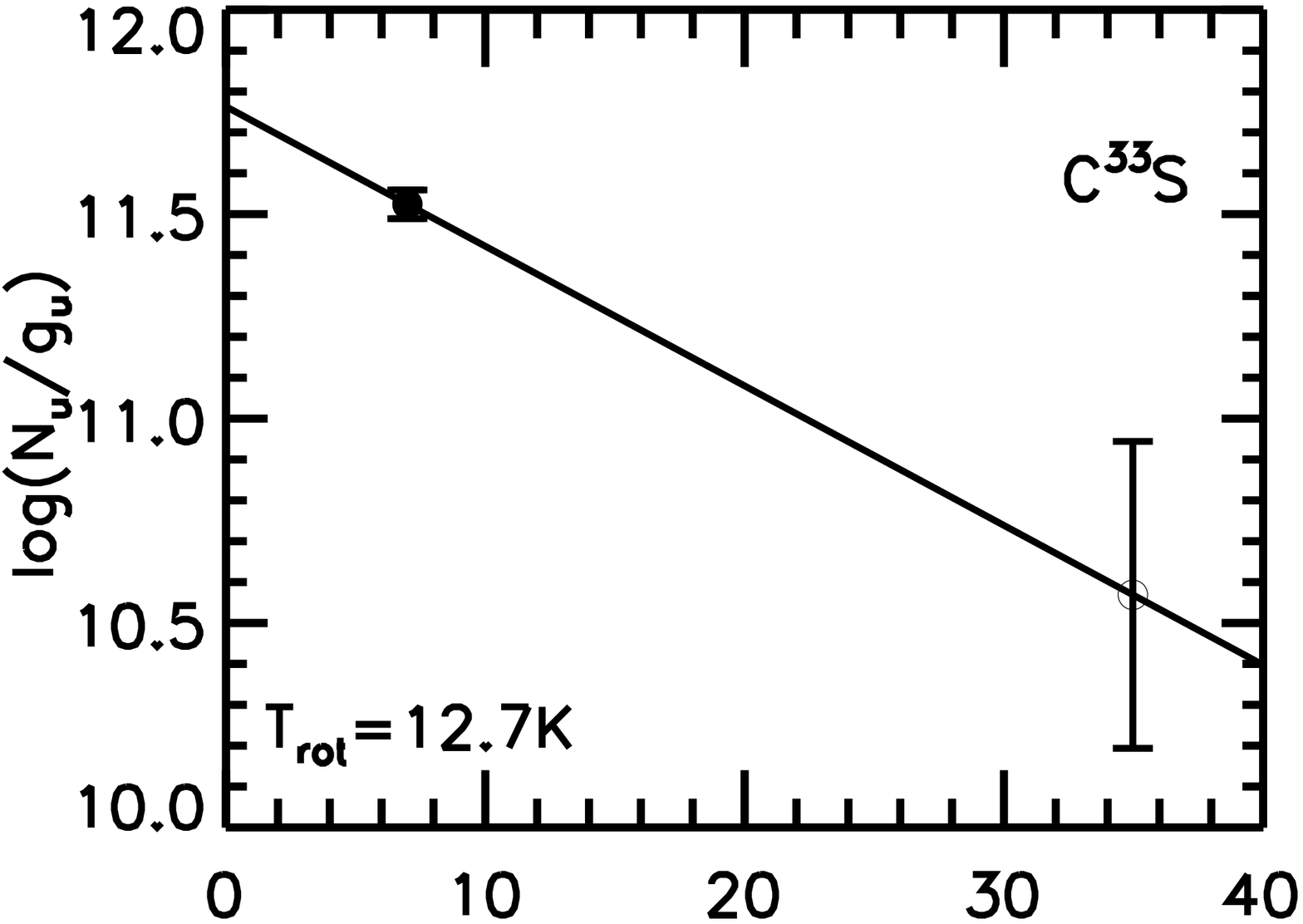}{0.45\textwidth}{}
          \fig{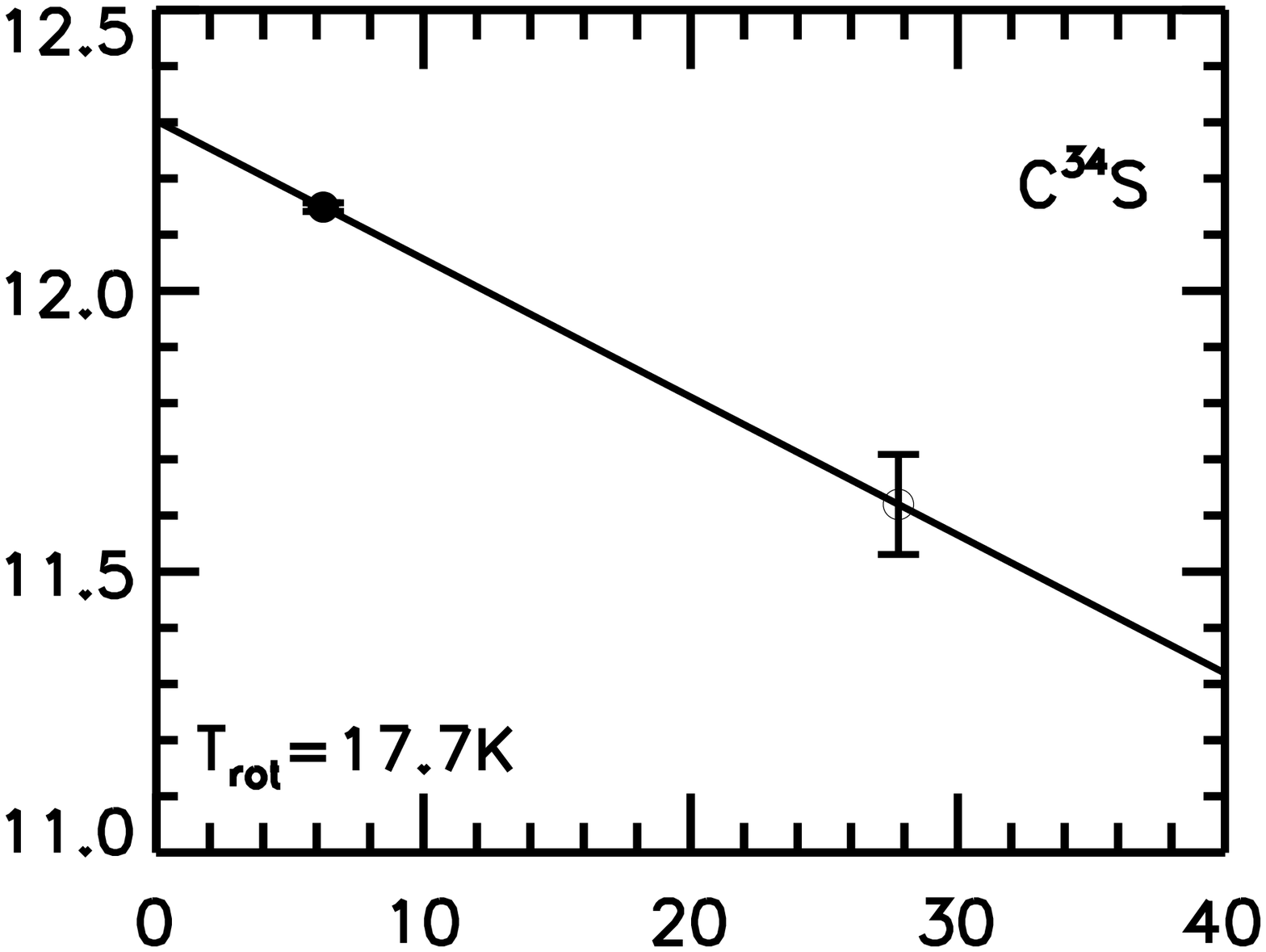}{0.45\textwidth}{}
         }
\vspace{-0.5cm}
\gridline{\fig{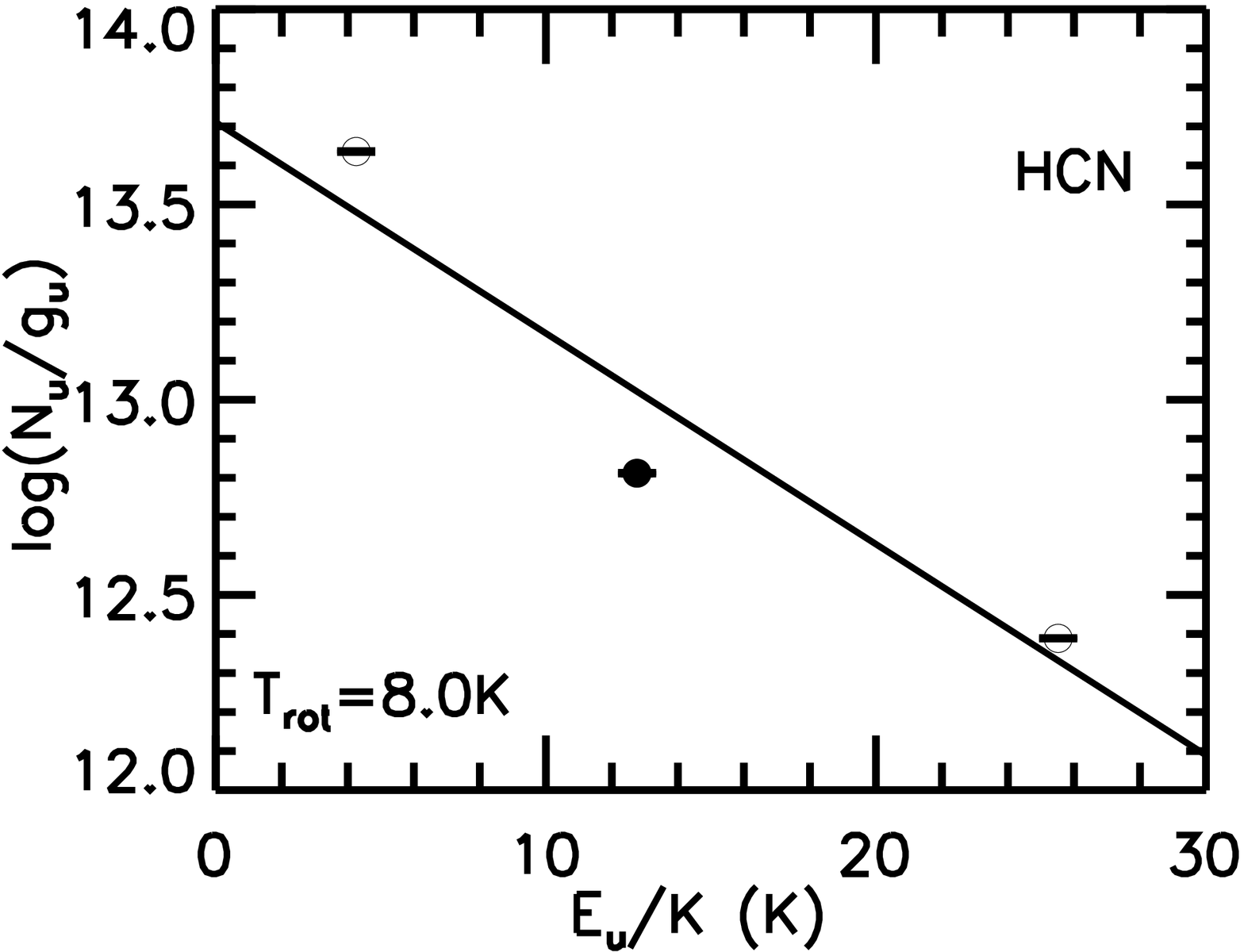}{0.45\textwidth}{}
          \fig{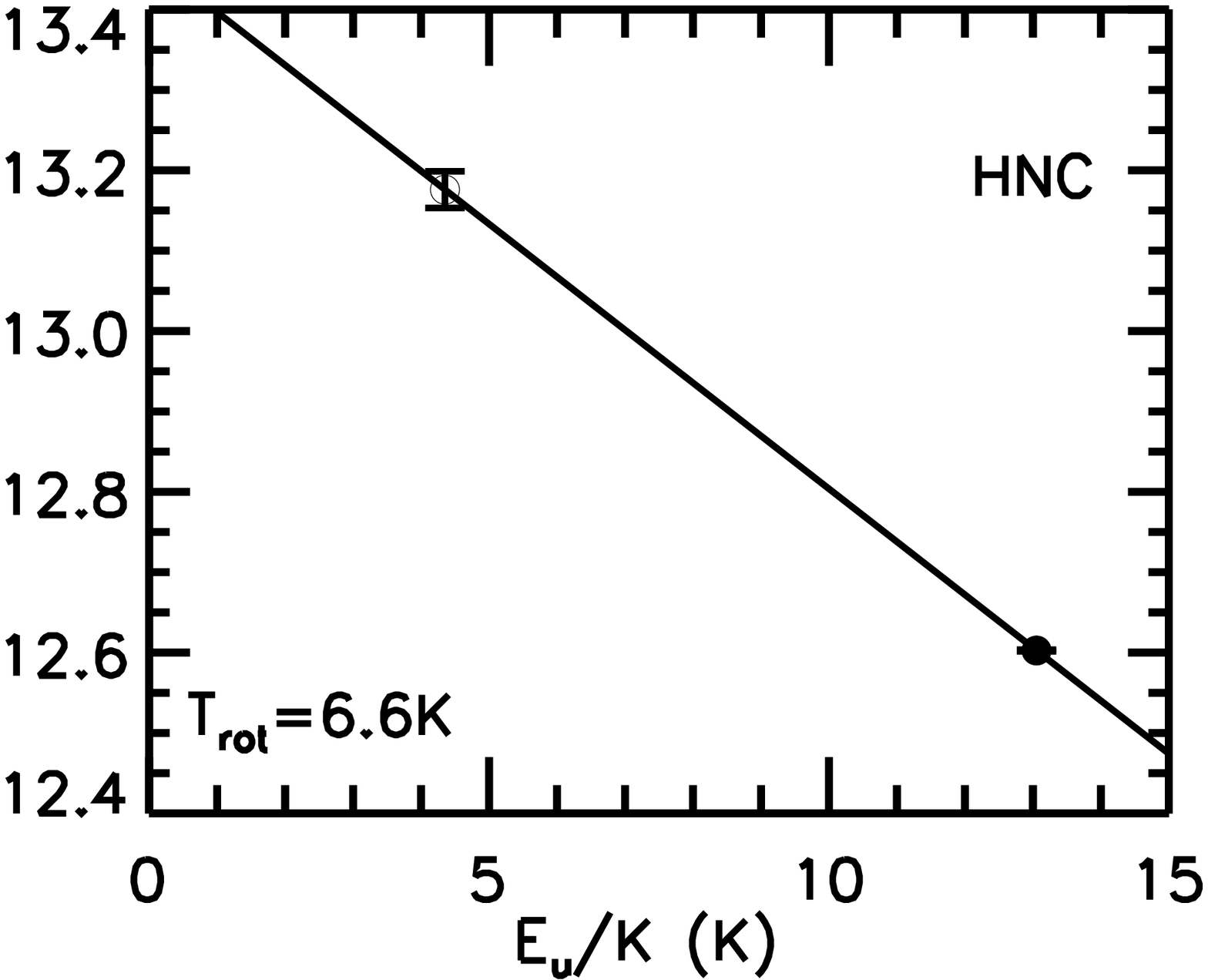}{0.45\textwidth}{}
         }
\caption{    
The rotation diagrams. The filled and open cycles represent the data obtained in this work and those taken from the
literature (see the text for details), respectively.
\label{Figure5}
}
\end{figure}

\renewcommand{\thefigure}{\arabic{figure} (Cont.)}
\addtocounter{figure}{-1}

\begin{figure}
\gridline{\fig{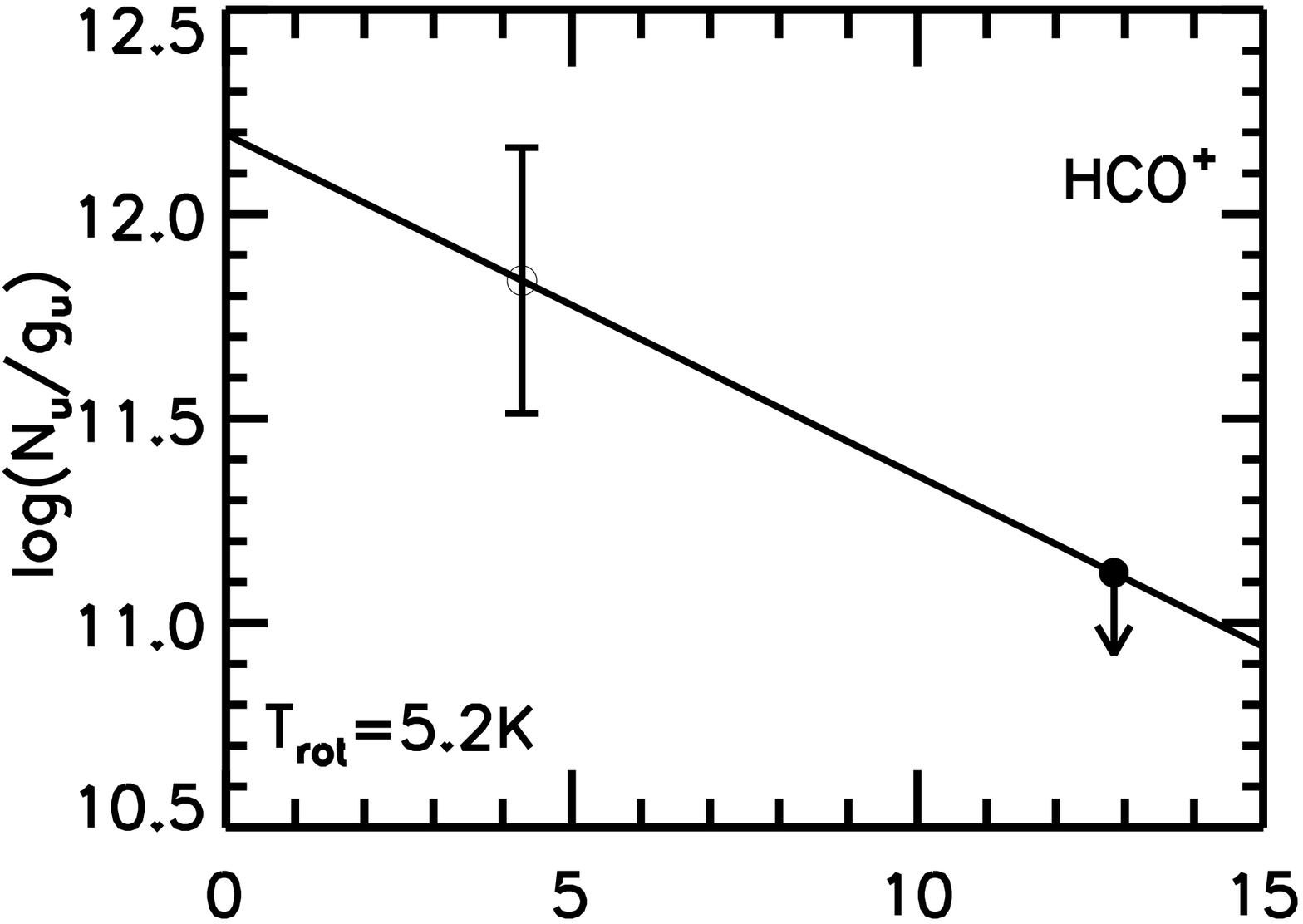}{0.45\textwidth}{}
          \fig{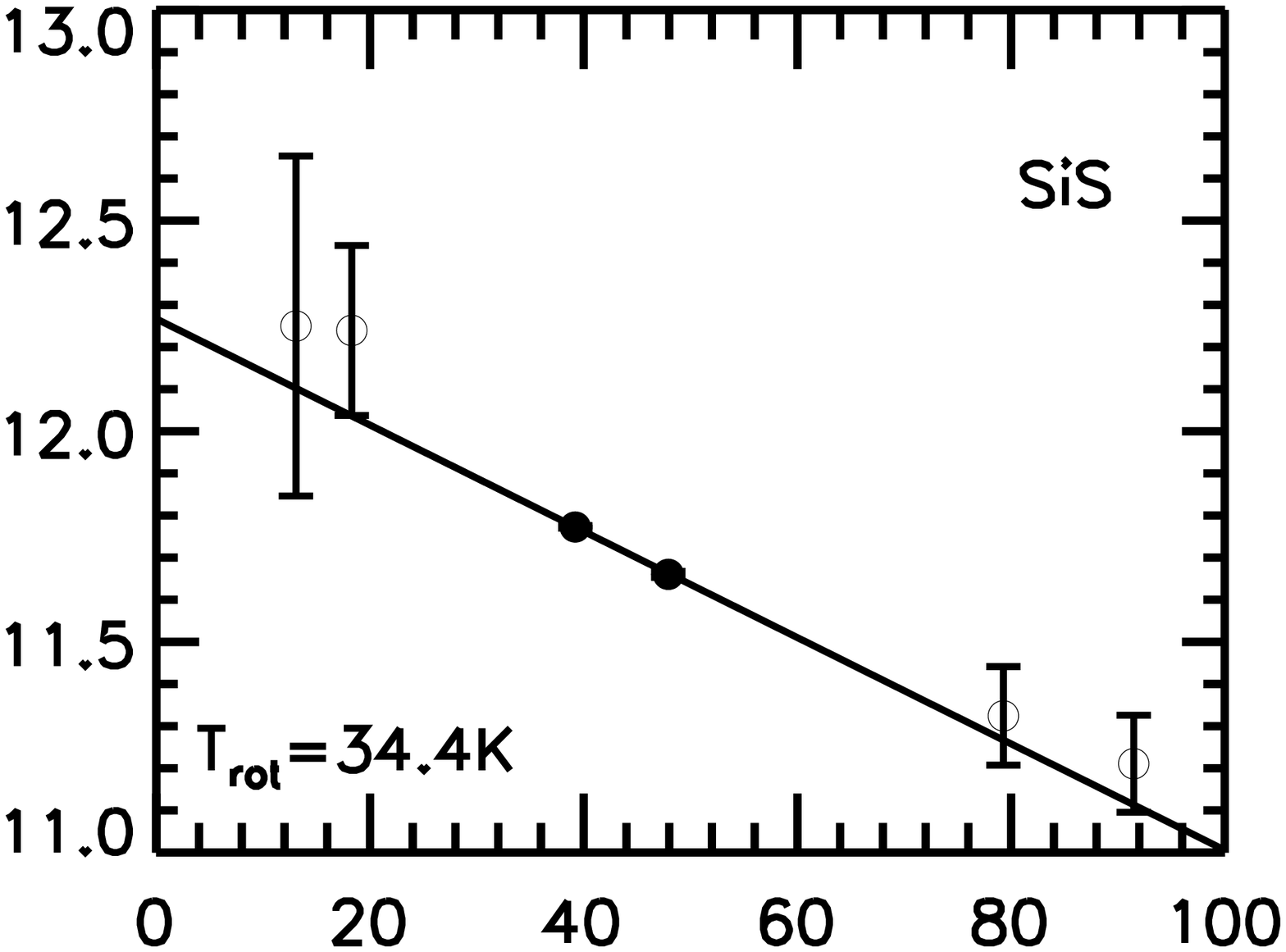}{0.45\textwidth}{}
         }
\vspace{-0.5cm}
\gridline{\fig{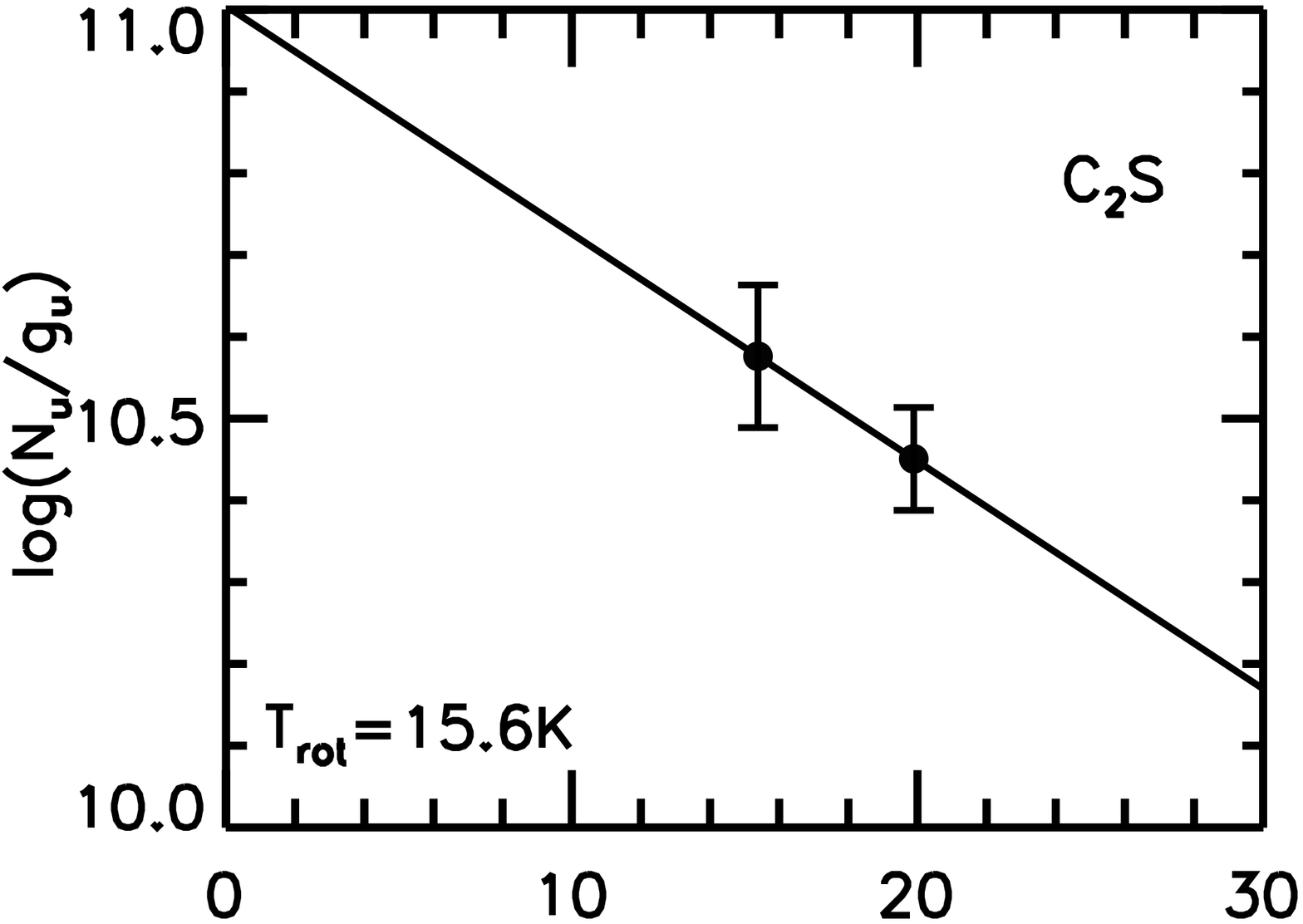}{0.45\textwidth}{}
          \fig{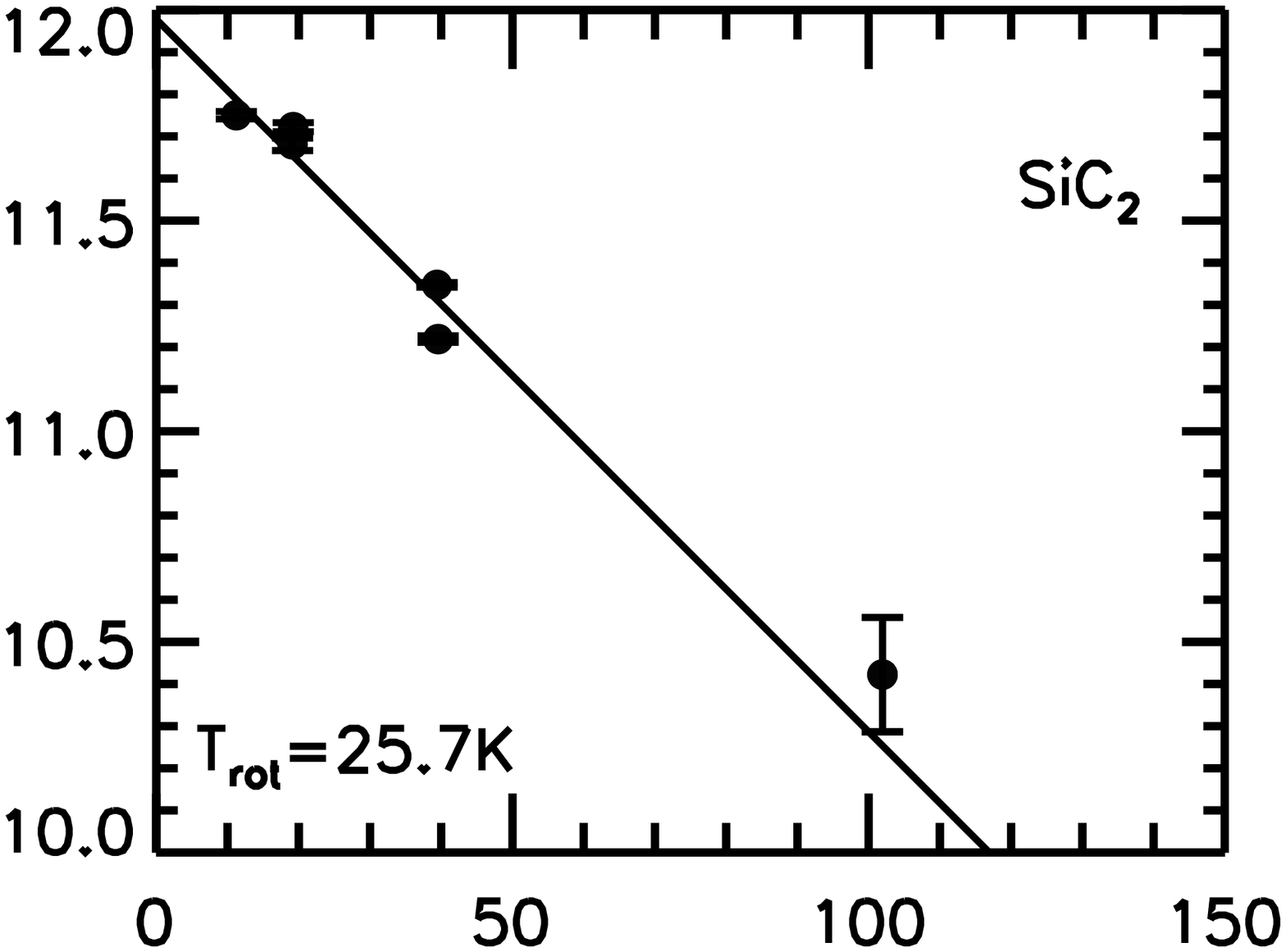}{0.45\textwidth}{}
         }
\vspace{-0.5cm}
\gridline{\fig{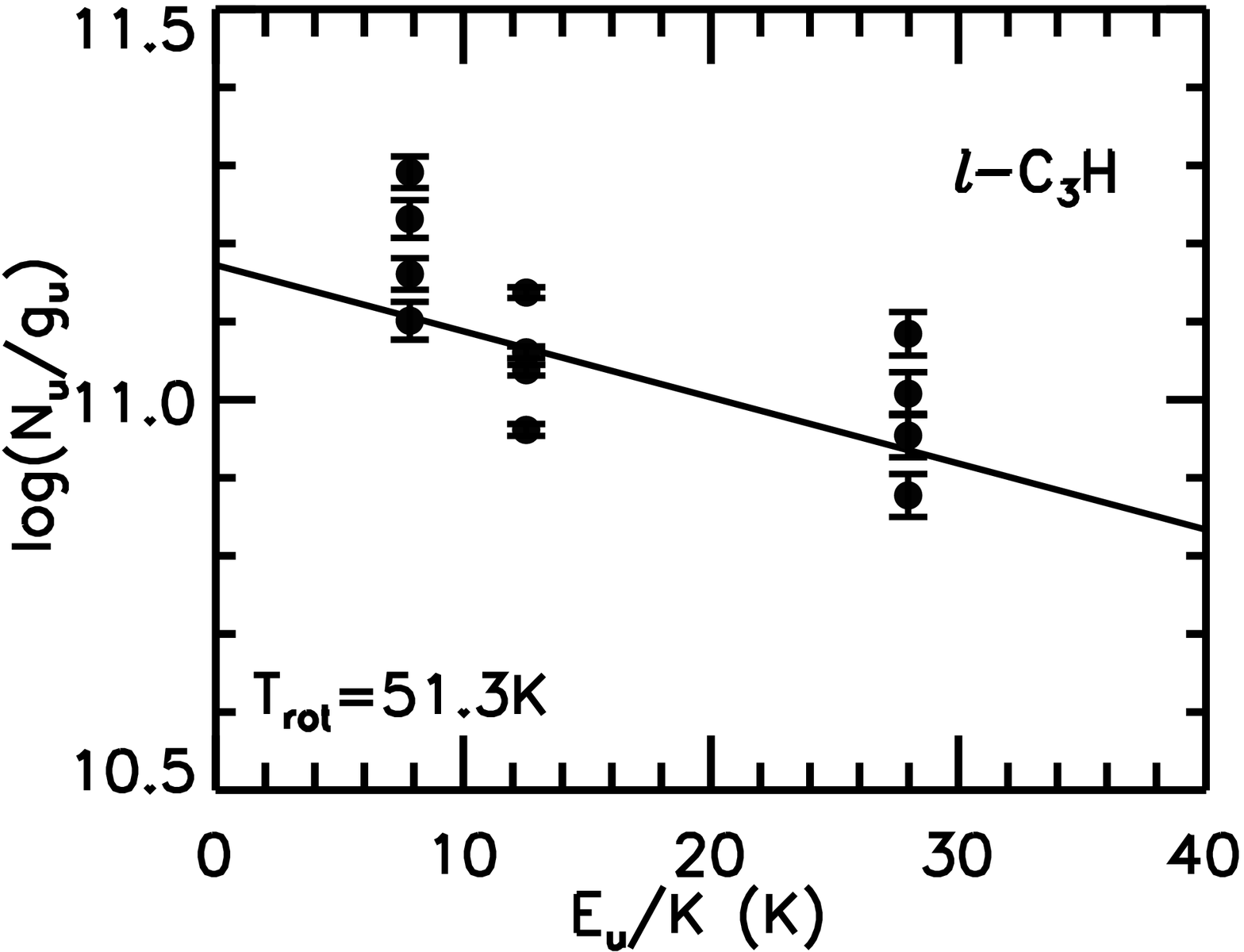}{0.45\textwidth}{}
          \fig{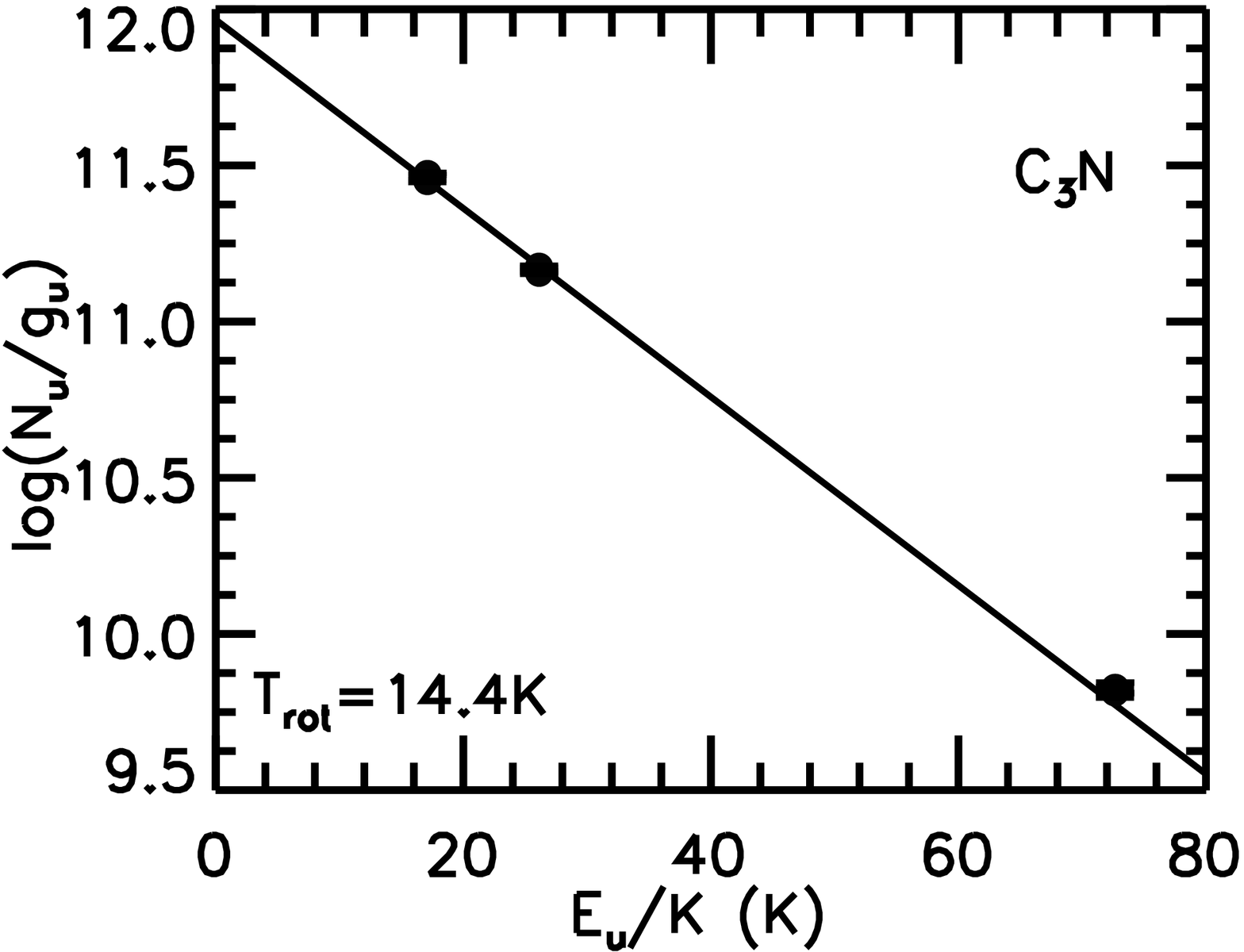}{0.45\textwidth}{}
         }
\caption{}
\end{figure}

\renewcommand{\thefigure}{\arabic{figure} (Cont.)}
\addtocounter{figure}{-1}

\begin{figure}
\gridline{\fig{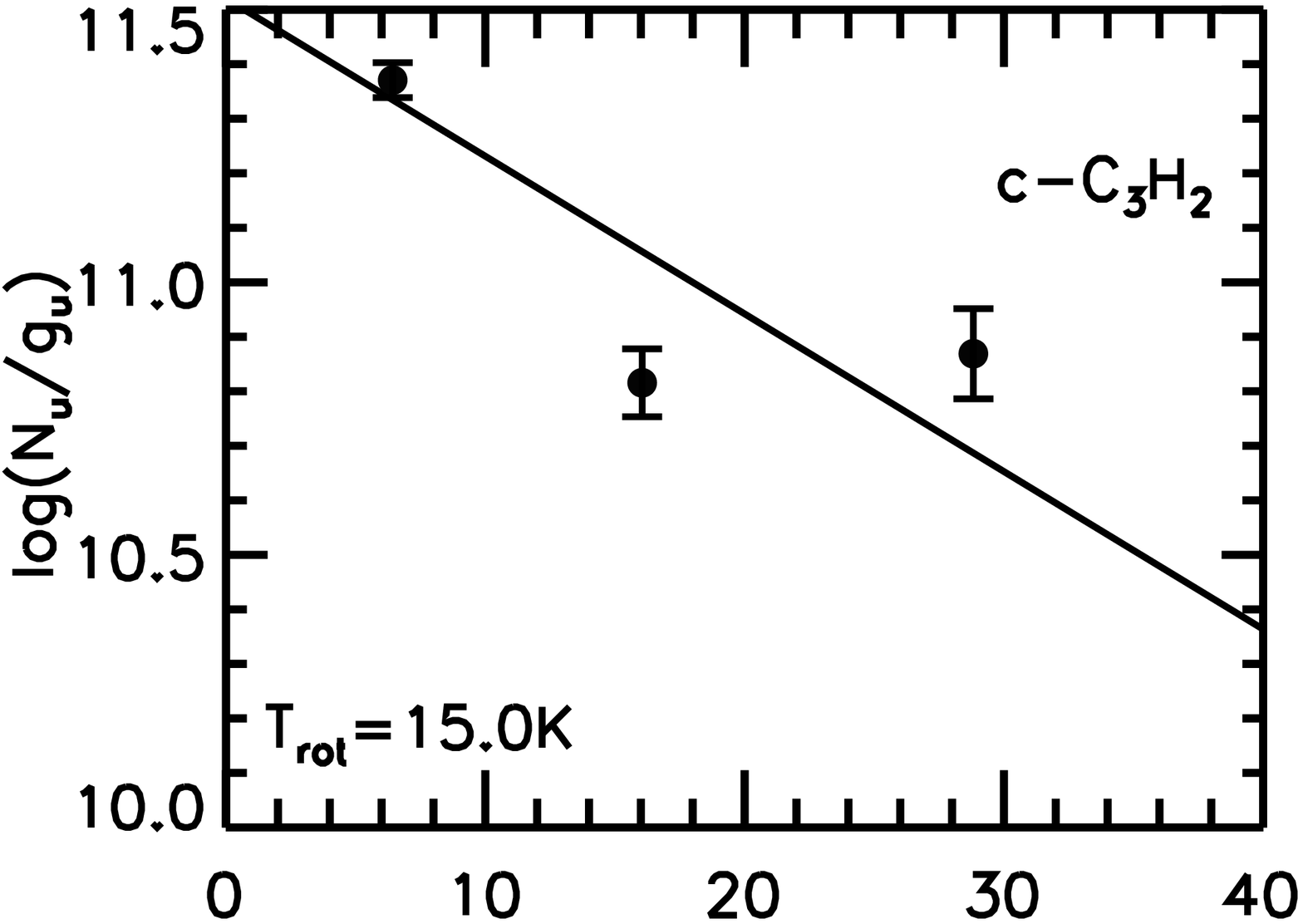}{0.45\textwidth}{}
          \fig{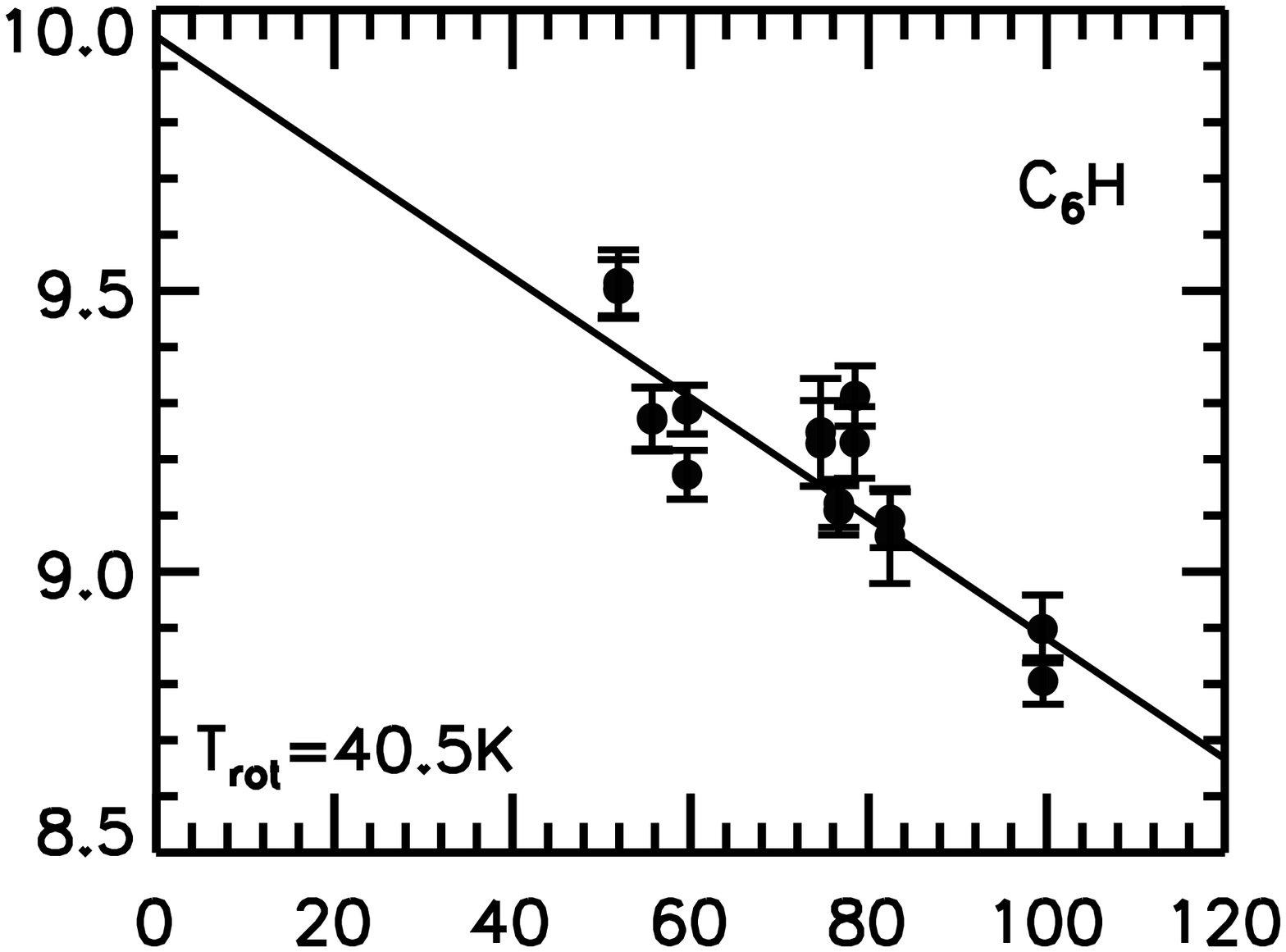}{0.45\textwidth}{}
         }
\vspace{-0.5cm}
\gridline{\fig{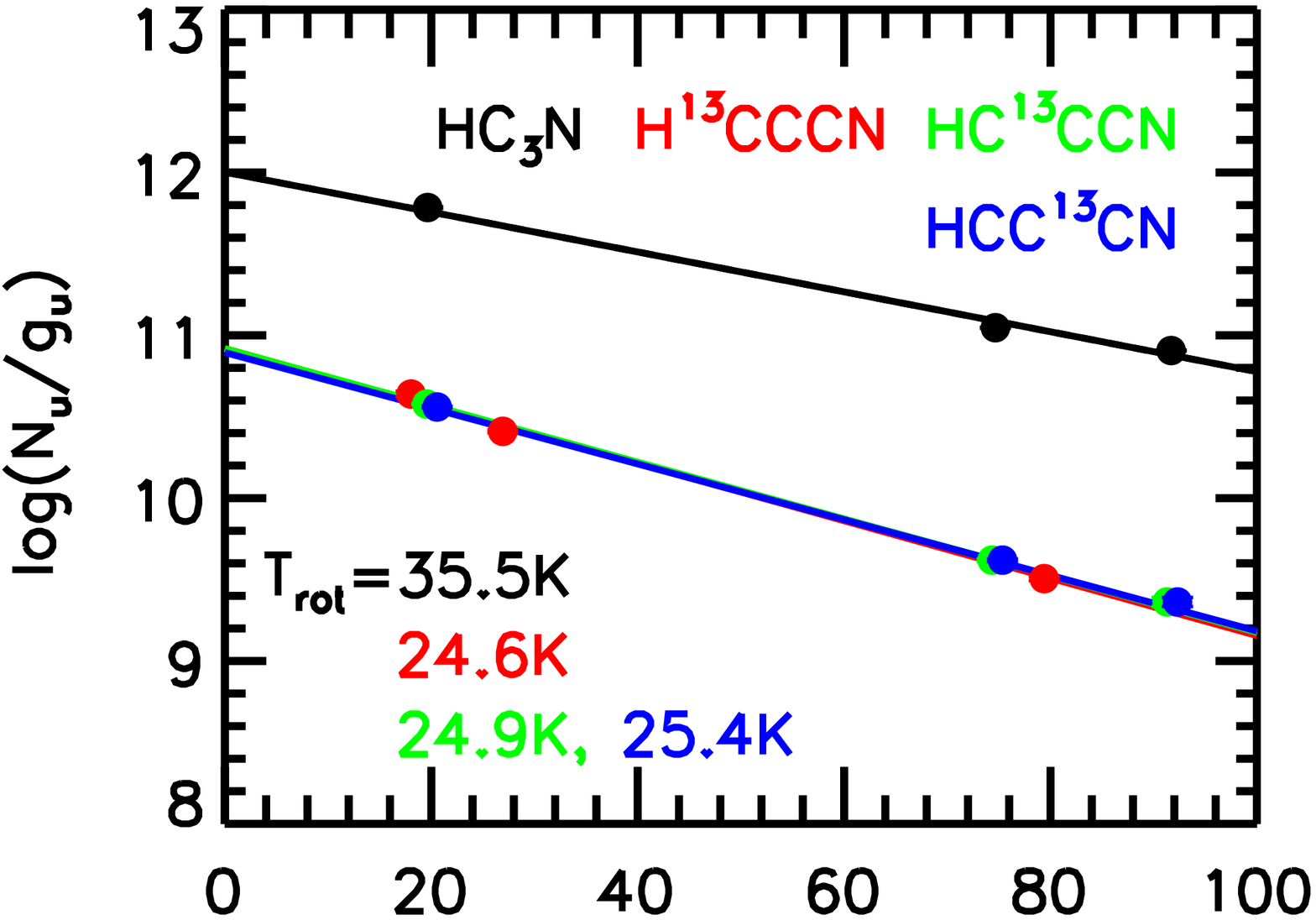}{0.45\textwidth}{}
          \fig{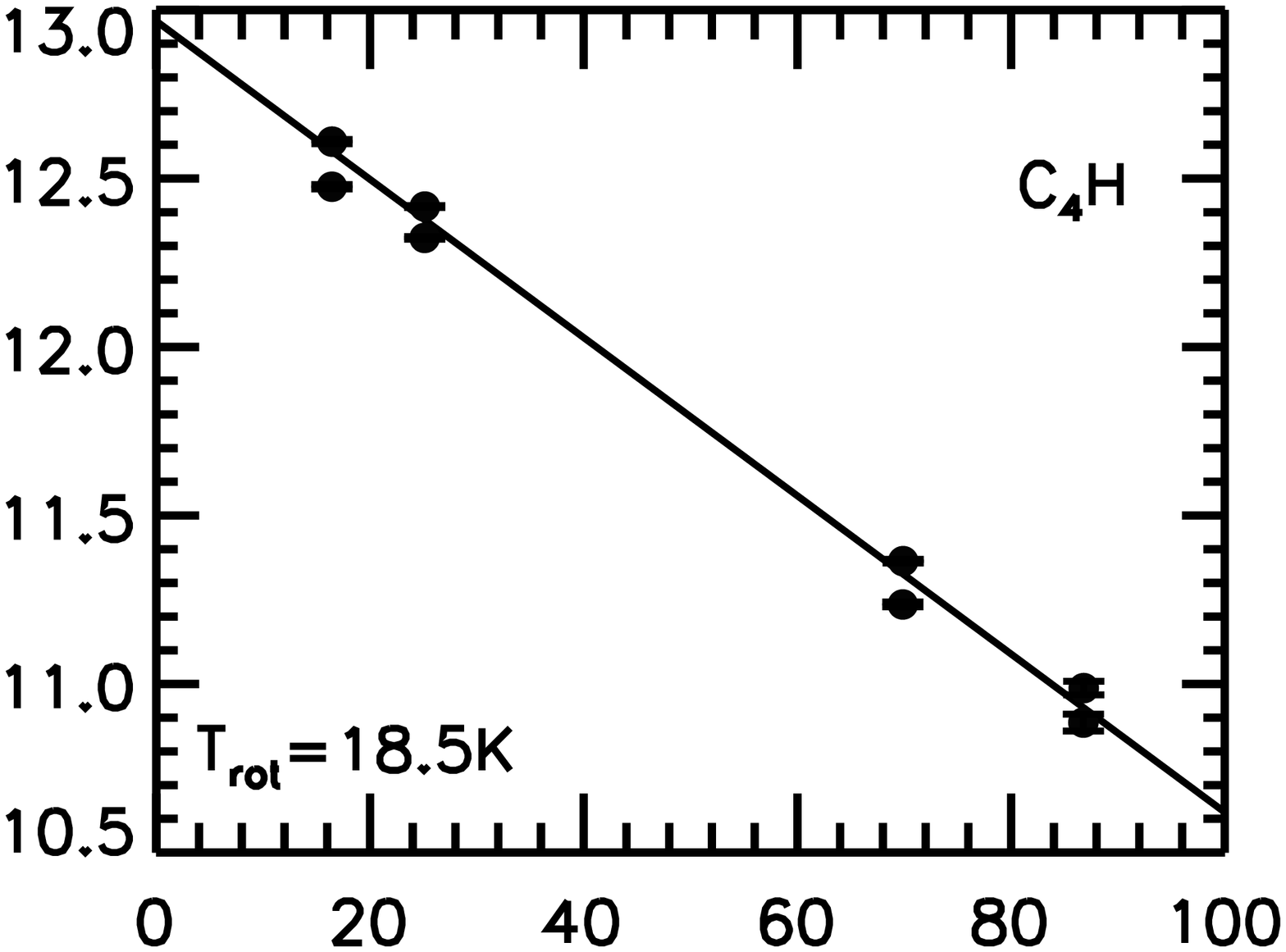}{0.45\textwidth}{}
         }
\vspace{-0.5cm}
\gridline{\fig{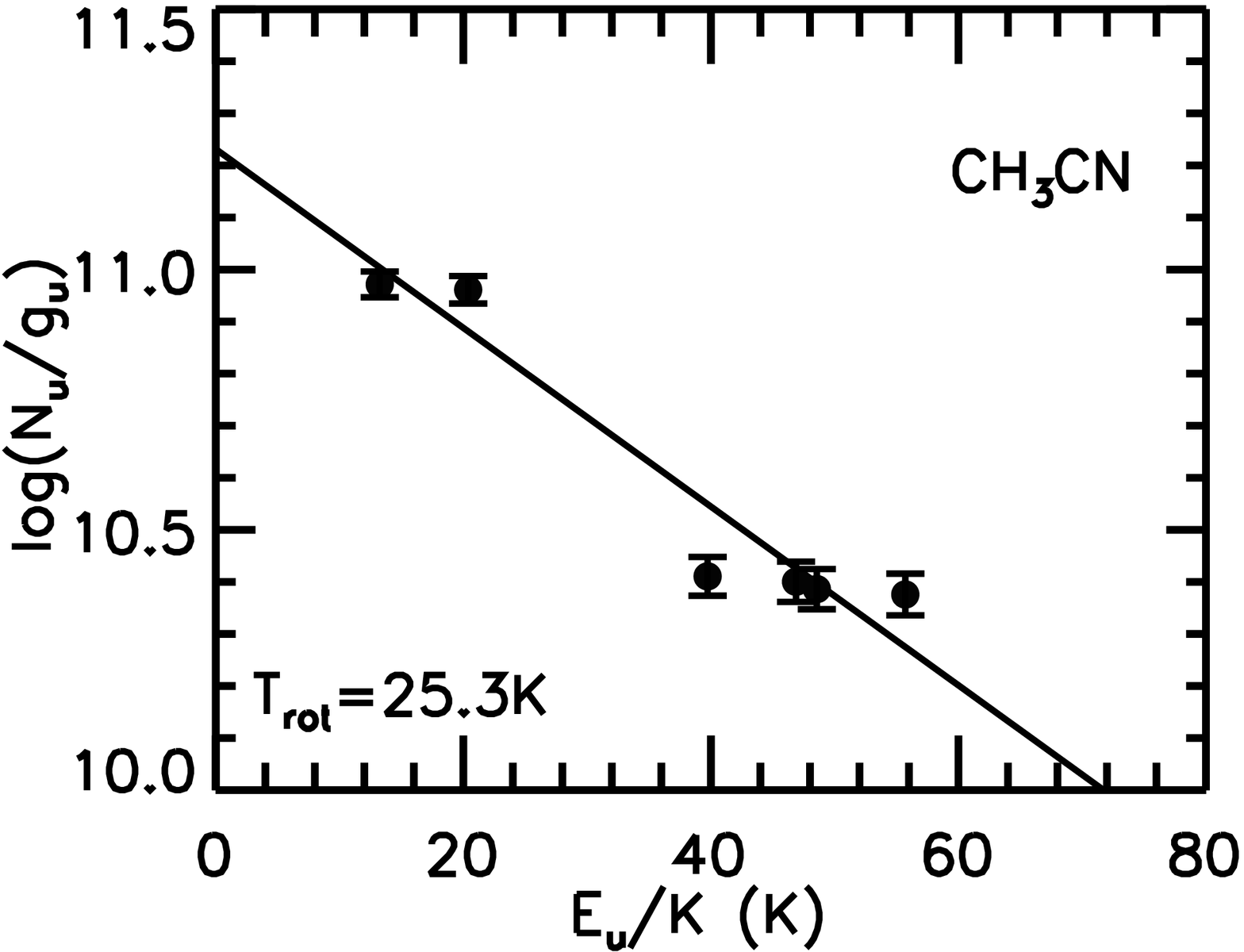}{0.45\textwidth}{}
          \fig{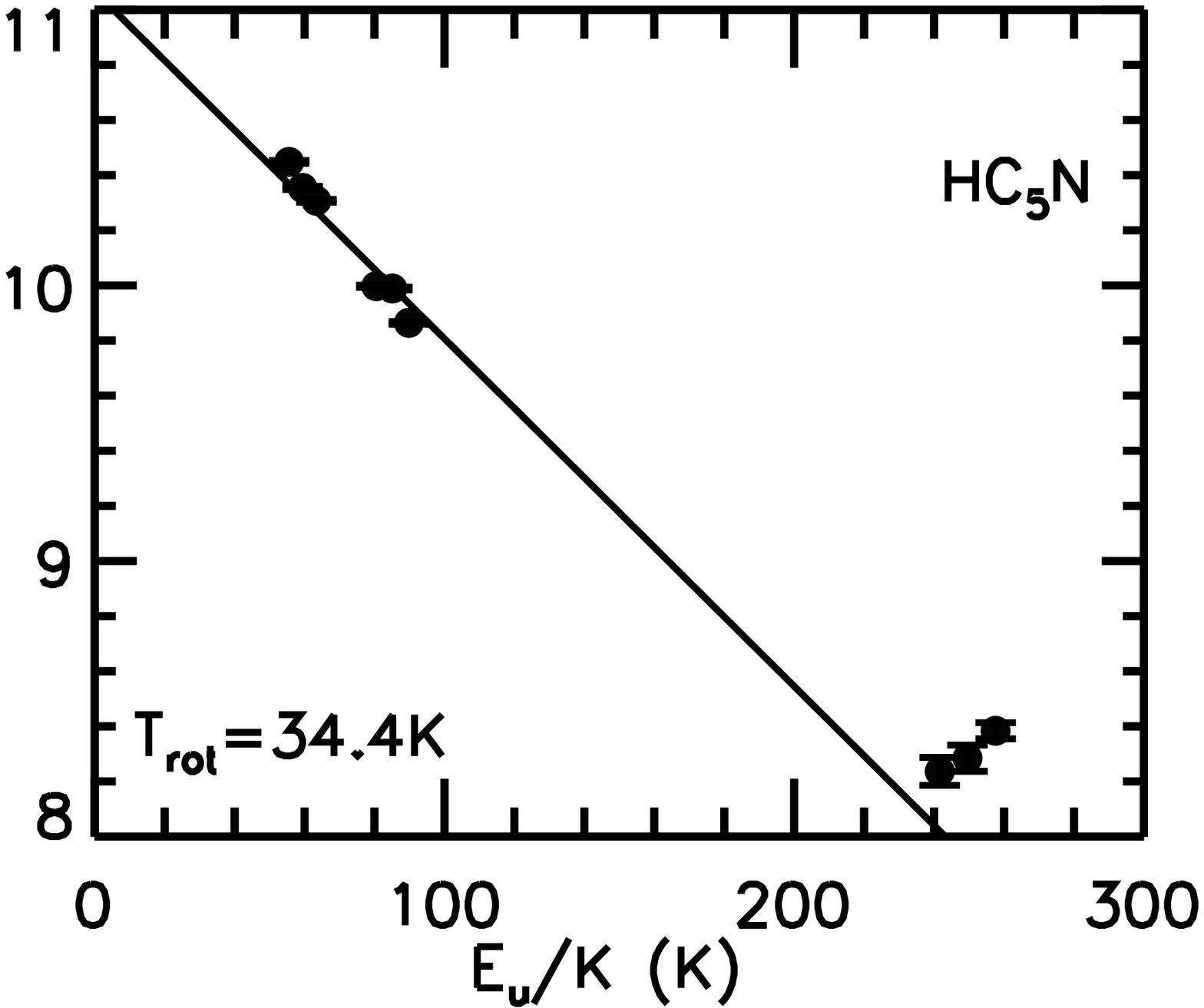}{0.45\textwidth}{}
         }
\caption{}
\end{figure}

\renewcommand{\thefigure}{\arabic{figure} (Cont.)}
\addtocounter{figure}{-1}

\begin{figure}
\gridline{\fig{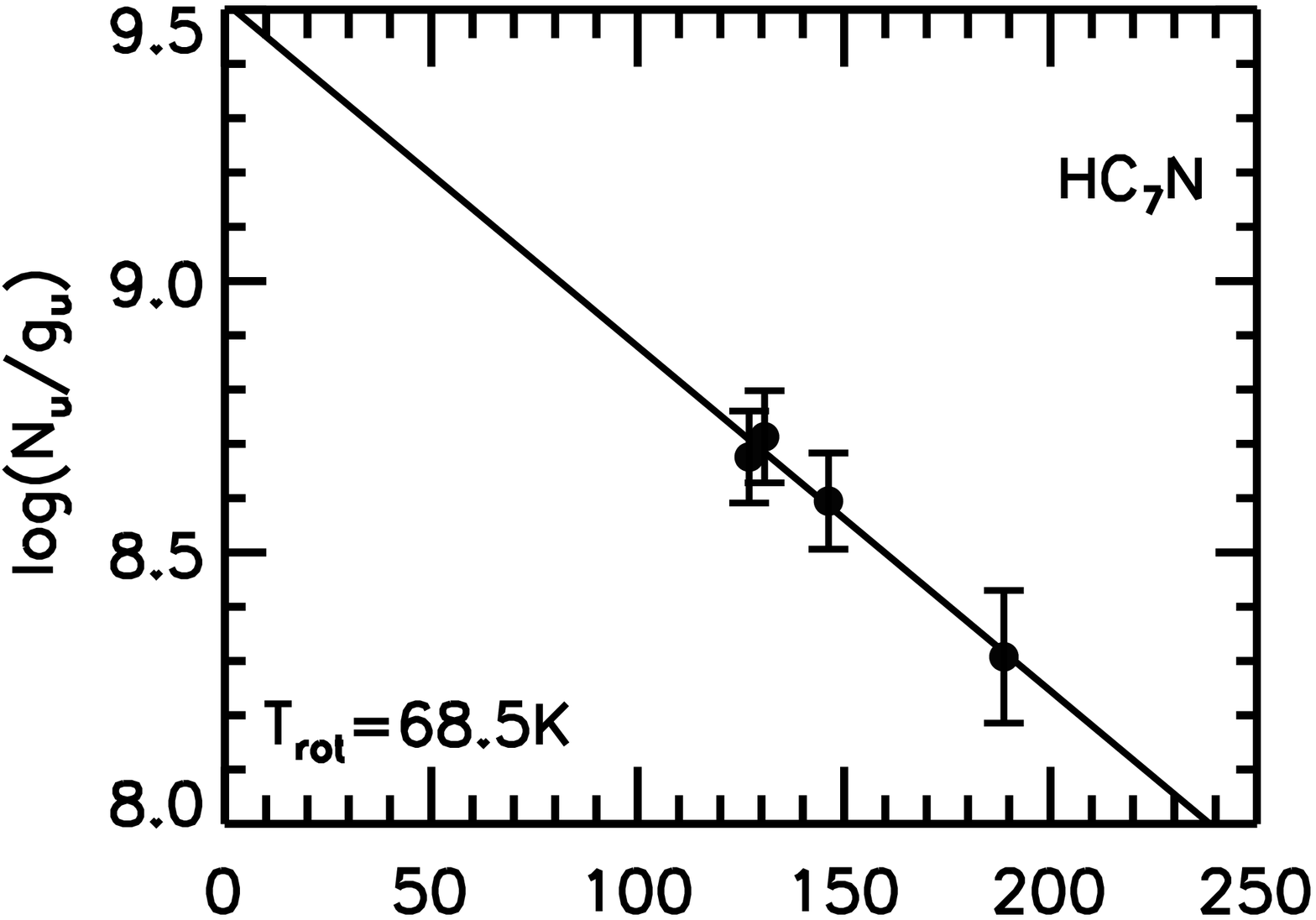}{0.45\textwidth}{}
          \fig{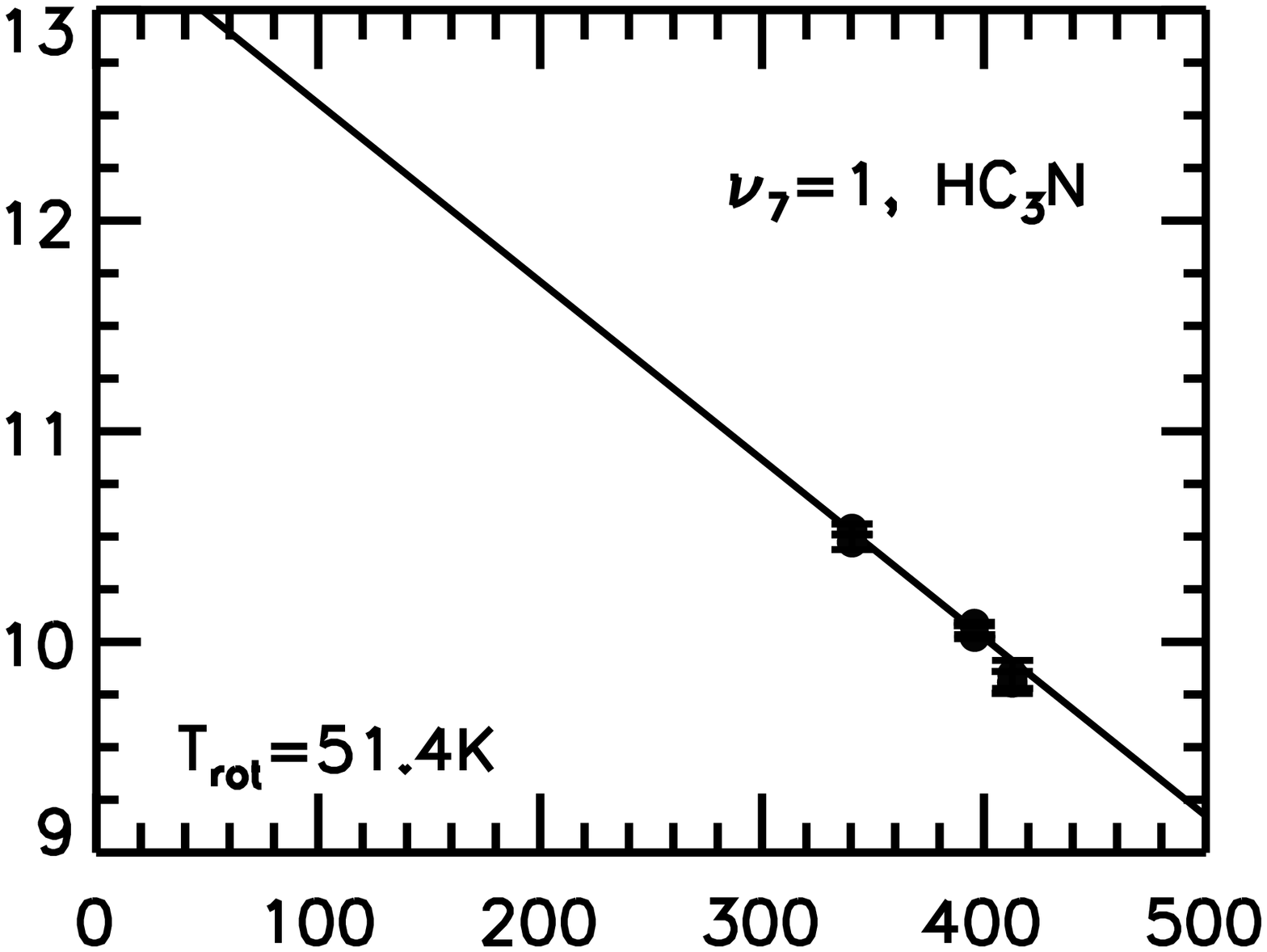}{0.45\textwidth}{}
         }
\vspace{-0.5cm}
\gridline{\fig{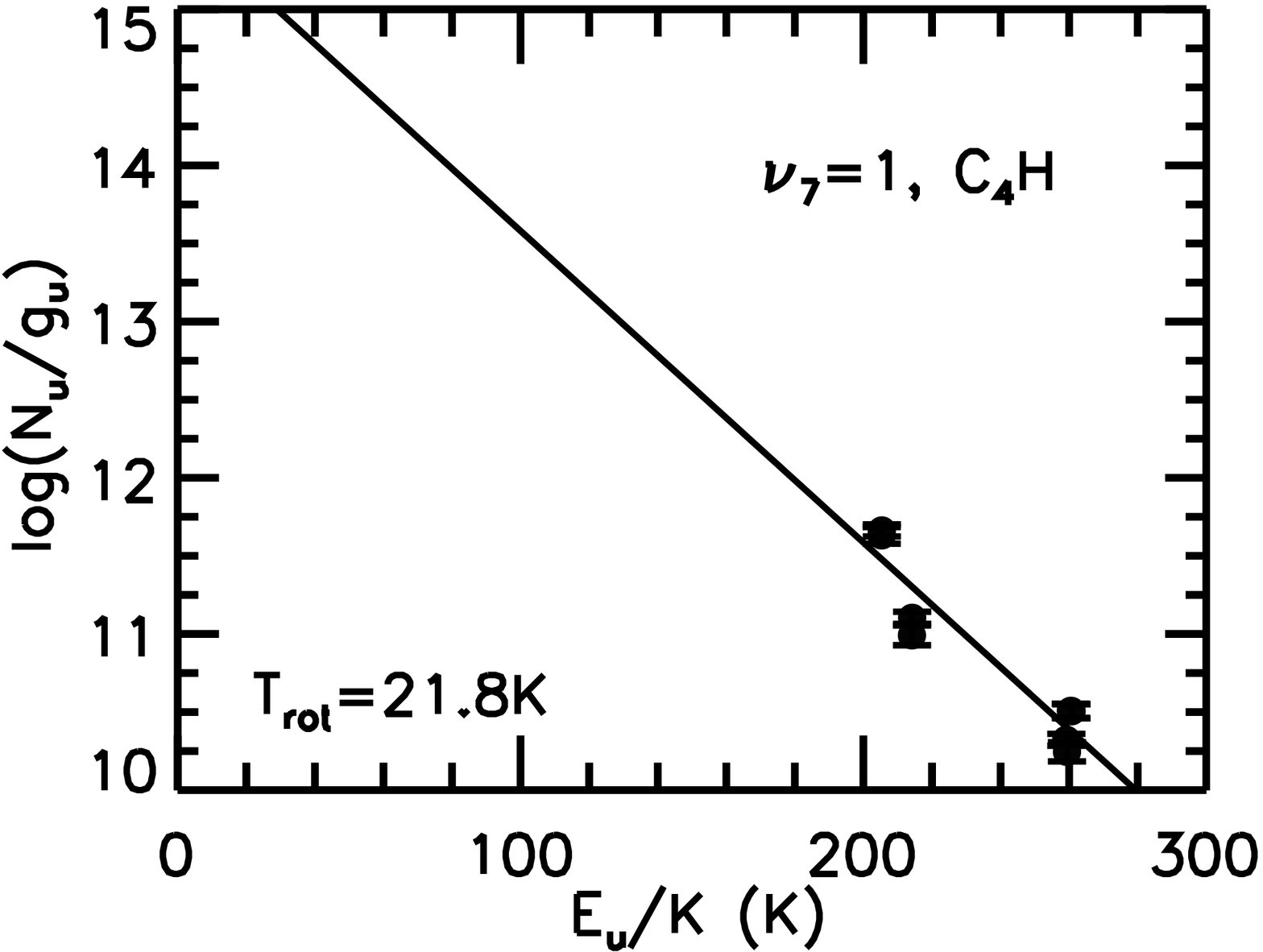}{0.45\textwidth}{}
         }
\caption{}
\end{figure}

\renewcommand{\thefigure}{\arabic{figure}}

\begin{figure}
\gridline{\fig{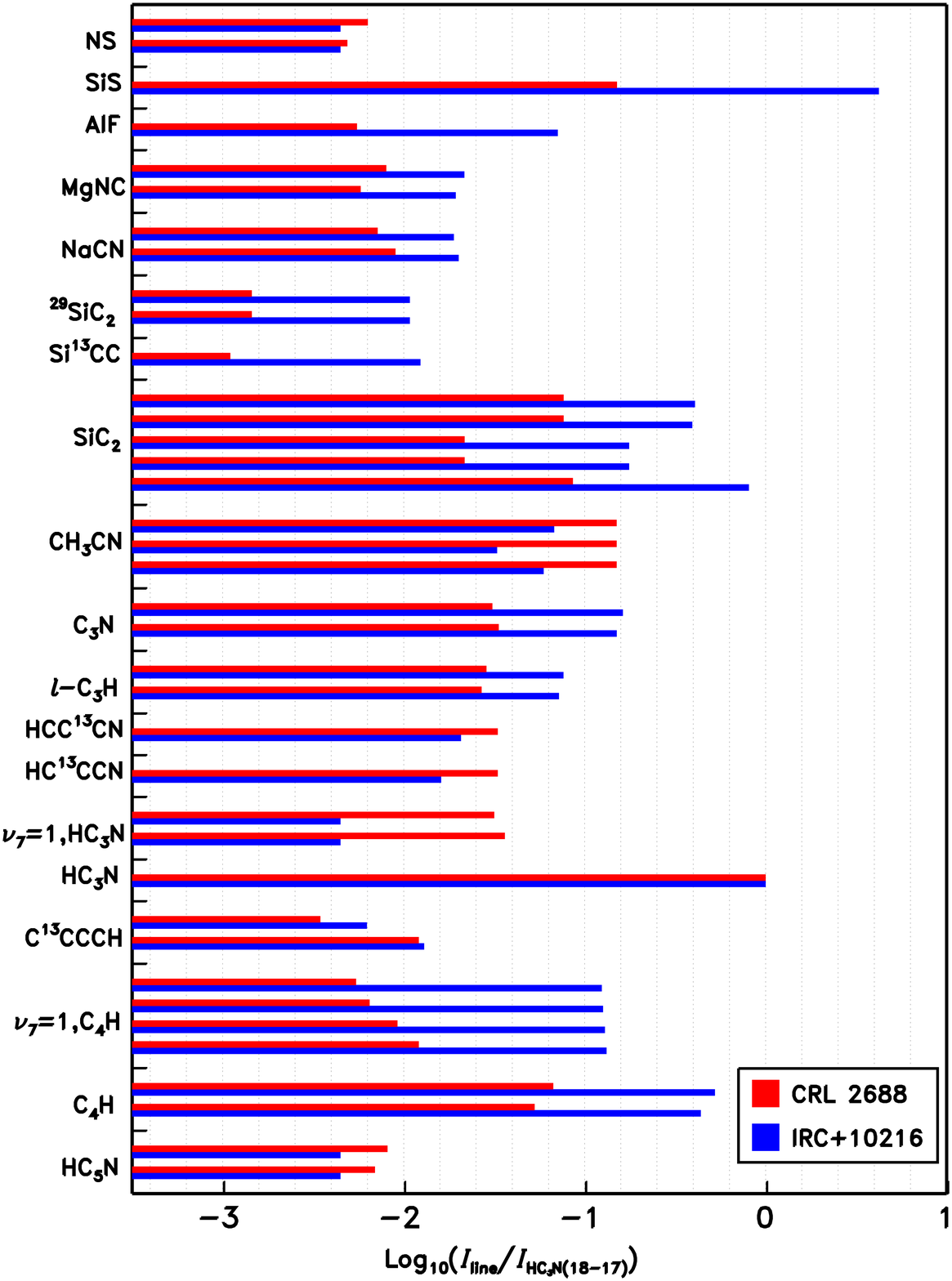}{0.8\textwidth}{}
         }
\caption{ Comparison of molecular line intensities between CRL\,2688 (red) and IRC+10216 (blue),
which is ordered by increasing carbon atoms from top to bottom. 
\label{Figure6}
}
\end{figure}

\begin{figure}
\gridline{\fig{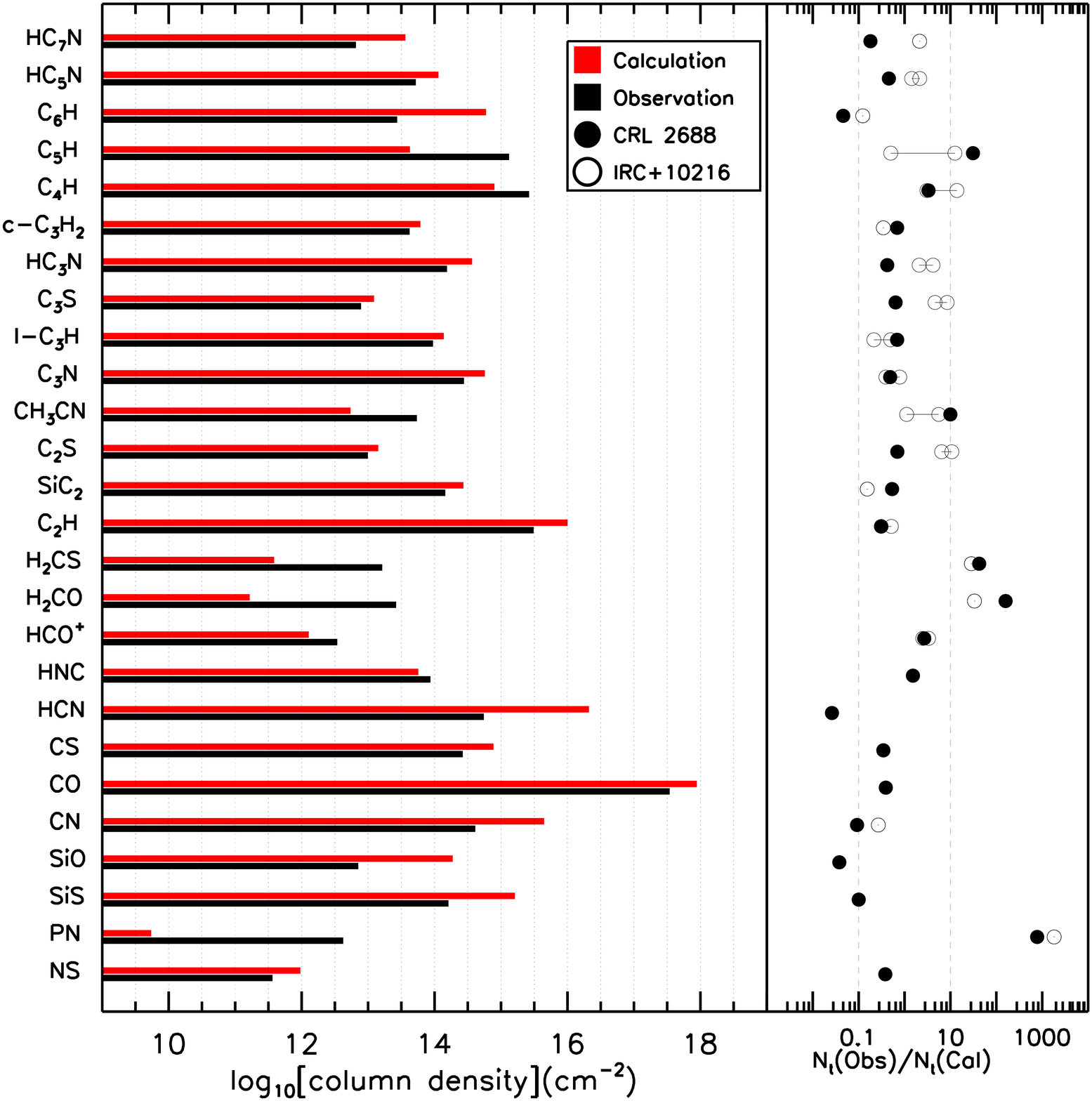}{0.8\textwidth}{}
         }
\caption{Comparison between the observed (black) and 
modelled (red)  column densities  of the molecules in CRL\,2688 (left panel), and 
the observed-to-modelled column density ratios of the molecules in 
CRL\,2688 and IRC+10216 (right panel). For some molecules in IRC+10216, diverse abundances have been reported in different observations, which are denoted by open circles linked by straight lines. 
The vertical lines in the right panel mark the ranges within which the observed and modelled values are in agreement 
within one order of magnitude.
\label{Figure7}
}
\end{figure}

\begin{figure}
\gridline{\fig{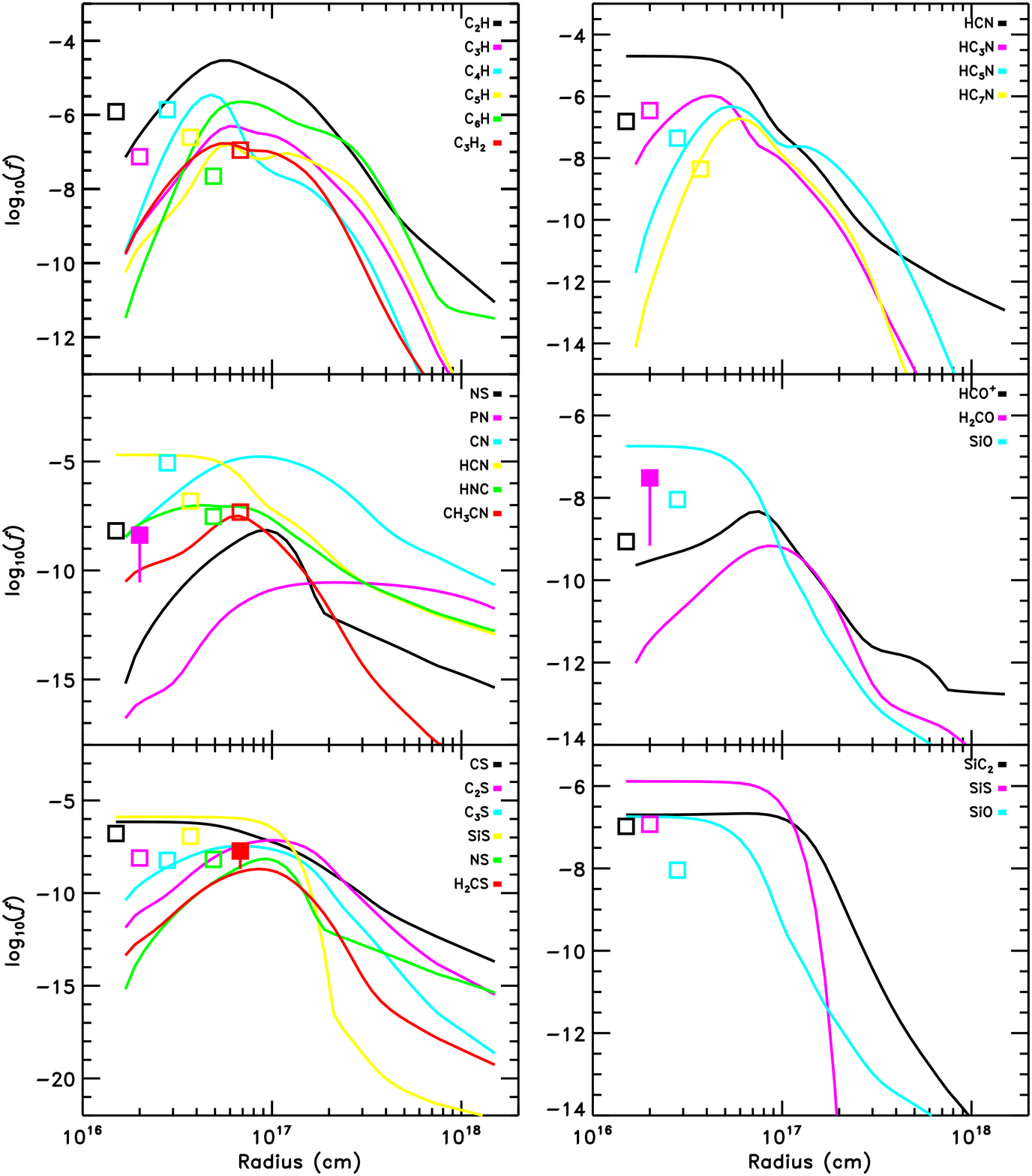}{0.95\textwidth}{}
         }
\caption{The modelled   
abundances of molecules in CRL\,2688  as a function of radius. 
The observed abundances are denoted by squares, which are located 
at different radii for the convenience of view. The open and solid squares represent those within and beyond the modelling results 
in the radius range from  1.5 $\times$ 10$^{16}$--1.5 $\times$ 10$^{18}$ cm, 
respectively. The vertical lines show the level of disagreement between the observed and the maximum modelled abundances.
\label{Figure8}
}
\end{figure}

\begin{figure}
\gridline{\fig{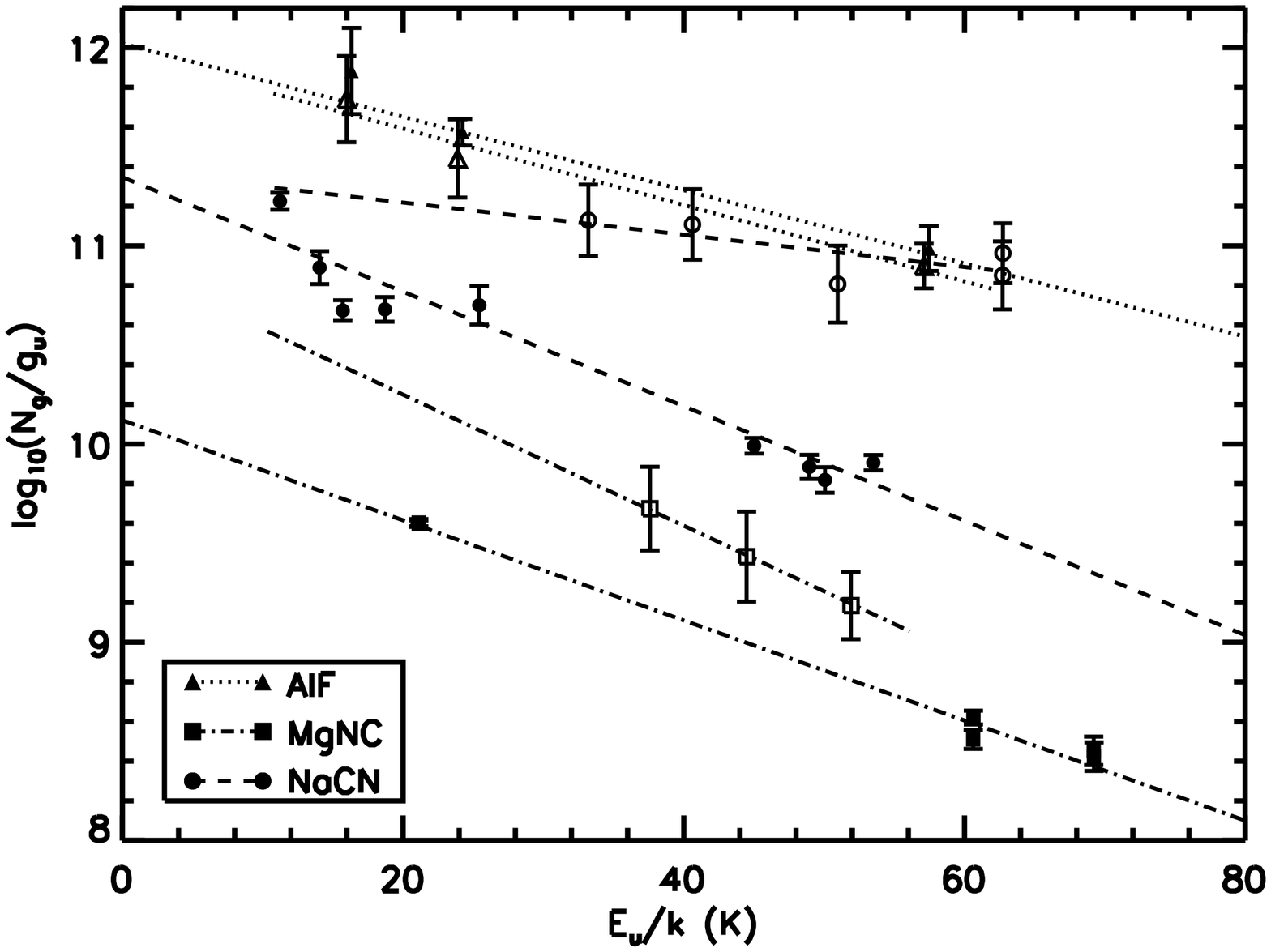}{0.8\textwidth}{}
         }
\caption{  
Rotation diagram of AlF, NaCN, and MgNC, for which
the source sizes have been assumed
to be  5$\arcsec$, 5$\arcsec$, and 30$\arcsec$, respectively.
The measurements of AlF (triangles), NaCN (cycles), and MgNC (squares) taken by us and those by \citet{Highberger01} are denoted by filled and open symbols, respectively.
\label{Figure9}
}
\end{figure}

\begin{figure}
\gridline{\fig{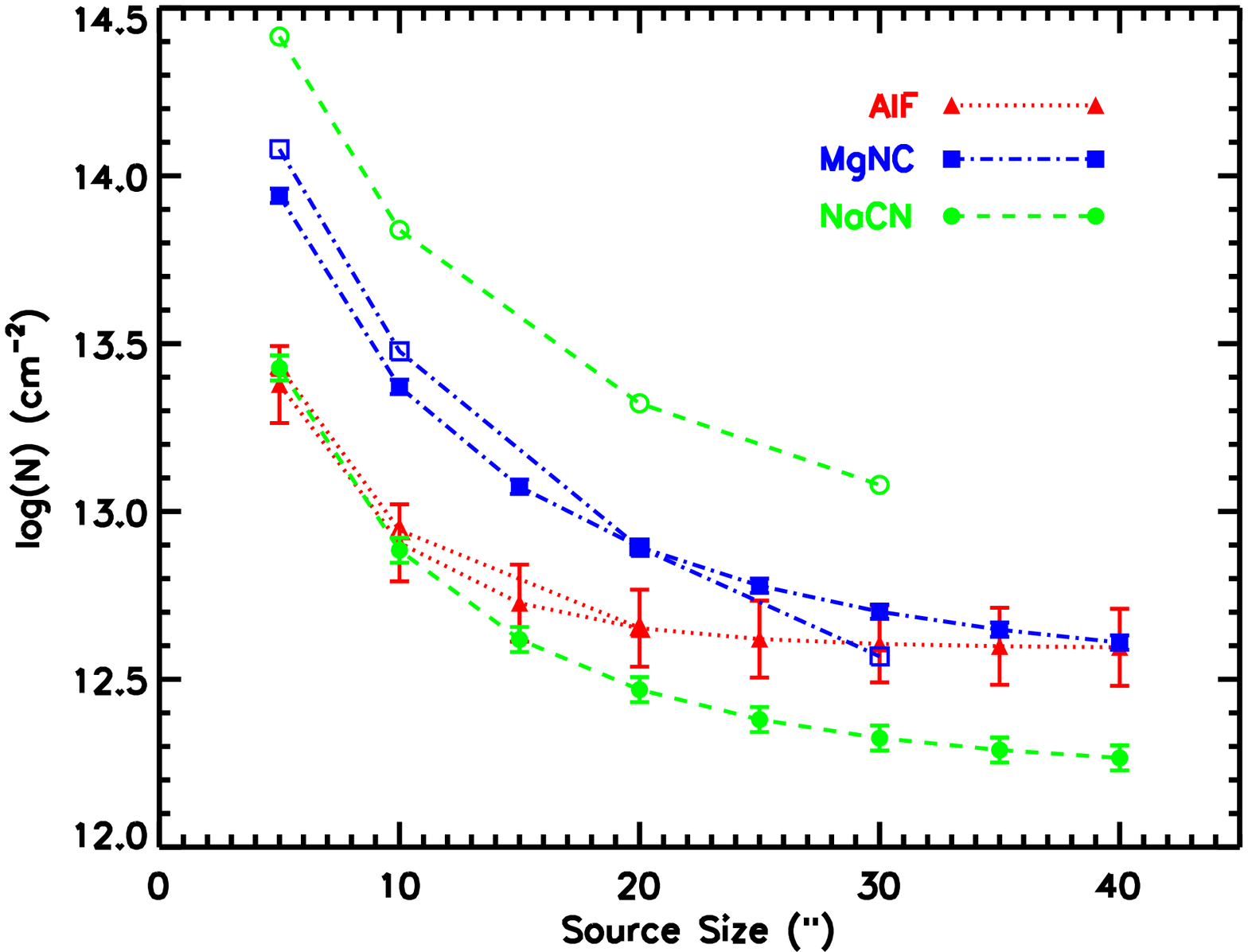}{0.8\textwidth}{}
         }
\caption{Column densities as a function of assumed source sizes.
The measurements of AlF (triangles), NaCN (cycles), and MgNC (squares) taken by us and those by \citet{Highberger01} are denoted by filled and open symbols, respectively.
\label{Figure10}
}
\end{figure}

\clearpage

\begin{deluxetable}{lccccc}
\tiny
\tablecaption{Stellar and circumstellar parameters of CRL\,2688.
\label{Table1}}
\tabletypesize{\small}
\tablewidth{0pt}
\tablehead{ 
 \colhead{Parameter Description}          & \colhead{Parameter}           & \colhead{Unit}                & \colhead{Value}                 & \colhead{Reference}     & \colhead{Note}                                      
}
\startdata
Stellar radius                                           & $R_{\ast}$                           & $R_{\odot}$                   & 51 $\pm$ 5                        &  (1)       &  1                                  \\ 
Stellar effective temperature                   & $T_{\rm eff}$                       & K                                    & 7000                                  &  (1)       &  2                                  \\ 
Stellar spectrum type                              & \nodata                                & \nodata                          & F5 Ia                                  & (2)        &  3                                  \\ 
Distance                                                  & $D$                                     & pc                                  & 1500                                   & (3)        &  4                                  \\ 
                                                                &                                             &                                      & 420                                     & (4)        &  5                                  \\ 
                                                                &                                             &                                      & 300--350                             & (1)        &  6                                  \\ 
                                                                &                                             &                                      & 1000                                   & (2)         &  7                                  \\ 
                                                                &                                             &                                      & 350 $\pm$ 53                     & (5)         &  8                                  \\
\enddata
\end{deluxetable}


\begin{deluxetable}{cccc}
\caption{Summary of the observations.} 
\tablehead{ 
 \colhead{Band Range} & \colhead{$F_{\rm eff}$$^a$}   &   \colhead{$B_{\rm eff}$$^a$}   &   \colhead{HPBW}                \\          
\colhead{(GHz)}                                   &                           &                            &   \colhead{(\arcsec)}                  }
\startdata
75.70--83.48                                 &  95                    &   81                     &   32.5--29.5            \\          
91.38--99.16                              &  95                    &   81                     &   26.9--24.8            \\  
160.84--168.62                         &  93                    &   73                     &   15.3--14.6             \\ 
176.52--184.30                          &  94                    &   63                     &   13.9--13.3             \\ 
\enddata
\tablenotetext{a}{Taken from the IRAM website. }
\label{Table2}
\end{deluxetable}

\begin{longrotatetable}
\startlongtable
\begin{deluxetable*}{rcccccccccccc}
\tablecaption{A summary of the lines in CRL\,2688 detected by our observations and in the literature.}
\tabletypesize{\scriptsize}
\tablehead{
\colhead{Frequency}   & \colhead{Species}     & \colhead{Transition}                      & \colhead{$V_{\rm exp}$}   & \colhead{Horn/Center}  & \colhead{$T_{\rm mb}$}   & \colhead{$\int T_{\rm mb}$d$V$}      &  \colhead{RMS}  &\colhead{Dv}                   &  \colhead{FWHM}      &  \colhead{Telescope}     & \colhead{Ref.}   & \colhead{Mark}   \\
\colhead{(MHz)}          &                                   & \colhead{(upper--lower)}               & \colhead{(km s$^{-1}$)}     & \colhead{}                      & \colhead{(mK)}                  & \colhead{(mK km s$^{-1}$)}               &  \colhead{(mK)}  &\colhead{(km s$^{-1}$)}  &  \colhead{(\arcsec)}   &   \colhead{}                    &                            &                     \\
\colhead{(1)}          &       \colhead{(2)}       & \colhead{(3)}               & \colhead{(4)}     & \colhead{(5)}                      & \colhead{(6)}                  & \colhead{(7)}               &  \colhead{(8)}  &\colhead{(9)}  &  \colhead{(10)}   &   \colhead{(11)}                    &     \colhead{(12)}     &        \colhead{(13)} 
} 
\startdata
113144.2      & CN                                 & $N$=1--0                                                                                    & ?                                 & ?                            & ?                  & 97000 $\pm$ 2000             & ?                   & ?              & 21.7 & IRAM-30m     & 1  & a  \\ 
113144.2      & CN                                 & $N$=1--0, $J$=1/2--1/2, $F$=1/2--3/2                                       & ?                                 & ?                            & 111.0            & 3575.9 $\pm$ 999.5           & 44.4             & ?              & 47.7 & TRAO-14m     & 2  &   \\ 
113170.5      & CN                                 & $N$=1--0, $J$=1/2--1/2, $F$=3/2--1/2                                       & ?                                 & ?                            & 266.6            & 5997.5 $\pm$ 733.0           & 44.4             & ?              & 47.7 & TRAO-14m     & 2  &   \\ 
113191.3      & CN                                 & $N$=1--0, $J$=1/2--1/2, $F$=3/2--3/2                                       & ?                                 & ?                            & 177.7            & 4020.9 $\pm$ 688.6           & 44.4             & ?              & 47.7 & TRAO-14m     & 2  &   \\ 
113488.1      & CN                                 & $N$=1--0, $J$=3/2--1/2, $F$=3/2--1/2                                       & ?                                 & ?                            & 400.4            & 9342.0 $\pm$ 311.4           & 66.7             & ?              & 47.5 & TRAO-14m     & 2  &  b  \\ 
113491.0      & CN                                 & $N$=1--0, $J$=3/2--1/2, $F$=5/2--3/2                                       & $\times$                     & $\times$                 & $\times$       & $\times$                             & $\times$       & $\times$  & 47.5 & TRAO-14m     & 2  &  b  \\
\enddata
\end{deluxetable*}
\begin{flushleft}
\textbf{Notes}.\\
\footnotesize{The temperature scale used in this table is given in term of main beam temperature. Columns 1, 2, and 3 give the rest frequency, the species, and the transition, respectively. 
Columns 4 and 5 give the expansion velocity and Horn-to-Center ratio of the the line profile, respectively, obtained from the stellar-shell model fitting. 
Columns 6 and 7 give the central temperature of the line profile and velocity-integrated intensity, respectively. 
Columns 8, 9, and 10 give the rms noise of the base line, the corresponding velocity resolution, and
the FWHM of each line, respectively. The telescopes used to perform the observations and the corresponding
references are listed in Columns 11 and 12, respectively.  
Column 13 provides comments for the details of lines. Boldface denotes newly detected lines in this work.
The rows are ordered by increasing carbon-atom number. 
} \\
\textbf{Symbol:}\\
\footnotesize{? means the values that can not be derived from our detection or are not listed in the archive data.}\\
\footnotesize{$\times$ means the values that can not be obtained because of unsolved hyperfine structures or two blended spin-rotation components.}\\ 
\footnotesize{: means uncertain measurements.}\\
\textbf{Reference:}\\
\footnotesize{
 (1) \cite{Bachiller97}; (2) \citet{Park08}; (3) \citet{Zhang13}; (4) \cite{Jaminet92}; (5) \cite{Bujarrabal88}; (6) \citet{Sopka89}; (7) this work; (8) \cite{Highberger01}; (9) \citet{Fukasaku94}; (10) \citet{Highberger03a}; (11) \citet{Milam08}; (12) \cite{Lucas86}; (13) \cite{Huggins84}; (14) \cite{Zhang20}; (15) \citet{Nguyen84};  (16) \cite{Truong88}; (17) \citet{Gupta09}; (18) \cite{Truong93}.  
}\\
\textbf{Mark:}\\
\tablenotetext{a}{Blended fine-structure group ($J=3/2 \rightarrow 1/2$).} 
\tablenotetext{b}{Unsolved hyperfine structure lines.}
\tablenotetext{c}{Blended fine-structure group ($J=5/2 \rightarrow 3/2$).}
\tablenotetext{d}{Blended fine-structure group ($J=7/2 \rightarrow 5/2$).}
\tablenotetext{e}{The CS $J=2 \rightarrow 1$ and $l$-C$_{3}$H $^{2}\Pi_{1/2}~ J=9/2 \rightarrow 7/2~ l=e~ F=5 \rightarrow 4$ and $F=4 \rightarrow 3$ lines are blended with each other. The CS $J=2 \rightarrow 1$ line is stronger.}
\tablenotetext{f}{The $^{13}$CS $J=2 \rightarrow 1$ and C$_{3}$S $J=16 \rightarrow 15$ lines are blended with each other. The former is stronger.}
\tablenotetext{g}{Blended line.}
\tablenotetext{h}{The PN $J=3 \rightarrow 2$ and $^{30}$SiC$_{2}$ $J_{K_a,K_c}=6_{2,4} \rightarrow 5_{2,3}$ lines are blended with each other.}
\tablenotetext{i}{The AlF $J=4 \rightarrow 3$ and SiC$_{4}$ $J=43 \rightarrow 42$ lines are blended with each other. The former is stronger.}
\tablenotetext{j}{The HC$_{3}$N $J=29 \rightarrow 28$ and AlF $J=8 \rightarrow 7$ lines are blended with each other. The former is stronger.}
\tablenotetext{k}{The NaCN $J_{K_a,K_c}=5_{0,5} \rightarrow 4_{0,4}$ and HC$_{7}$N $J=69 \rightarrow 68$ lines are blended with each other. The former is stronger.}
\tablenotetext{l}{The NaCN $J_{K_a,K_c}=10_{3,7} \rightarrow 9_{3,6}$ and U-line\,(156647 GHz) lines are blended with each other. The former is stronger.}
\tablenotetext{m}{Two blended spin-rotation components.}
\tablenotetext{n}{The MgNC $N=11-10$ and $l$-C$_{5}$H $^2\Pi_{1/2}~ J=65/2 \rightarrow 63/2~ l=e$ and $l=f$ lines are blended with each other. The MgNC $N=11 \rightarrow 10$ line is stronger.}
\tablenotetext{o}{The line intensity was obtained by integrating the main beam temperature over the velocity coverage from -80 to -20 km s$^{-1}$.}
\tablenotetext{p}{The MgNC $N=12 \rightarrow 11~ J=23/2 \rightarrow 21/2$  and $J=25/2 \rightarrow 23/2$ and Si$^{33}$S $J=8 \rightarrow 7$ lines are blended with each other. The MgNC $N=12 \rightarrow 11$ line is stronger.}
\tablenotetext{q}{The line intensity was obtained by integrating the main beam temperature over the velocity coverage from -110 to -20 km s$^{-1}$.}
\tablenotetext{r}{The SiC$_{2}$ $J_{Ka,Kc}=7_{4,4} \rightarrow 6_{4,3}$ and 7$_{4,3} \rightarrow 6_{4,2}$ and CH$_{3}$CN $J_{K}=9_{4} \rightarrow 8_{4}$ and $9_3 \rightarrow 8_{3}$ lines are blended with each other.}
\tablenotetext{s}{The C$_2$H $N=3 \rightarrow 2~ J=7/2 \rightarrow 5/2$ and SiC$_2$ $J_{K_a,K_c}=12_{0,12} \rightarrow 11_{0,11}$ lines are blended with each other. The former is stronger.}
\tablenotetext{t}{The line intensity was obtained by integrating the main beam temperature over the velocity coverage from -60 to -10 km s$^{-1}$.}
\tablenotetext{u}{The C$_{2}$S $N=7 \rightarrow 6~ J=8 \rightarrow 7$ and C$_{4}$H $\nu_{7}=1~ J=19/2 \rightarrow 17/2~ ^{2}\Pi_{1/2}~ l=f$ lines are blended with each other.}
\tablenotetext{v}{The HCN $J=2 \rightarrow 1$ and HCN $\nu_{2}=1~ J=2 \rightarrow 1~ l=e$ lines are blended with each other. The former is stronger.}
\tablenotetext{w}{The HCO$^{+}$ $J=2 \rightarrow 1$ and HC$_{5}$N $J=67-66$ lines are blended with each other. The former is stronger.}
\tablenotetext{x}{The line intensity was obtained by integrating the main beam temperature over the velocity coverage from -70 to 0 km s$^{-1}$.}
\tablenotetext{y}{Probably suffering from a contamination from the C$_{2}$N $^{2}\Pi_{1/2}~ J=23/2 \rightarrow 21/2~ l=e$ and $l=f$ line.}
\tablenotetext{z}{The line intensity was obtained by integrating the main beam temperature over the velocity coverage from -90 to 10 km s$^{-1}$.}
\tablenotetext{\alpha}{The HC$_3$H $J=12 \rightarrow 11$ and HC$_5$H $J=41 \rightarrow 40$ lines are blended with each other. The former is stronger.}
\tablenotetext{\beta}{Blended with the corresponding transition lines from HCC$^{13}$CN.}
\tablenotetext{\gamma}{The  HC$^{13}$CCN and HCC$^{13}$CN
lines arising from the same  $J$ transition lines are blended.}
\tablenotetext{\delta}{The C$_{4}$H $\nu_{7}=1~ ^{2}\Pi_{1/2}~ J=33/2 \rightarrow 31/2~ l=e$ and C$^{13}$CCCH $N=17 \rightarrow 16~ J=35/2 \rightarrow 33/2$ lines are blended with each other.}
\label{Table3}
\end{flushleft}
\end{longrotatetable}


\begin{longrotatetable}
\begin{deluxetable*}{ccccccccccccccc}
\rotate
\tablecaption{Excitation temperatures, column densities, and molecular abundances.}
\tablewidth{700pt}
\tabletypesize{\scriptsize}
\tablehead{
\colhead{Species}        & \colhead{$T_{\rm ex}$$^{a}$}      & \multicolumn{5}{c}{$N$ (cm$^{-2}$)}                                                                                                                                               &    & \multicolumn{2}{c}{$f_{\rm X}$}                                                  &  & \multicolumn{2}{c}{$f_{\rm X}/f_{\rm HC_3N}$}\\
                                                                    \cline{3-7}                                                                                                                                                                                                                             \cline{9-10}                                                                                \cline{12-13}
                                     &                             & \multicolumn{2}{c}{CRL\,2688}                                                                      &   &   \multicolumn{2}{c}{IRC+10216}                                               &     & CRL2688                              &  IRC+10216                      &  & CRL2688 & IRC+10216         \\
                                                                    \cline{3-4}                                                                                                               \cline{6-7}                                                                                                                               
                                     & (K)                       & Observed$^{b,c}$                                      & Modeled                                  &    & Observed$^{d}$                    &  Modeled                       &     & Observed$^{b}$                   & Observed$^{d}$                &   & Observed$^{b}$  & Observed
} 
\startdata 
NS                                & \nodata                & (3.64 $\pm$ 8.08) $\times$ 10$^{11}$        & 9.56 $\times$ 10$^{11}$        &    & \nodata                                  & \nodata                                     &     &$6.60\times10^{-9}$               & \nodata	                              &   & 0.019       &  \nodata    \\
PN                                & \nodata                & (5.10 $\pm$ 0.54) $\times$ 10$^{12}$        & 5.44 $\times$ 10$^{9}$          &    & 1 $\times$ 10$^{13}$            & 5.5 $\times$ 10$^{9}$              &     &$4.20\times10^{-9}$               & \nodata	                              &   & 0.012       &  \nodata    \\
SiS                               & 36.6 $\pm$ 2.4    & (1.62 $\pm$ 0.12) $\times$ 10$^{14}$        & 1.61 $\times$ 10$^{15}$        &    & \nodata                                  & \nodata                                     &     &$1.18\times10^{-7}$               & $1.05\times10^{-6}$          &   & 0.34         &  0.84         \\
$^{29}$SiS                   & \nodata                & (5.97 $\pm$ 1.38) $\times$ 10$^{12}$        & \nodata                                   &    & \nodata                                  & \nodata                                    &     &$7.26\times10^{-9}$               & \nodata	                              &   & 0.021       &  \nodata    \\
Si$^{34}$S                   & \nodata                & (6.23 $\pm$ 0.70) $\times$ 10$^{12}$        & \nodata                                   &    & \nodata                                  & \nodata                                    &     &$8.05\times10^{-9}$               & \nodata	                              &   & 0.023       &  \nodata    \\
SiO                               & 25.1                     & 7.11 $\times$ 10$^{12}$                             & 1.88 $\times$ 10$^{14}$         &    & \nodata                                  & \nodata                                    &     &$9.07\times10^{-9}$               & $1.30\times10^{-7}$         &    & 0.026       &  0.10          \\
CN                               & \nodata                & 4.10 $\times$ 10$^{14}$                             & 4.45 $\times$ 10$^{15}$         &    & 1.1 $\times$ 10$^{15}$          & 4.1 $\times$ 10$^{15}$           &     &$8.67\times10^{-6}$               & $2.18\times10^{-6}$         &    & 25.06        &  1.74         \\
$^{13}$CN                   & \nodata                & 2.38 $\times$ 10$^{13}$                             & \nodata                                    &    & \nodata                                  & \nodata                                    &     &$2.85\times10^{-6}$               & $7.60\times10^{-8}$         &    & 8.24          &  0.061       \\
CO                               & \nodata                & 3.45 $\times$ 10$^{17}$                             & 8.84 $\times$ 10$^{17}$         &    & \nodata                                  & \nodata                                    &     &\nodata                                   & \nodata	                             &    &  \nodata   &  \nodata    \\
$^{13}$CO                   & \nodata                & 2.45 $\times$ 10$^{16}$                             & \nodata                                    &    & \nodata                                  & \nodata                                    &     &\nodata                                   & \nodata	                             &    &  \nodata   &  \nodata    \\
C$^{17}$O                   & \nodata                & 2.47 $\times$ 10$^{16}$                             & \nodata                                    &    & \nodata                                  & \nodata                                    &     &\nodata                                   & \nodata	                             &    &  \nodata   &  \nodata    \\
C$^{18}$O                   & \nodata                & 2.18 $\times$ 10$^{16}$                             & \nodata                                    &    & \nodata                                  & \nodata                                    &     &\nodata                                   & \nodata	                             &    &  \nodata   &  \nodata    \\
CS                                & 13.3 $\pm$ 0.5    & (2.66 $\pm$ 0.07) $\times$ 10$^{14}$        & 7.70 $\times$ 10$^{14}$        &    & \nodata                                  & \nodata                                    &     &$1.64\times10^{-7}$               & $1.10\times10^{-6}$          &   & 0.47         &  0.88         \\
$^{13}$CS                    & 20.3 $\pm$ 4.5    & (1.40 $\pm$ 0.14) $\times$ 10$^{13}$        & \nodata                                   &    & \nodata                                  & \nodata                                    &     &$9.99\times10^{-9}$               & $2.20\times10^{-8}$          &   & 0.029       &  0.018       \\ 
C$^{33}$S                    & 12.2 $\pm$ 4.6    & (5.85 $\pm$ 1.39) $\times$ 10$^{12}$        & \nodata                                   &    & \nodata                                  & \nodata                                    &     &$3.29\times10^{-9}$               & \nodata	                              &   & 0.0095     & \nodata     \\
C$^{34}$S                    & 16.7 $\pm$ 2.7    & (2.76 $\pm$ 0.17) $\times$ 10$^{13}$        & \nodata                                   &    & \nodata                                  & \nodata                                    &     &$1.94\times10^{-8}$               & $3.70\times10^{-8}$          &   & 0.056       &  0.030       \\
HCN                             & 8.5 $\pm$ 0.1      & (5.51 $\pm$ 0.04) $\times$ 10$^{14}$        & 2.11 $\times$ 10$^{16}$        &    & \nodata                                  & \nodata                                    &     &$1.54\times10^{-7}$               & $1.40\times10^{-5}$           &   & 0.45         &  11.20       \\        
H$^{13}$CN                 & \nodata                & 1.52 $\times$ 10$^{14}$                             & \nodata                                    &    & \nodata                                 & \nodata                                    &     &$5.34\times10^{-7}$               & $3.10\times10^{-7}$          &   & 1.54         &  0.25         \\
HC$^{15}$N                 & \nodata                & 1.45 $\times$ 10$^{13}$                             & \nodata                                    &    & \nodata                                 & \nodata                                    &     &$2.91\times10^{-8}$               & \nodata	                             &    & 0.084       &  \nodata    \\
H$^{13}$C$^{15}$N     & \nodata                & (4.81 $\pm$ 1.09) $\times$ 10$^{11}$        & \nodata                                   &    & \nodata                                  & \nodata                                    &     &$4.25\times10^{-10}$             & \nodata                               &   & 0.0012     &  \nodata    \\
HNC                             & 9.9 $\pm$ 0.6      & (8.67 $\pm$ 0.68) $\times$ 10$^{13}$        & 5.69 $\times$ 10$^{13}$        &    & \nodata                                  & \nodata                                    &     &$3.01\times10^{-8}$               & $7.20\times10^{-8}$           &   & 0.087       &  0.058       \\        
HN$^{13}$C                 & \nodata                & 1.60 $\times$ 10$^{12}$                             & \nodata                                    &    & \nodata                                 & \nodata                                    &     &$4.79\times10^{-9}$               & $<$$1.90\times10^{-9}$    &   & 0.014       &  0.0015     \\
HCO$^{+}$                   & $<$7.1                & $>$3.44 $\times$ 10$^{12}$                        & 1.28 $\times$ 10$^{12}$        &    & 3--4 $\times$ 10$^{12}$        & 1.2 $\times$ 10$^{12}$           &     &$8.63:\times10^{-10}$            &  \nodata	                      &   & 0.0025     &  \nodata    \\        
H$_2$CO                     & \nodata                & 2.64 $\times$ 10$^{13}$                             & 1.66 $\times$ 10$^{11}$         &    & 5 $\times$ 10$^{12}$            & 1.5 $\times$ 10$^{11}$            &     &$3.05\times10^{-8}$               & \nodata	                             &    & 0.088       &  \nodata    \\
H$_{2}$CS                   & \nodata                & 1.63 $\times$ 10$^{13}$                             & 3.86 $\times$ 10$^{11}$         &    & 1 $\times$ 10$^{13}$            & 3.5 $\times$ 10$^{11}$            &     &$1.87\times10^{-8}$               & \nodata	                             &    & 0.054       &  \nodata    \\
SiC$_2$                       & 32.4 $\pm$ 0.7    & (1.45 $\pm$ 0.03) $\times$ 10$^{14}$        & 2.71 $\times$ 10$^{14}$        &    & 2 $\times$ 10$^{14}$             & 1.3 $\times$ 10$^{15}$           &     &$1.05\times10^{-7}$               & $2.40\times10^{-7}$           &   & 0.30         &  0.19         \\
$^{29}$SiC$_2$           & \nodata                & (2.67 $\pm$ 1.26) $\times$ 10$^{13}$        & \nodata                                   &    & \nodata                                  & \nodata                                    &     &$3.74\times10^{-8}$               & \nodata	                              &   & 0.11         &  \nodata     \\
Si$^{13}$CC                & \nodata                & (6.23 $\pm$ 0.58) $\times$ 10$^{12}$        & \nodata                                   &    & \nodata                                  & \nodata                                    &     &$8.80\times10^{-9}$               & \nodata	                              &   & 0.025       &  \nodata    \\
C$_{2}$H                     & 6.6                       & 3.10 $\times$ 10$^{15}$                             & 1.00 $\times$ 10$^{16}$        &    & 3--5 $\times$ 10$^{15}$       & 9.7 $\times$ 10$^{15}$            &     &$1.22\times10^{-6}$               & $2.80\times10^{-6}$         &    & 3.53         &  2.24          \\
$^{13}$CCH                 & \nodata                & (2.85 $\pm$ 0.57) $\times$ 10$^{13}$        & \nodata                                   &    & \nodata                                  & \nodata                                    &     &$3.68\times10^{-8}$               & \nodata	                              &   & 0.11         &  \nodata    \\
C$_2$S                        & 18.7 $\pm$ 19.5  & (9.91 $\pm$ 10.17) $\times$ 10$^{12}$      & 1.42 $\times$ 10$^{13}$        &    & 9--15 $\times$ 10$^{13}$      & 1.4 $\times$ 10$^{13}$           &     &$7.85\times10^{-9}$               & $<4.35\times10^{-9}$         &   &  0.023      &  0.0034     \\
C$_3$N                        & 15.7 $\pm$ 0.1    & (2.77 $\pm$ 0.03) $\times$ 10$^{14}$        & 5.70 $\times$ 10$^{14}$        &    & 2--4 $\times$ 10$^{14}$        & 5.1 $\times$ 10$^{14}$           &     &$1.79\times10^{-7}$               & $4.45\times10^{-7}$           &   & 0.52         &  0.36         \\
c-C$_3$H                     & \nodata                & (2.82 $\pm$ 0.32) $\times$ 10$^{13}$        & \nodata                                   &    & \nodata                                  & \nodata                                    &     &$3.61\times10^{-8}$               & \nodata	                              &   & 0.10         &  \nodata     \\
l-C$_3$H                      & 48.0 $\pm$ 4.5    & (9.49 $\pm$ 0.25) $\times$ 10$^{13}$        & 1.38 $\times$ 10$^{14}$        &    & 3--7 $\times$ 10$^{13}$        & 1.4 $\times$ 10$^{14}$            &     &$7.44\times10^{-8}$               & $5.50\times10^{-8}$          &   & 0.22         &  0.044       \\
C$_3$S                        & \nodata                & (7.83 $\pm$ 0.77) $\times$ 10$^{12}$        & 1.23 $\times$ 10$^{13}$        &    & 6--11 $\times$ 10$^{13}$       & 1.3 $\times$ 10$^{13}$           &     &$5.92\times10^{-9}$               & $<$$1.20\times10^{-8}$    &   & 0.017       &  0.0096      \\
C$_4$H                        & 20.2 $\pm$ 0.1    & (2.64 $\pm$ 0.02) $\times$ 10$^{15}$        & 7.95 $\times$ 10$^{14}$        &    & 2--9 $\times$ 10$^{15}$        & 6.5 $\times$ 10$^{14}$            &     &$1.38\times10^{-6}$               & $3.20\times10^{-7}$          &   & 3.99         &  0.26         \\
C$_4$H, $\nu_7=1$     & 24.2 $\pm$ 0.9    & (5.39 $\pm$ 1.86) $\times$ 10$^{17}$        & \nodata                                   &    & \nodata                                  & \nodata                                    &     &$7.74\times10^{-4}$               & \nodata	                      &   & 2236.99   &  1.00         \\
C$^{13}$CCCH            & \nodata                & $<$3.67 $\times$ 10$^{13}$                       & \nodata                                   &    & \nodata                                  & \nodata                                     &     &$<$$1.30\times10^{-7}$        & \nodata	                               &   & 0.38        &  \nodata    \\
HC$_3$N                     & 41.1 $\pm$ 0.3    & (1.54 $\pm$ 0.02) $\times$ 10$^{14}$        & 3.67 $\times$ 10$^{14}$        &    & 1--2 $\times$ 10$^{15}$        & 4.8 $\times$ 10$^{14}$            &     &$1.15\times10^{-7}$               & $1.25\times10^{-6}$           &   & 1.00        &  1.00         \\
HC$_3$N, $\nu_7=1$  & 66.5 $\pm$ 3.8    & (1.47 $\pm$ 0.49) $\times$ 10$^{15}$        & \nodata                                   &    & \nodata                                  & \nodata                                    &     &$1.04\times10^{-6}$                & \nodata	                      &   & 3.01         &  \nodata    \\
H$^{13}$CCCN            & 27.3 $\pm$ 0.5    & (8.72 $\pm$ 0.27) $\times$ 10$^{12}$        & \nodata                                   &    & \nodata                                  & \nodata                                    &     &$6.82\times10^{-9}$                & \nodata	                      &   & 0.060       &  \nodata    \\
HC$^{13}$CCN            & 28.1 $\pm$ 0.3    & (8.46 $\pm$ 0.08) $\times$ 10$^{12}$        & \nodata                                   &    & \nodata                                  & \nodata                                    &     &$7.61\times10^{-9}$                & $3.25\times10^{-8}$          &   & 0.066      &  0.026        \\
HCC$^{13}$CN            & 28.7 $\pm$ 0.3    & (8.18 $\pm$ 0.12) $\times$ 10$^{12}$        & \nodata                                   &    & \nodata                                  & \nodata                                    &     &$6.82\times10^{-9}$                & $3.25\times10^{-8}$          &   & 0.059      &  0.026        \\
c-C$_3$H$_{2}$           & 15.0 $\pm$ 1.9    & (4.21 $\pm$ 0.46) $\times$ 10$^{13}$        & 6.11 $\times$ 10$^{13}$        &    & 2 $\times$ 10$^{13}$            & 5.8 $\times$ 10$^{13}$            &     &$1.12\times10^{-7}$               & $3.25\times10^{-8}$          &   & 0.32        &  0.026        \\
C$_5$H                        & \nodata                & 1.32 $\times$ 10$^{15}$                             & 4.26 $\times$ 10$^{13}$         &    & 2--50 $\times$ 10$^{13}$      & 4 $\times$ 10$^{13}$              &     &$2.47\times10^{-7}$                & \nodata	                      &   & 0.71        &  \nodata    \\
C$_6$H                        & 42.6 $\pm$ 3.6    & (2.74 $\pm$ 0.40) $\times$ 10$^{13}$        & 5.96 $\times$ 10$^{14}$        &    & 7 $\times$ 10$^{13}$             & 5.7 $\times$ 10$^{14}$           &     &$2.21\times10^{-8}$                & \nodata	                      &   & 0.064      & \nodata     \\
CH$_3$CN                   & 30.6 $\pm$ 1.8    & (5.41 $\pm$ 0.35) $\times$ 10$^{13}$        & 5.43 $\times$ 10$^{12}$        &    & 6--30 $\times$ 10$^{12}$      & 5.4 $\times$ 10$^{12}$           &     &$4.75\times10^{-8}$                & $6.60\times10^{-8}$          &   & 0.14        &  0.053        \\
HC$_5$N                     & 36.6 $\pm$ 0.3    & (5.22 $\pm$ 0.90) $\times$ 10$^{13}$        & 1.15 $\times$ 10$^{14}$        &    & 2--3 $\times$ 10$^{14}$         & 1.4 $\times$ 10$^{14}$           &     &$4.42\times10^{-8}$                & $7.03\times10^{-6}$          &   & 1.19        & 5.62          \\
HC$_7$N                     & 73.9 $\pm$ 28.3  & (6.56 $\pm$ 4.88) $\times$ 10$^{12}$        & 3.63 $\times$ 10$^{13}$        &    & 1 $\times$ 10$^{14}$             & 4.7 $\times$ 10$^{13}$           &     &$4.40\times10^{-9}$                & $1.20\times10^{-9}$           &   &  0.038     & \nodata     \\
\enddata
\begin{flushleft}
\tablenotetext{a}{
If the rotation diagram cannot be constructed for a given molecule,
a constant excitation temperature of 40 K is assumed for
the calculations of its column density and abundance.
 }
\tablenotetext{b}{For the species with optically thick emission, 
the lower limits are given.}
\tablenotetext{c}{The values obtained by \cite{Zhang13} are complemented here.}
\tablenotetext{d}{Taken from \citet{Woods03b}.}
\label{Table4}
\end{flushleft}
\end{deluxetable*}
\end{longrotatetable}

\clearpage

\begin{deluxetable}{ccccccccc}
\tablecaption{The fractional abundances of selected molecules derived
using the LIME radiative transfer code.
\label{Table5}}
\tabletypesize{\small}
\tablewidth{0pt}
\tablehead{
\colhead{Species}        & \colhead{Transition}                & \multicolumn{2}{c}{$R_i$}               & & \multicolumn{2}{c}{$R_e$}                & \colhead{$f_{\rm LIME}$}   & \colhead{$f_{\rm LIME}$/$f_{\rm X}$}\\
\cline{3-4} \cline{6-7} 
\colhead{}                     & \colhead{}                               & \colhead{(cm)}     & \colhead{(\arcsec)}     &  & \colhead{(cm)}     & \colhead{(\arcsec)}       & 
}     
\startdata
CS                                 & $2\rightarrow1$                      &  $2.0 \times 10^{15}$             & 0.1          &    &  $4.5 \times 10^{17}$        & 30.0               & $1.2 \times 10^{-7}$                &  0.7       \\ 
SiS                                & $10\rightarrow9$                    &  $0.8 \times 10^{17}$             & 0.5          &    &  $4.5 \times 10^{17}$        & 30.0               & $5.0 \times 10^{-7}$                &  4.2       \\
PN                                 & $2\rightarrow1$                      &  $1.0 \times 10^{16}$             & 0.7          &    &  $4.5 \times 10^{16}$        & 3.0                 & $2.5 \times 10^{-9}$                &  0.6       \\ 
SiC$_{2}$                      & $4_{0,4}\rightarrow3_{0,3}$   &  $1.7 \times 10^{17}$             & 11.3        &    &  $4.5 \times 10^{17}$        & 30.0               & $5.0 \times 10^{-6}$                &  1.0       \\
HCN                              & $2\rightarrow1$                      &  $1.0 \times 10^{16}$             & 0.7          &    &  $1.5 \times 10^{17}$        & 10.0               & $7.1 \times 10^{-7}$                &  4.6       \\
HNC                              & $2\rightarrow1$                      &  $1.0 \times 10^{16}$             & 0.7          &    &  $4.5 \times 10^{17}$        & 30.0               & $5.0 \times 10^{-8}$                &  1.7       \\
HCO$^{+}$                    & $2\rightarrow1$                      &  $1.5 \times 10^{17}$             & 10.0        &    &  $4.5 \times 10^{17}$        & 30.0               & $2.0 \times 10^{-9}$                &  2.3       \\
$c$-C$_{3}$H$_{2}$     & $2_{0,2}\rightarrow1_{1,1}$   &  $2.5 \times 10^{17}$             & 16.7        &    &  $4.5 \times 10^{17}$        & 30.0               &  $2.0 \times 10^{-8}$               &  0.2       \\
CH$_{3}$CN                 & $5_{3}\rightarrow4_{3}$         &  $0.8 \times 10^{17}$             & 5.3          &    &  $4.5 \times 10^{17}$        & 30.0               &  $5.0 \times 10^{-8}$                &  1.1       \\
\hline
\enddata
\end{deluxetable}

\clearpage

\begin{deluxetable}{lccccc}
\tablecaption{Isotopic abundance ratios.
\label{Table6}}
\tabletypesize{\small}
\tablewidth{0pt}
\tablehead{
\colhead{Isotopic ratio}                      & \multicolumn{3}{c}{AFGL\,2688}                                                         & \colhead{IRC+10216$^a$} & \colhead{Solar$^b$}\\
\cline{2-4} 
                                                          & \colhead{Species}                                          & Transition             & \colhead{Value}
}
\startdata
$^{12}$C/$^{13}$C                            & $^{12}$CS/$^{13}$CS                                    &  $2 \rightarrow 1$             & 30.4 $\pm$ 1.7                  &                               &                  \\
                                                          & H$^{12}$C$_3$N/H$^{13}$CCCN                 &  $9 \rightarrow 8$             & 17.2 $\pm$ 0.7                  & 45$\pm3$              & 89.3          \\ 
                                                          & H$^{12}$C$_3$N/HC$^{13}$CCN                 &  $9 \rightarrow 8$             & 20.3 $\pm$ 0.7                  &                               &                  \\
                                                          & H$^{12}$C$_3$N/HCC$^{13}$CN                 &  $9 \rightarrow 8$             & 15.4 $\pm$ 0.5                  &                               &                  \\
                                                          & H$^{12}$C$_3$N/HC$^{13}$CCN                 &  $18 \rightarrow 17$         & 28.9 $\pm$ 0.5                  &                               &                  \\
                                                          & H$^{12}$C$_3$N/HCC$^{13}$CN                 &  $18 \rightarrow 17$         & 28.9 $\pm$ 0.5                  &                               &                  \\
                                                          & H$^{12}$C$_3$N/HC$^{13}$CCN                 &  $20 \rightarrow 19$         & 37.6 $\pm$ 2.4                  &                               &                  \\
                                                          & H$^{12}$C$_3$N/HCC$^{13}$CN                 &  $20 \rightarrow 19$         & 37.6 $\pm$ 2.4                  &                               &                  \\
                                                          & H$^{12}$C$_5$N/H$^{13}$CCCCCN            &  $30 \rightarrow 29$         & 23.9 $\pm$ 7.6                  &                               &                  \\
$^{14}$N/$^{15}$N                           & H$^{12}$C$^{14}$N/H$^{13}$C$^{15}$N       &  $2 \rightarrow 1$             & 32.5 $\pm$ 4.5$^c$          & $>$4400               & 272.0         \\
$^{32}$S/$^{34}$S                            & C$^{32}$S/C$^{34}$S                                    &  $2 \rightarrow 1$             & 9.5 $\pm$ 0.2                    & 21.8 $\pm$2.6       & 22.5           \\
                                                          & Si$^{32}$S/Si$^{34}$S                                   &  $10 \rightarrow 9$           & 14.1 $\pm$ 1.6                  &                               &                  \\
$^{33}$S/$^{34}$S                            & C$^{33}$S/C$^{34}$S                                    &  $2 \rightarrow 1$             & 0.2 $\pm$ 0.02                  & 0.18 $\pm$0.1       & 0.18           \\
$^{28}$Si/$^{29}$Si                          & $^{28}$SiC$_{2}$/$^{29}$SiC$_{2}$              &  $7 \rightarrow 6$             & 26.8 $\pm$ 3.9                  & $>$15.4                 & 19.7          \\
                                                          & $^{28}$SiS/$^{29}$SiS                                   &  $10 \rightarrow 9$           & 17.1 $\pm$ 2.7                  &                               &                   \\
$^{24}$Mg/$^{26}$Mg                      & $^{24}$MgNC/$^{26}$MgNC                          &  $8 \rightarrow 7$             & 3.8 $\pm$ 1.9                    & 6.4 $\pm$ 0.7        & 7.2            \\
\hline
\enddata
\tablenotetext{a}{Taken from \citet{Kahane00}.}
\tablenotetext{b}{Taken from \citet{Lodders03}.}
\tablenotetext{c}{Assuming the $^{12}$C/$^{13}$C ratio of 30.}
\end{deluxetable}

\clearpage


\begin{deluxetable}{ccccccccc}
\tablecaption{Results of the rotation-diagram analysis of metal-bearing molecules.
\label{Table7}}
\tabletypesize{\small}
\tablewidth{0pt}
\tablehead{\\
                          &                                & \multicolumn{2}{c}{\citet{Highberger01}}                          &     & \multicolumn{2}{c}{This work$^a$}                                   &                                                          \\
                          \\
                                                              \cline{3-4}                                       \cline{6-7}                                                                                                                          
\colhead{Source size}   & \colhead{Molecule}   & \colhead{$N$}    & \colhead{$T_{\rm ex}$}  &     & \colhead{$N$}   & \colhead{$T_{\rm ex}$}            \\
\colhead{(\arcsec)}       &                      & \colhead{(cm$^{-2}$)}    & \colhead{(K)}           &         & \colhead{(cm$^{-2}$)}   & \colhead{(K)}               
}
\startdata
MB$^b$              & AlF                         & $3.8 \times 10^{12}$      & 79                                     &     & $4.0(1.2) \times 10^{12}$    & 81.5(58.0)                   \\ 
                           & MgNC                    & $7.0 \times 10^{11}$      & 17                                     &     & $2.9(0.1) \times 10^{12}$    & 21.0(0.7)                     \\
                           & NaCN                    & $5.5 \times 10^{12}$      & 88                                     &     & $1.6(0.1) \times 10^{12}$    & 34.6(2.8)                     \\
30                       & AlF                         &  \nodata                          & \nodata                             &     & $4.0(1.2) \times 10^{12}$    & 59.1(30.5)                  \\ 
                           & MgNC                    & $3.7 \times 10^{12}$      & 13                                     &     & $5.0(0.3) \times 10^{12}$    & 17.9(0.5)                    \\
                           & NaCN                    & $1.2 \times 10^{13}$      & 61                                      &     & $2.1(0.2) \times 10^{12}$    & 24.1(1.4)                    \\
20                       & AlF                         & $4.5 \times 10^{12}$      & 42                                     &     & $4.5(1.4) \times 10^{12}$    & 47.6(19.7)                  \\  
                           & MgNC                    & $7.8 \times 10^{12}$      & 12                                     &     & $7.8(0.4) \times 10^{12}$    & 16.4(0.4)                     \\
                           & NaCN                    & $2.1 \times 10^{13}$      & 57                                      &     & $2.9(0.3) \times 10^{12}$    & 20.6(1.0)                     \\
10                       & AlF                         & $8.8 \times 10^{12}$      & 28                                     &     & $8.1(2.4) \times 10^{12}$    & 32.0(8.9)                     \\  
                           & MgNC                    & $3.0 \times 10^{13}$      & 12                                     &     & $2.4(1.2) \times 10^{13}$    & 14.4(0.3)                     \\
                           & NaCN                    & $6.9 \times 10^{13}$      & 54                                      &     & $7.7(0.7) \times 10^{12}$    & 16.8(0.7)                      \\
5                         & AlF                         & $2.7 \times 10^{13}$      & 23                                     &     & $2.4(0.7) \times 10^{13}$    & 25.2(5.6)                      \\  
                           & MgNC                    & $1.2 \times 10^{14}$      & 12                                     &     & $8.7(0.4) \times 10^{13}$    & 13.6(0.3)                      \\
                           & NaCN                    & $2.6 \times 10^{14}$      & 53                                     &     & $2.7(0.2) \times 10^{13}$    & 15.4(0.6)                       \\
\hline
\enddata
\tablenotetext{a}{Values in brackets represent the uncertainties.}
\tablenotetext{b}{The main-beam size is assumed.}
\end{deluxetable}


\begin{deluxetable}{lcclcr}
\tablecaption{The fractional abundance of metal-bearing molecules relative to H$_2$
 in CRL\,2688 and IRC + 10216.
\label{Table8}}
\tabletypesize{\small}
\tablewidth{0pt}
\tablehead{ 
 \colhead{Object}     & \colhead{Distance}   & \colhead{Mass-loss rate}                                          & \colhead{Molecule}       & \colhead{Source Size}                    & \colhead{${f_{\rm X}}^a$}                     \\
 \colhead{     }          & \colhead{(kpc)}         & \colhead{($M_{\odot}$\,yr$^{-1}$)}                            & \colhead{}                     & \colhead{(\arcsec)}                         & \colhead{}                               
}
\startdata
CRL\,2688$^{b}$   & 1                                & $1.7 \times 10^{-4}$                                                   & AlF                                & 5                                                     & $5.4 \times 10^{-9}$                             \\ 
                              &                                   &                                                                                    &                                      & 10                                                   & $3.5 \times 10^{-9}$                             \\ 
                              &                                   &                                                                                    & MgNC                           & 10--30                                             & $4.1 \times 10^{-9}$                             \\ 
                              &                                   &                                                                                    & NaCN                           & 5                                                     & $5.2 \times 10^{-8}$                             \\ 
                              &                                   &                                                                                    &                                      & 10                                                   & $2.7 \times 10^{-8}$                             \\ 
CRL\,2688$^{c}$  & 1                                 & $1.7 \times 10^{-4}$                                                   & NaCN                           & 11--25                                             & $5.2 \times 10^{-9}$                             \\ 
CRL\,2688$^{d}$   & 1                               & $3.0 \times 10^{-5}$                                                   & AlF                                & 5                                                     & $2.7(0.8) \times 10^{-8}$                      \\ 
                             &                                   &                                                                                    &                                      & 10                                                    & $1.8(0.5) \times 10^{-8}$                      \\ 
                             &                                   &                                                                                    & MgNC                           & 10--30                                             & $4.0(0.6) \times 10^{-8}$                       \\ 
                             &                                   &                                                                                    & NaCN                           & 11--25                                              & $1.2(0.1) \times 10^{-8}$                       \\ 
IRC+10216$^{b}$ & 0.15                           & $3.0 \times 10^{-5}$                                                   & AlF                                & 5                                                     & $1.5 \times 10^{-7}$                               \\ 
                             &                                   &                                                                                    &                                      & 10                                                    & $8.4 \times 10^{-8}$                               \\ 
                             &                                   &                                                                                    & MgNC                           & 20--40                                             & $4.8 \times 10^{-8}$                               \\ 
                             &                                   &                                                                                    & NaCN                            & 20--40                                             & $1.3 \times 10^{-8}$                               \\ 
\enddata
\tablenotetext{a}{Values in brackets represent the uncertainties.}
\tablenotetext{b}{Taken from \cite{Highberger01}.}
\tablenotetext{c}{Taken from \cite{Highberger03a}.}
\tablenotetext{d}{This work.}
\end{deluxetable}

\end{CJK*}
\end{document}